\PassOptionsToPackage{numbers,sort&compress}{natbib}
\documentclass[preprintnumbers,amsmath,amssymb,prd, notitlepage,nofootinbib,twocolumn, superscriptaddress]{revtex4}

\usepackage{amsfonts,amssymb,amsmath}
\usepackage{gensymb}
\usepackage{color}
\usepackage{graphicx}
\usepackage{multirow}
\usepackage[utf8]{inputenc}
\usepackage{aas_macros}
\usepackage{hyperref}

\graphicspath{{./Images/}}
\usepackage[normalem]{ulem}

\newcommand{\greaterthanapprox}{\mathrel{\vcenter{
  \offinterlineskip\halign{\hfil$##$\cr
    >\cr\noalign{\kern2pt}\sim\cr\noalign{\kern-2pt}}}}}
    
    \newcommand{\lessthanapprox}{\mathrel{\vcenter{
  \offinterlineskip\halign{\hfil$##$\cr
    <\cr\noalign{\kern2pt}\sim\cr\noalign{\kern-2pt}}}}}
    
\newcommand{\lb}{\left(}
\newcommand{\rb}{\right)}

\newcommand{\Planck}{{\it Planck}~}

\newcommand{\be}{\begin{equation}}        
\newcommand{\ee}{\end{equation}}

\newcommand{\fnlocal}{f_{\rm NL}^{\mathrm{local}}}
\newcommand{\fnl}{f_{\rm NL}}

\newcommand{\fnlconfidenceinterval}{$-87<\fnl<19$}

\begin{document}

\title{Constraints on primordial non-Gaussianity from halo bias measured through CMB lensing cross-correlations}

\author{Fiona McCarthy}
\email{fmccarthy@flatironinstitute.org}

\affiliation{Center for Computational Astrophysics, Flatiron Institute, New York, NY, USA 10010}
\affiliation{Perimeter Institute for Theoretical Physics, Waterloo, Ontario N2L 2Y5, Canada}
\affiliation{Department of Physics and Astronomy, University of Waterloo, Waterloo, Ontario, N2L 3G1, Canada}

\author{Mathew S.~Madhavacheril}
\affiliation{Department of Physics and Astronomy, University of Pennsylvania, Philadelphia, PA 19104, USA}
\affiliation{Perimeter Institute for Theoretical Physics, Waterloo, Ontario N2L 2Y5, Canada}

\author{Abhishek S.~Maniyar}
\affiliation{Center for Cosmology and Particle Physics, Department of Physics,
New York University, New York, NY 10003, USA}

\date{\today}

\begin{abstract}
Local non-Gaussianities in the initial conditions of the Universe, parameterized by $f_{\rm NL}$, induce a scale-dependence in the large-scale bias of halos in the late Universe. This effect is a promising path to constrain multi-field inflation theories that predict non-zero $f_{\rm NL}$. While most existing constraints from the halo bias involve auto-correlations of the galaxy distribution, cross-correlations with probes of the matter density provide an alternative channel with fewer systematics. We present the  strongest large-scale structure constraint on local primordial non-Gaussianity that utilizes cross-correlations alone. We use the cosmic infrared background (CIB) consisting of dusty galaxies as a halo tracer and cosmic microwave background (CMB) lensing as a probe of the underlying matter distribution, both from \Planck data. Milky Way dust is one of the key challenges in using the large-scale modes of the CIB. Importantly, the cross-correlation of the CIB with CMB lensing is far less affected by Galactic dust compared to the auto-spectrum of the CIB, since the latter picks up an additive bias from Galactic dust. We find no evidence for primordial non-Gaussianity and obtain  \fnlconfidenceinterval, with a Gaussian $\sigma(f_{\rm NL})\approx 41$, assuming universality of the halo mass function. We find that future CMB lensing data from Simons Observatory and CMB-S4 could achieve $\sigma(f_{\rm NL})$ of 23 and 20 respectively. The constraining power of such an analysis is limited by current Galactic dust cleaning techniques which introduce a multiplicative bias on very large scales, requiring us to choose a minimum multipole of $\ell=70$. If this challenge is overcome with improved analysis techniques or external data, constraints as tight as $\sigma(f_{\rm NL})=4$ can be achieved through the cross-correlation technique. More optimistically, constraints better than $\sigma(f_{\rm NL})=2$ could be achieved if the CIB auto-spectrum is dust-free down to the largest scales.
 \end{abstract}
\maketitle

\section{Introduction}

The search for non-Gaussianities in the initial conditions of the Universe (``primordial non-Gaussianities'', or PNG) is a key goal of the cosmology community. Of particular interest is primordial non-Gaussianity of the local type, parameterized by $\fnlocal$, with $\fnlocal=0$ indicating exact Gaussianity. Multi-field inflation models predict $\fnlocal$ of $\mathcal{O}(1)$ (e.g. \cite{2017PhRvD..95l3507D}), and so a detection of $\fnlocal$ will be key for discriminating between inflation models. To date, all measurements are consistent with Gaussian initial conditions, with the strongest constraint coming from measurements of the early-universe bispectrum (or three-point function) through the cosmic microwave background (CMB) as measured by \textit{Planck}: $\fnlocal=-0.9 \pm 5.1$~\cite{2020A&A...641A...9P}. This constraint is not expected to improve enough to probe multi-field inflation with future measurements of the primary CMB fluctuations (e.g. up to $\sigma(\fnlocal)=2$ with the Simons Observatory\cite{Ade:2018sbj}).

 The late-universe large-scale structure (LSS) bispectrum is perhaps the next obvious probe of non-Gaussianities; although, as gravitational evolution induces non-Gaussianities in an initially non-Gaussian field, these must first be disentangled before constraining the primordial Universe from a measurement of the bispectrum of LSS~\cite{1011.1513,2010.04567,2206.01619,2206.15450}. However, there exists a well-known signature of $\fnlocal$ (henceforth $\fnl$) in the two-point {power spectrum} of biased objects such as halos. In particular, non-zero  $\fnl$ induces a scale-dependence in the bias of these objects with respect to dark matter, a signal that is strongest on the largest scales~\cite{0710.4560}: 
 \be
 \Delta b\sim \frac{\fnl}{k^2}(b^G-1),
 \ee
  where $b^G$ is the Gaussian bias (which is scale-independent on large scales), and $\Delta b$ is the change in bias induced by $\fnl$. Constraints from the bias of quasars in the SDSS/BOSS surveys~\cite{2008JCAP...08..031S,2013MNRAS.428.1116R,2014MNRAS.441L..16G,2014PhRvD..89b3511G,2014PhRvL.113v1301L,2022MNRAS.514.3396M} have used this signal to constrain $\fnl$, with the strongest finding $\fnl=-12 \pm 21$~\cite{2022MNRAS.514.3396M}, and recent combined constraints from the BOSS bispectrum and power spectrum in fact get most of their constraining power on $\fnl$ from the effect on the power spectrum ~\cite{2022arXiv220111518D,2022PhRvD.106d3506C}. Forecasts have indicated that future LSS surveys such as Rubin Observatory's  Legacy Survey of Space and Time (LSST)~\cite{lsst} and SPHEREx~\cite{2014arXiv1412.4872D}, a high-number density galaxy clustering survey, will be able to reach the  $\sigma(\fnl)\sim1$ regime if systematics are well-controlled.

Many of the aforementioned LSS constraints on $\fnl$ involve multiple powers of the halo overdensity field (two in the power spectrum, three in the bispectrum). On the other hand, constraints from cross-correlations with probes of the unbiased matter distribution---like those of~\cite{2014MNRAS.441L..16G,2014PhRvD..89b3511G}---offer advantages: (1) an analysis involving a cross-correlation of one power of the halo overdensity typically does not suffer from additive systematics in measurements of the LSS survey (e.g. selection effects and Milky Way dust); and (2) a joint analysis of all cross- and auto-spectra can significantly improve the bias measurement through sample variance cancellation~\cite{Seljak:2008xr}. Such measurements have been proposed using unbiased tracers of mass such as CMB lensing convergence maps~\cite{Schmittfull:2017ffw} or velocity such as the kinetic Sunyaev--Zel'dovich (kSZ) effect~\cite{meyer:2018eey}; ~\cite{2014MNRAS.441L..16G,2014PhRvD..89b3511G} use the integrated Sachs Wolfe (ISW) effect as well as CMB lensing.

In this work, we present the strongest constraint on $\fnl$ through cross-correlation alone, the previous strongest being $\fnl=46\pm68$ from the cross correlation of the ISW effect and galaxies~\cite{2014MNRAS.441L..16G}. We use (1) the cosmic infrared background (CIB) as our halo tracer and (2) weak lensing of the CMB as our probe of the unbiased matter distribution.
\begin{enumerate}

\item The CIB is sourced by the thermal radiation of dust grains in distant galaxies; these dust grains absorb ultraviolet (UV) starlight, which heats them up and is re-emitted in the infrared (IR). The star formation rate (SFR) of our Universe peaked at around $z\sim2$~\cite{2014ARA&A..52..415M}, and the CIB is thus sourced from galaxies at around this redshift and higher, although it is a diffuse field with contributions from all redshifts up to reionization at $z\sim7$. The CIB anisotropies that we measure trace the clustering of these objects~\cite{2001ApJ...550....7K}. For this reason, it might be considered a promising candidate for constraining $\fnl$: the $\fnl$ signal increases with bias, and galaxies at high redshift such as those sourcing the CIB are more highly biased than galaxies at lower redshifts. As well as this, it is highly correlated with the CMB lensing convergence field $\kappa$, giving a potential opportunity to improve the $\fnl$ measurement by using a simultaneous measurement  of $\kappa$ and the CIB intensity to exploit sample variance cancellation.
\item The CMB lensing convergence field is a map of all the matter between us and the surface of last scattering, projected along the line of sight~\cite{1987A&A...184....1B}. As the CMB has been traveling through the Universe, it has interacted gravitationally with this matter in a well-understood way~\cite{Lewis:2006fu}. The result is that the CMB we see has been weakly lensed, an effect which can be detected statistically, and has been done with high statistical significance by the \textit{Planck} satellite~\cite{1303.5077,1502.01591,1807.06210,2206.07773} and  high-resolution ground-based CMB experiments such as the Atacama Cosmology Telescope (ACT)~(e.g. \cite{Darwish:2020fwf,Sherwin:2016tyf,2011PhRvL.107b1301D}) and the South Pole Telescope (SPT)~(e.g. \cite{2021ApJ...922..259M,2019ApJ...884...70W,2017ApJ...849..124O,2015ApJ...810...50S,2012ApJ...756..142V}).
\end{enumerate}

Previous work has shown that the information contained in the auto-power spectrum of the CIB anisotropies could in principle yield a measurement with $\sigma (\fnl)<1$~\cite{1606.02323}. However, as indicated earlier, there are significant difficulties associated with using auto-spectra for $\fnl$ measurements, and this is especially true for the CIB. The  signal of interest is  mostly sourced at large scales, where it is difficult to separate the cosmological CIB signal from the emission from dust in our own Milky Way galaxy. The Galactic dust signal is also scale-dependent with significant power on large scales; even in maps post-processed through component separation or foreground cleaning techniques, any spurious dust power will bias the inference of $\fnl$. For this reason, we do not use the large-scale CIB auto-power spectrum\footnote{We use ``CIB auto-power spectrum'' to refer to the cross-power spectra between the different frequency channels at which the CIB is measured. As described later, we do include small-scale CIB auto-spectra to help constrain the CIB model itself.} in this work and instead focus on constraining $\fnl$ from its cross-power spectrum with the CMB lensing convergence field $C_\ell^{\nu\kappa}$  alone, as this statistic does not suffer from the same additive dust bias. There is however a multiplicative bias associated with the dust cleaning procedure that prevents us from accessing all scales \cite{1905.00426}; this is discussed later in this work.

The paper is organized as follows. In Section~\ref{sec:theory} we discuss the relevant theory, including the scale-dependence induced in the bias by $\fnl$, and the formalism we use to model the CIB and the the CIB-CMB-lensing cross-correlation. In Section~\ref{sec:data} we discuss the data products used in our analysis and in Section~\ref{sec:pipeline} we present our pipeline for the extraction of $\fnl$. We present our results in Section~\ref{sec:results}. In Section~\ref{sec:future_constraints} we forecast constraints from future CMB lensing experiments. We conclude in Section~\ref{sec:discussion}.

Throughout, we use the cosmology of~\cite{Ade:2013zuv}: $\{H_0=67.11\, \mathrm{km/s/Mpc}, \Omega_c h^2 = 0.1209, \Omega_b h^2 =0.022068 , A_s =2.2\times 10^{-9} , n_s =0.9624 \}$ where $H_0$ is the Hubble constant today, $\Omega_c h^2$ is the physical cold dark matter density today, $\Omega_b h^2$ is the physical baryon density today, $A_s$ is the amplitude of scalar fluctuations, and $n_s$ is the spectral index (with a pivot scale of 0.05 $\mathrm{Mpc}^{-1}$). All matter power spectra and transfer functions are calculated with the Einstein-Boltzmann code \href{https://camb.info}{\texttt{CAMB}}\footnote{\href{https://camb.info}{https://camb.info}}~\cite{2000ApJ...538..473L}.

\section{Theory}\label{sec:theory}

In Part~\ref{sec:fnl_bias_theory} of this Section we discuss the induction of scale-dependence in halo bias from $\fnl$. In Part~\ref{sec:cib_theory} we present the theory model we use to model the CIB and the CIB-CMB lensing cross power spectrum.

\subsection{$\fnl$ from scale-dependent bias}\label{sec:fnl_bias_theory}

$\fnl$ parameterizes primordial non-Gaussianity of the local type as follows:
\be
\Phi(\boldsymbol {x}) = \phi(\boldsymbol{x})+\fnl\left(\phi^2(\boldsymbol{x})-\left<\phi^2\right>\right)\label{fnldef}
\ee 
where $\Phi(\boldsymbol {x})$ is the Newtonian potential at $\boldsymbol {x}$ and $\phi(\boldsymbol{x})$ is an underlying Gaussian field. On sub-horizon scales, $\Phi$ is related to the overdensity $\delta$ through the Poisson equation. 

While the overdensity field $\delta$ is continuous, in several situations the \textit{peaks} of $\delta$ are the objects of interest. This is because gravitational collapse happened only where $\delta$ was higher than a critical value $\delta_c$, and so these regions (with $\delta>\delta_c$) are those in which large scale structure formed. These peaks of $\delta$ are biased with respect to $\delta$:
\be
\delta_h = b_h \delta,
\ee
where $\delta_h$ is the overdensity of the peaks (the ``halo overdensity''), and $b_h$ is their bias (the ``halo bias''). This leads to them following a different power spectrum to that of the underlying dark matter:
\be
P_{hh}(k) = b_h^2 P_{mm}(k)
\ee
where $P_{hh}(k)$ is the halo power spectrum and $P_{mm}(k)$ is the matter power spectrum. For Gaussian initial conditions, $b_h$ is scale  independent on large scales---i.e., it does not depend on $k$. However, non-Gaussianity of the form of Equation~\eqref{fnldef} serves to induce a scale dependence~\cite{0710.4560}:
\be
b_h^{NG}= b_h^{G}+\fnl\frac{3\Omega_m H_0^2}{ k^2T(k)D(z)} \delta_c(b_h^{G}-1)\label{fnlbias}
\ee
where $\Omega_m$ is the mean density of matter today; $H_0$ is the Hubble constant; $T(k)$ and $D(z)$ are the transfer and growth functions of the density field, respectively, with $T(k)$ normalized to 1 at low $k$ and $D(z)$ normalized such that $D(z)=\frac{1}{1+z}$ during matter domination;  and $\delta_c=1.686$  is the critical overdensity above which objects undergo gravitational collapse. $b_h^{G}$ refers to the \textit{Gaussian} bias, i.e. the bias in the absence of $\fnl$. 

\subsection{The CIB-CMB lensing cross correlation}\label{sec:cib_theory}

\subsubsection{The CIB}

The CIB is sourced by thermal emission of dust in star-forming galaxies. As the physics of star-formation is not well understood, we lack a first-principles model for the CIB. Instead several parametric models of various physical motivation have been proposed (see, e.g.~\cite{2012MNRAS.421.2832S,Wu:2016vpb,Maniyar:2018hfp,2021A&A...645A..40M}).

The CIB intensity at frequency $\nu$ $I_\nu$ is given by
\be
I_\nu(\hat{\boldsymbol{ n}}) = \int_0^{\chi_{\rm{re}}} d\chi a(\chi) j_\nu(\chi, \hat{\boldsymbol{ n}}),\label{cibemiss}
\ee
where $j_\nu$ is the comoving CIB emissivity density, $a(\chi)$ is the scale factor, and the integral over comoving distance $\chi$ is done out to reionization at $\chi_{\rm{re}}$. $j_\nu(\chi, \hat{\boldsymbol{ n}})$ can be separated into its mean value and fluctuations:
\be
 j_\nu(\chi, \hat{\boldsymbol{ n}}) = \bar j_\nu(\chi) \lb 1 +\frac{\delta j_\nu(\chi, \hat{\boldsymbol{ n}})}{ j_\nu(\chi)}\rb.
\ee
CIB models generally include a model for the mean emissivity $\bar j_\nu$ as well as a prescription for the clustering of the fluctuations, in particular the three-dimensional emissivity power spectrum $P_{jj}^{\nu\nu^\prime}(k,z,z^\prime)$, which is defined as follows:
\be
\frac{\left<\delta j_\nu(\boldsymbol{k},z)\delta j_{\nu^\prime}(\boldsymbol{k}^\prime,z^\prime)\right>}{\bar j_\nu(z) \bar j_{\nu^\prime}(z^\prime)}\equiv \lb2\pi\rb^3 P_{jj}^{\nu\nu^\prime}(k,z,z^\prime)\delta^3(\boldsymbol{k}-\boldsymbol{k}^\prime).\label{Pjnunuprime}
\ee
The angular CIB power spectrum can then be integrated directly according to
\begin{align}
&C_\ell^{\nu\nu^\prime} =\frac{2}{\pi}\int d\chi  d\chi^\prime \int k^2 d k \label{cl_full}\\
&  a(\chi)a(\chi^\prime) \bar j_\nu(\chi) \bar j_{\nu^\prime}(\chi^\prime)P_{jj}^{\nu\nu^\prime}(k,z,z^\prime) j_\ell(k\chi) j_{\ell}(k\chi^\prime)\nonumber
\end{align}
where $ j_{\ell}(x)$ are the spherical Bessel functions of degree $\ell$. As $j_\nu(\chi)$ has support on a very wide range of $\chi$, in most cases the Limber approximation~\cite{1953ApJ...117..134L} is valid and we can simplify Equation~\eqref{cl_full} to reduce to the more standard expression:
\be
C_\ell^{\nu\nu^\prime} =\int \frac{d\chi}{\chi^2}  a^2(\chi)\bar j_\nu(\chi)\bar j_{\nu^\prime}(\chi)P_{jj}^{\nu\nu^\prime}\lb k=\frac{\ell}{\chi},z\rb \label{lin_cib_power}
\ee
where $P_{jj}(k,z)\equiv P_{jj}(k,z,z)$ is the equal-time emissivity power spectrum. However, at the lowest values of $\ell$ ($\ell\lessthanapprox40$), we should integrate the full expression~\eqref{cl_full}.

In this work, we use the linear CIB model of~\cite{2018A&A...614A..39M} to model the CIB. In this model, the mean CIB emissivity is related directly to the mean star formation rate density (SFRD) with the Kennicutt relation~\cite{1998ARA&A..36..189K}:
\be
\bar j_\nu(z) = \frac{\rho_{SFR}(z)(1+z) S_{\nu,\mathrm{eff}}(z)\chi^2}{K}
\ee
where $K$ is the Kennicutt constant $K=1.7\times 10^{-10} M_\odot \mathrm{yr}^{-1}L_\odot^{-1}$ and $ S_{\nu,\mathrm{eff}}(z)$ is the mean effective spectral energy distribution (SED), calculated using the method of~\cite{Bethermin_SFR} using SEDs calibrated with \textit{Herschel} data~\cite{Bethermin_SEDs,2017A&A...607A..89B}\footnote{ These are available at \href{https://github.com/abhimaniyar/halomodel_cib_tsz_cibxtsz}{this URL}~\cite{2021A&A...645A..40M}}. The SFRD is parameterized according to
\be
\rho_{SFR}(z) = \alpha\frac{\lb1+z\rb^\beta}{1+\lb\frac{1+z}{\gamma}\rb^\delta}
\ee
with $\alpha,\beta,\gamma,\delta$ free parameters of the model.
As this is a linear model, the CIB fluctuations can be parameterized directly by defining the CIB bias $b^{\rm CIB}(z)$:
\be
P^{\nu\nu^\prime}_{jj}{}^{\mathrm{lin}}(k,z,z^\prime) = b^{\rm CIB}(z)b^{\rm CIB}(z^\prime) P^{\mathrm{lin}}_{mm}(k,z,z^\prime)
\ee
where $P^{\mathrm{lin}}_{mm}(k,z,z^\prime)$ is the linear matter power spectrum;  $b^{\rm CIB}(z)$ is parameterized as
\be
b^{\rm CIB}(z) = b_0+b_1 z +b_2 z^2
\ee
with $b_0$, $b_1$, $b_2$ free parameters of the model (note that $P^{\nu\nu^\prime}_{jj}$ as defined in Equation~\eqref{Pjnunuprime} is thus independent of $\nu$ and $\nu^\prime$, with frequency-dependence of $C_\ell^{\nu\nu^\prime}$ coming from the SFRD alone). We expect this linear model to be sufficient since we restrict our analysis to relatively large scales  ($\ell\le610$). 

The parameters $\{\alpha,\beta,\gamma,\delta,b_0,b_1,b_2\}$ were fit to the \textit{Planck} CIB auto and CIB-lensing power spectra at $\nu=\{217,353,545,857\}  \,\mathrm{GHz}$ in Ref.~\cite{2018A&A...614A..39M}; their values are given in Table~\ref{tab:fiducial_params}. In our analysis, we marginalize over all of these parameters, with a prior of $b_0=0.83\pm0.11$. We note that we do not vary any cosmological parameters, since these are very well determined by primary CMB measurements.

There is also a small contribution to the CIB power from the small-scale regime (1-halo term) and the  shot noise (as the CIB is intrinsically sourced by discrete objects), which is constant in $\ell$. We include these contributions to the power by using the prescription presented in~\cite{2018A&A...614A..39M,2021A&A...645A..40M}\footnote{Again, see \href{https://github.com/abhimaniyar/halomodel_cib_tsz_cibxtsz}{this URL} for the pre-computed 1-halo term.}. However, in practice we will marginalize over the values of the shot noise, which we expect to allow for model uncertainty in both the shot noise and the 1-halo term which are very degenerate on the linear scales we use, as the 1-halo term is only very mildly scale dependent in this regime.

\begin{table}

\begin{tabular}{|c|c|c|}
\hline
&Parameter & Value\\\hline\hline
&$\alpha$ & 0.007\\\cline{2-3}
$\rho_{SFR}(z)$&$\beta$ & 3.590\\\cline{2-3}
Evolution&$\gamma$ & 2.453\\\cline{2-3}
&$\delta$ &6.578\\\hline\hline
CIB&$b_0$ &0.83\\\cline{2-3}
bias&$b_1$ &0.742\\\cline{2-3}
evolution&$b_2$ &0.318\\\hline
\end{tabular}
\caption{The fiducial values for the parameters of the CIB model, from~\cite{2018A&A...614A..39M}. }\label{tab:fiducial_params}
\end{table}

Thus, in total, the full model for the CIB power is
\begin{align}
C_\ell^{\nu \nu^\prime} = C_\ell^{\nu\nu^\prime}{}^{\mathrm{linear}}
+  C_\ell^{\nu\nu^\prime}{}^{\mathrm{one-halo}}+S_{\nu\nu^\prime}
\end{align}
where $C_\ell^{\nu\nu^\prime}{}^{\mathrm{linear}}$ is the linear term that we model by calculating Equation\eqref{lin_cib_power} using $P_{jj}^{\nu \nu^\prime}{}^{\mathrm{lin}}$ and $\bar j_\nu$ as described above; $ C_\ell^{\nu\nu^\prime}{}^{\mathrm{one-halo}}$ is the (almost-constant) one-halo contribution, which we pre-compute; and $S_{\nu\nu^\prime}$ is the constant shot-noise (over which we will marginalize in our analysis).

\subsubsection{CMB lensing}

Gravitational lensing induces a specific form of statistical anisotropy in the CMB allowing the use of quadratic estimators to reconstruct the line-of-sight gravitational potential $\phi$~\cite{astro-ph/0111606}  integrated all the way to the surface of last scattering. The contribution to the lensing potential peaks at redshifts around $z\sim2$.  As the CIB is sourced mostly at the same redshifts where the CMB lensing efficiency peaks, the two fields are expected to be highly correlated with each other; indeed, their correlation has been detected by \textit{Planck}~\cite{Ade:2013aro}, SPT\cite{1303.5048} and ACT~\cite{1412.0626,Darwish:2020fwf}. Going forward, we may interchangeably refer to both the lensing potential $\phi$ and the lensing \textit{convergence} field $\kappa$ (proportional to the projected matter density), which are straightforwardly related through $\nabla^2 \phi = - 2 \kappa$.

The CMB lensing potential $\phi$ is given by
\be
\phi(\hat{\boldsymbol{n}}) = -2\int_0^{\chi_S} d\chi \frac{\chi_S-\chi}{\chi_S\chi}\Phi(\chi, \hat{\boldsymbol{n}})
\ee
where $\chi_S$ is the comoving distance to the surface of last scattering, where the CMB was released, and $\Phi(\chi, \hat{\boldsymbol{n}})$ is the Newtonian potential. $\Phi$ can be related directly to the matter overdensity $\delta$ on sub-horizon scales with the Poisson equation
\be
\nabla^2 \Phi = -\frac{3}{2}\lb\frac{H_0}{c}\rb^2\frac{\Omega_m H_0}{a}\delta.
\ee 
As a result of this, in harmonic space the lensing potential is related to the lensing convergence $\kappa$ by
\be
\phi_\ell=\frac{2}{\ell\lb\ell+1\rb}\kappa_\ell,
\ee
where
\be
\kappa = \int_0^{\chi_S} d\chi W^\kappa(\chi) \delta(\chi, \hat{\boldsymbol{n}}),
\ee
with the lensing convergence kernel $W^\kappa(\chi)$ given by
\be
W^\kappa(\chi) = \frac{3}{2}\lb\frac{H_0}{c}\rb^2\frac{\Omega_m}{a}\chi\lb1-\frac{\chi}{\chi_S}\rb.
\ee
The angular power spectrum of the CMB lensing convergence field is
\begin{align}
C_\ell^{\kappa\kappa}=&\frac{2}{\pi}\int d\chi d\chi^\prime\int k^2 dk \\
&W^\kappa(\chi)W^\kappa(\chi^\prime) P_{mm}(k,z,z^\prime) j_\ell(k \chi)   j_\ell(k \chi^\prime),\nonumber
\end{align}
which in the Limber approximation reduces to
\be
C_\ell^{\kappa\kappa}=\int \frac{d\chi}{\chi^2}   W^\kappa(\chi)^2P_{mm}\lb k=\frac{\ell}{\chi},z\rb .
\ee
As we work on linear scales, we use the linear matter power spectrum $P_{mm}^{\mathrm{lin}}(k)$ in place of $P_{mm}(k)$.

\subsubsection{The CIB-CMB lensing cross correlation}

On linear scales, the CIB-CMB lensing cross-power spectrum is given by
\begin{align}
C^{\nu\kappa}_\ell=&\frac{2}{\pi}\int d\chi d\chi^\prime\int k^2 dk\\
 &a(\chi) \bar j_\nu(\chi)   W^\kappa(\chi^\prime)P_{jm}^\nu(k,z,z^\prime) j_\ell(k \chi)   j_\ell(k \chi^\prime)\nonumber
\end{align}
with the cross-power spectrum $ P_{jm}^\nu$ given on linear scales by
\be
 P_{jm}^\nu{}^{\mathrm{lin}}(k,z,z^\prime) = b^{\rm CIB}(z)P_{mm}^{\mathrm{lin}}(k,z,z^\prime).\label{linear_pjm}
\ee
Except for on the largest scales, this reduces to the standard expression with the Limber approximation:
\be
C_\ell^{\nu \kappa}=\int \frac{d\chi}{\chi^2} a(\chi)  \bar j_\nu(\chi) W^\kappa(\chi)P_{jm}^\nu\lb k=\frac{\ell}{\chi},z\rb.\label{cellcross}
\ee
As we restrict our analysis to linear scales, we use the linear expression~\eqref{linear_pjm} when calculating Equation~\eqref{cellcross}.

\subsubsection{Including $\fnl$}

\begin{figure}
\includegraphics[width=\columnwidth]{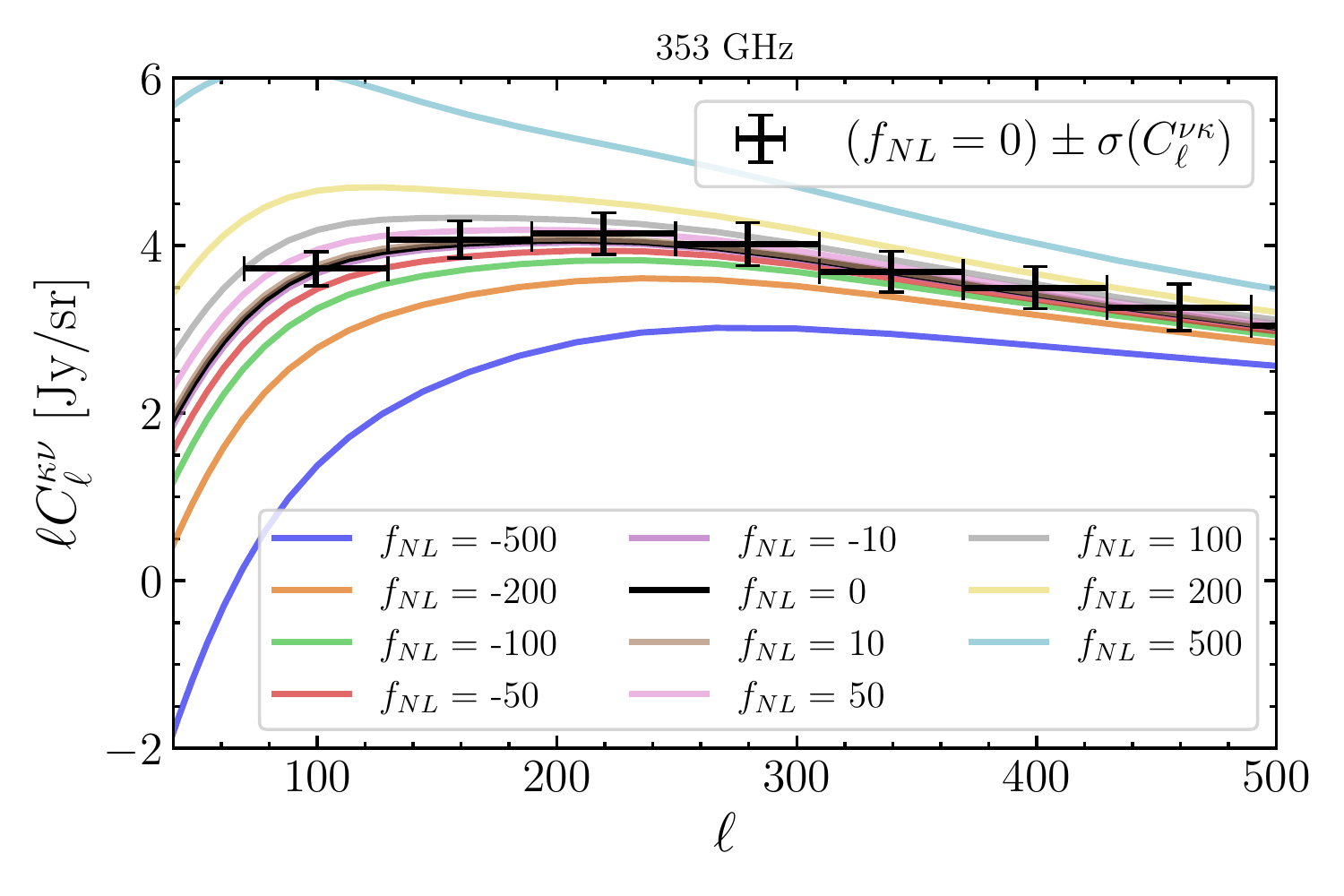}
\includegraphics[width=\columnwidth]{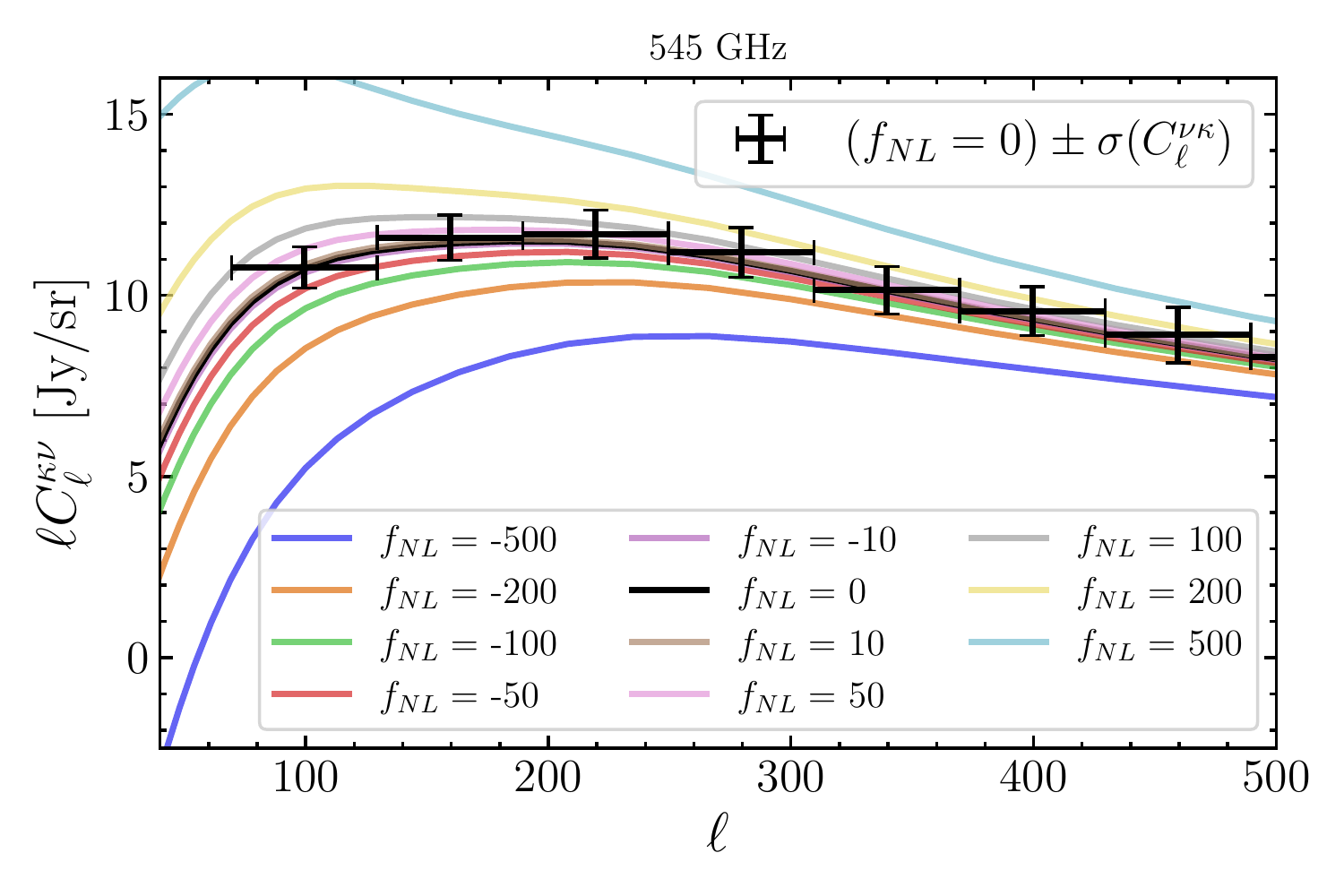}
\includegraphics[width=\columnwidth]{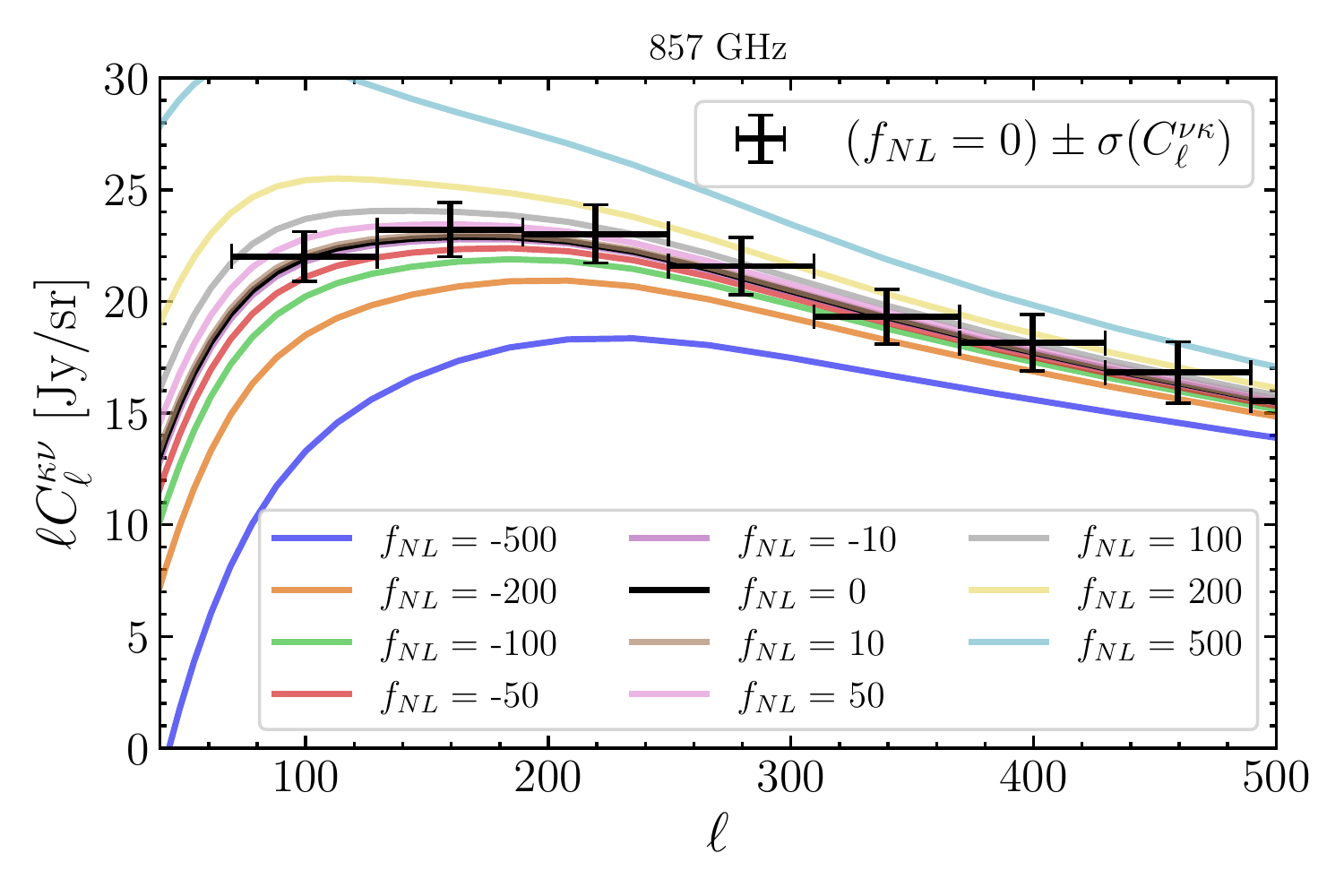}
\caption{The effect of various values of $\fnl$ on the CIB-CMB lensing power spectra. Also indicated is the size of the $1\sigma$ uncertainty on the measurement $C_\ell^{\kappa\nu}$ from the maps we are using ($N_{HI}<2.5\,\mathrm{cm}^{-2}$), when binned linearly with bins of width $\Delta\ell=60$.}\label{fig:fnl_power_examples}
\end{figure}

To allow for dependence on $\fnl$, we directly promote the CIB bias to be scale-dependent according to Equation~\eqref{fnlbias}. The CIB-CMB lensing power spectra for various values of $\fnl$ are shown in Fig.~\ref{fig:fnl_power_examples}. Note that on the largest scales $\ell\lessthanapprox40$, the Limber approximation is not valid, and in principle we must directly integrate the three-dimensional power-spectrum to find $C_\ell$; however, as we restrict our analysis to $\ell>70$ in this work, we employ the Limber approximation throughout.

\subsubsection{Color correction}

Our model is constructed for the $\nu I_\nu=\mathrm{constant}$ photometric convention. In practice, this means that we must colour-correct our model to compare appropriately with the data measured with the \textit{Planck} bandpasses:
\begin{equation}
C_\ell^{\nu X}{}^{\mathrm{color-corrected}  }=\mathrm{cc}_\nu C_\ell^{\nu X}
\end{equation}
where $\mathrm{cc}_{353}=1.097$,  $\mathrm{cc}_{545}=1.068$, and  $\mathrm{cc}_{857}=0.995$; note that this means that the $C_\ell^{\nu \nu^\prime}$ spectra are multiplied by two factors and $C_\ell^{\nu \kappa}$ only by one. 

\section{Data}\label{sec:data}

We measure the CIB-CMB lensing cross correlation with the CIB maps of~\cite{1905.00426} (constructed from \Planck HFI maps and HI4PI neutral hydrogen maps), and the CMB lensing reconstruction of \textit{Planck}~\cite{1807.06210}. In this section we briefly describe these data.

\subsection{CIB maps}

We use the CIB maps of~\cite{1905.00426} which were produced from high-frequency (353, 545, 857 GHz) data from the \textit{Planck} satellite's HFI instrument, with Galactic dust cleaned by using neutral hydrogen (HI) data collated from various radio surveys, in particular the  Effelsberg--Bonn HI Survey (EBHIS)~\cite{2010ASPC..438..381W,2011AN....332..637K,2016A&A...585A..41W}, and the Galactic All-Sky Survey (GASS)~\cite{2009ApJS..181..398M,2010A&A...521A..17K,2015A&A...578A..78K}, collected in the HI4PI Survey~\cite{1610.06175}.  The HI data is used to create a template for the Milky Way Galactic dust to be subtracted from the \textit{Planck} single-frequency maps. Going from HI data to dust templates requires the modeling of a dust-to-gas ratio; this is a spatially-dependent quantity, depending on the environment of the gas, and so local modeling is required. As there is a spatial limit to the size over which the dust-to-gas ratio can be modeled, there is a scale above which the maps cannot be properly cleaned. Due to this, the maps of~\cite{1905.00426} are not guaranteed to be unbiased below angular scales $\ell\sim70$, and so we restrict ourselves to $\ell>70$ in our analysis. This is a significant penalty on the extraction of information on $\fnl$, as most information is in the largest scales. Regardless of this multiplicative transfer function present in the maps, the CIB maps of \cite{1905.00426} are far more appropriate for our work than the raw intensity maps from \Planck; while those raw intensity maps do not have a multiplicative transfer function and could in principle be used for unbiased cross-correlations down to arbitrary scales, in practice, the presence of Galactic dust induces very large scatter on any measured cross-correlation. Thus, we proceed with the HI template-subtracted CIB maps from \cite{1905.00426}.

\subsection{CMB lensing map}

For our CMB lensing map, we use the minimum variance (MV) CMB lensing convergence ($\kappa$) reconstruction from the \textit{Planck} 2018 release~\cite{1807.06210}, available on the Planck legacy archive (PLA). This reconstructed map is reliable down to $\ell=8$ making it ideally suited for studying local primordial non-Gaussianity. The lensing map itself does not contain information on $\fnl$, but it provides an unbiased probe of the matter distribution that is highly correlated with the CIB, allowing the redshift distribution of the CIB to be constrained and the sample variance in the measurement to be reduced\cite{2010.16405}. The lensing map is reconstructed exploiting the fact that the small-scale anisotropies in the CMB (measured primarily at 100 and 150 GHz) are modulated by large-scale lenses in a well-understood way. The reconstruction uses a quadratic estimator dominated by information in the CMB temperature anisotropy at low frequencies (LF) $\hat{\kappa} \propto \langle T^{\rm LF}_{{\rm high-}\ell}T^{\rm LF}_{{\rm high-}\ell} \rangle$. As described in \cite{2010.16405}, crucially, this means that the large-scale lensing map that is produced primarily uses small-scale CMB data ($\ell>800$) at frequencies where Galactic dust contamination is much smaller than at the high frequencies at which the \Planck CIB measurements are made. In the cross-correlation of the CIB map with the CMB lensing map, we therefore {\it do not} expect a Galactic dust bias proportional to the power spectrum of the Galactic dust at high frequencies (HF) (where the dust is brighter) but rather a negligible bispectrum of the form $\langle D^{\rm HF}_{{\rm low-}\ell}D^{\rm LF}_{{\rm high-}\ell}D^{\rm LF}_{{\rm high-}\ell}\rangle$ for Galactic dust modes $D_{\rm HF}$ at high frequencies and the much dimmer modes $D_{\rm LF}$ at low frequencies (LF).

\section{Analysis Pipeline}\label{sec:pipeline}

We constrain $\fnl$ by maximizing a likelihood defined as
\be
-2\ln\mathcal L =\left(C(\Pi)-\hat C\right)^T \mathbb{C}^{-1} \left(C(\Pi)-\hat C\right)+\chi^2_{\mathrm{priors}}.\label{likelihood}
\ee
$\Pi$ is the parameter vector; $C(\Pi)$ is the theoretical data vector calculated from the parameters; $\hat C$ is the measured data (superscript $T$ denotes the transpose); and $\mathbb{C}$ is the covariance matrix. We include priors on the CIB mean, the CIB calibration parameters, and the bias at $z=0$; we will discuss each element below. 

\subsection{Parameter vector}
$\Pi$ is a 17-dimensional parameter vector, which contains $\fnl$ along with all the  parameters over which we marginalize: the CIB bias parameters $\{b_0,b_1,b_2\}$, the star formation rate density parameters $\{\alpha,\beta,\gamma,\delta\}$; and the \textit{Planck} instrument calibration parameters $\{f_{353},f_{545},f_{857}\}$; and the CIB shot noise values $\{S_{353,353}, S_{353,545}, S_{353,857}S_{545,545}S_{545,857},S_{857,857}\}$:
\be
\Pi = \left[\fnl;b_0,b_1,b_2,\alpha,\beta,\gamma,\delta,\{f_{\nu}\},\{S_{\nu\nu^\prime}\}\right].
\ee
The calibration parameters $f_\nu$ are nuisance parameters that we implement as multiplicative biases on the $C_\ell^{\nu X}$; the remaining parameters have been discussed in Section~\ref{sec:cib_theory}.

\subsection{Data vector}

We take as our data vector $C$ the cross-power spectrum $C_\ell^{\nu\kappa}$ binned in $\ell$-space between $\ell=70$ and $\ell=610$ in bins of equal (linear) extent in $\ell$. As we require some auto-power spectrum data to constrain our nuisance parameters (in particular the CIB SFRD parameters), we also include the CIB auto-spectrum between $\ell$ of 430 and 610. Thus we have
\be
C = \begin{cases}
C_\ell^{\nu \kappa} & 70\le\ell\le430\\
C_\ell^{\nu \nu^\prime},C_\ell^{\nu \kappa} & 430\le\ell\le610.
\end{cases}
\ee

It is important not to include CIB auto-power spectrum data at $\ell<430$, as we wish to avoid adding constraining power on $\fnl$ from the CIB auto-power spectrum. At lower $\ell$, there would be significant information on $\fnl$ in this data, but also significant potential for bias from any residual Galactic dust in the maps; this is not a problem for the $\nu\kappa$ power spectra as the residual Galactic dust will add noise to the measurement but not bias.

As we are using a linear model for the CIB, we never use any data from multipoles greater than $\ell=610$.

We will discuss  in detail in Section~\ref{sec:powerspecmeas} how we measure the $C_\ell$ from the maps.

\subsection{Covariance matrix}\label{sec:covmat}
 
 In Equation~\eqref{likelihood}, $\mathbb{C}$ is the covariance matrix of our data, which we take to be diagonal in $\ell$.

A theoretical expression for the covariance matrix is given by
\begin{align}
\mathbb{C}(\hat C_\ell^{\alpha\beta},\hat  C_{\ell^\prime}^{\gamma\delta})=&\frac{1}{\lb2\ell+1\rb f_{\mathrm{sky}}}\bigg{(} \lb C_\ell^{\alpha\gamma}+N_\ell^{\alpha\gamma} \rb \lb C_\ell^{\beta\delta}+N_\ell^{\beta\delta} \rb \nonumber\\
&+\lb C_\ell^{\alpha\delta}+N_\ell^{\alpha\delta} \rb \lb C_\ell^{\beta\gamma}+N_\ell^{\beta\gamma} \rb\bigg{)}\label{covm_defn}\delta_{\ell \ell^\prime}
\end{align}
where $C_\ell$ is a fiducial (theoretically calculated) power spectrum and $N_\ell$ contains any noise and foreground power; $f_{\mathrm{sky}}$ is the sky area on which the analysis is done. Instead of the theoretical covariance matrix, we simulate 170 Gaussian full-sky maps using  \texttt{healpy}\footnote{\href{http://healpix.sf.net}{http://healpix.sf.net}}'s \cite{Zonca2019,2005ApJ...622..759G} {\texttt synalm} function and apply our power spectrum estimation pipeline (see Section~\ref{sec:powerspecmeas}) to calculate the covariance matrix used in our analysis by directly measuring the covariance of these simulations; this accounts for effects not accounted for such as the mask apodization procedure.  

To simulate the sky, we need theoretical power spectra (auto and cross) and also a theoretical model for the noise power spectra $N_\ell$. For the theoretical power spectra, we use the fiducial values of our model. We must also include noise in these simulations; we include the noise in the CIB maps as beam-deconvolved white noise corresponding to the values in Table~\ref{tab:noise_cib}, which we take from~\cite{2014A&A...571A...6P}. For all auto power spectra we take the half-mission splits, so in practice when simulating the half-mission maps we multiply the noise power spectrum by 2. We use the following expression for the power spectrum of the beam-deconvolved noise:
\begin{equation}
N_\ell^{\rm CIB} =N_{\mathrm{white}}e^{\ell(\ell+1)\Theta^2/8\ln 2}.
\end{equation}

We also  include the CMB reconstruction noise in $N_\ell^{\kappa\kappa}$ provided with the \textit{Planck} 2018 release for the MV lensing reconstruction map.

\begin{table}[h!]
\begin{tabular}{|c| c| c|}\hline
Frequency & Noise & Beam (arcmin)\\\hline\hline
353 GHz& 305 $\mathrm{Jy}^2/\mathrm{sr}$&4.86\\\hline
545 GHz& 369 $\mathrm{Jy}^2/\mathrm{sr}$&4.84\\\hline
857 GHz& 369 $\mathrm{Jy}^2/\mathrm{sr}$&4.63\\\hline
\end{tabular}
\caption{The values we used to model the white noise in our Gaussian simulations, in $\mathrm{Jy}^2/\mathrm{sr}$, on the CIB maps at each frequency}\label{tab:noise_cib}
\end{table}

\begin{figure*}
\includegraphics[width=\columnwidth]{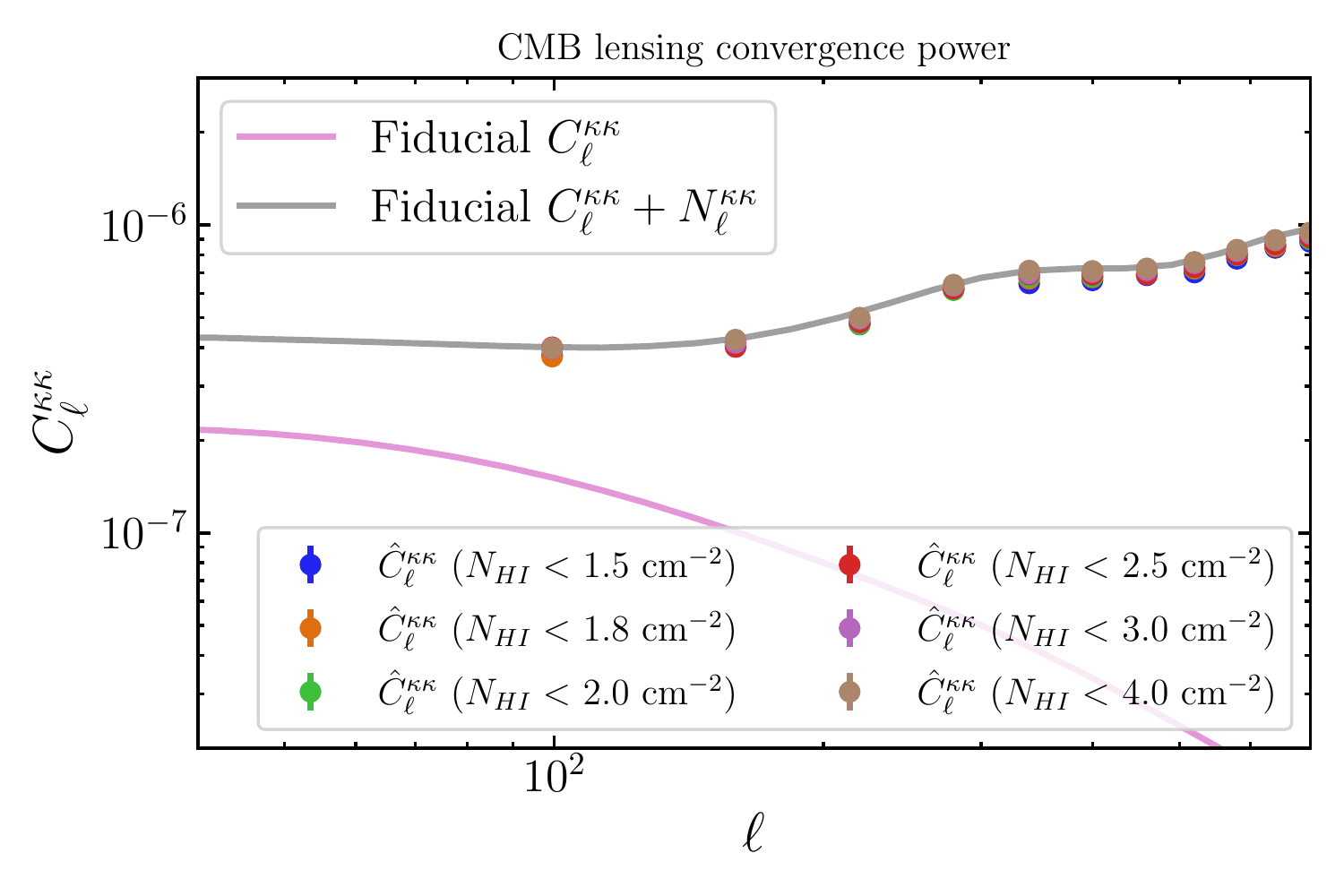}
\includegraphics[width=\columnwidth]{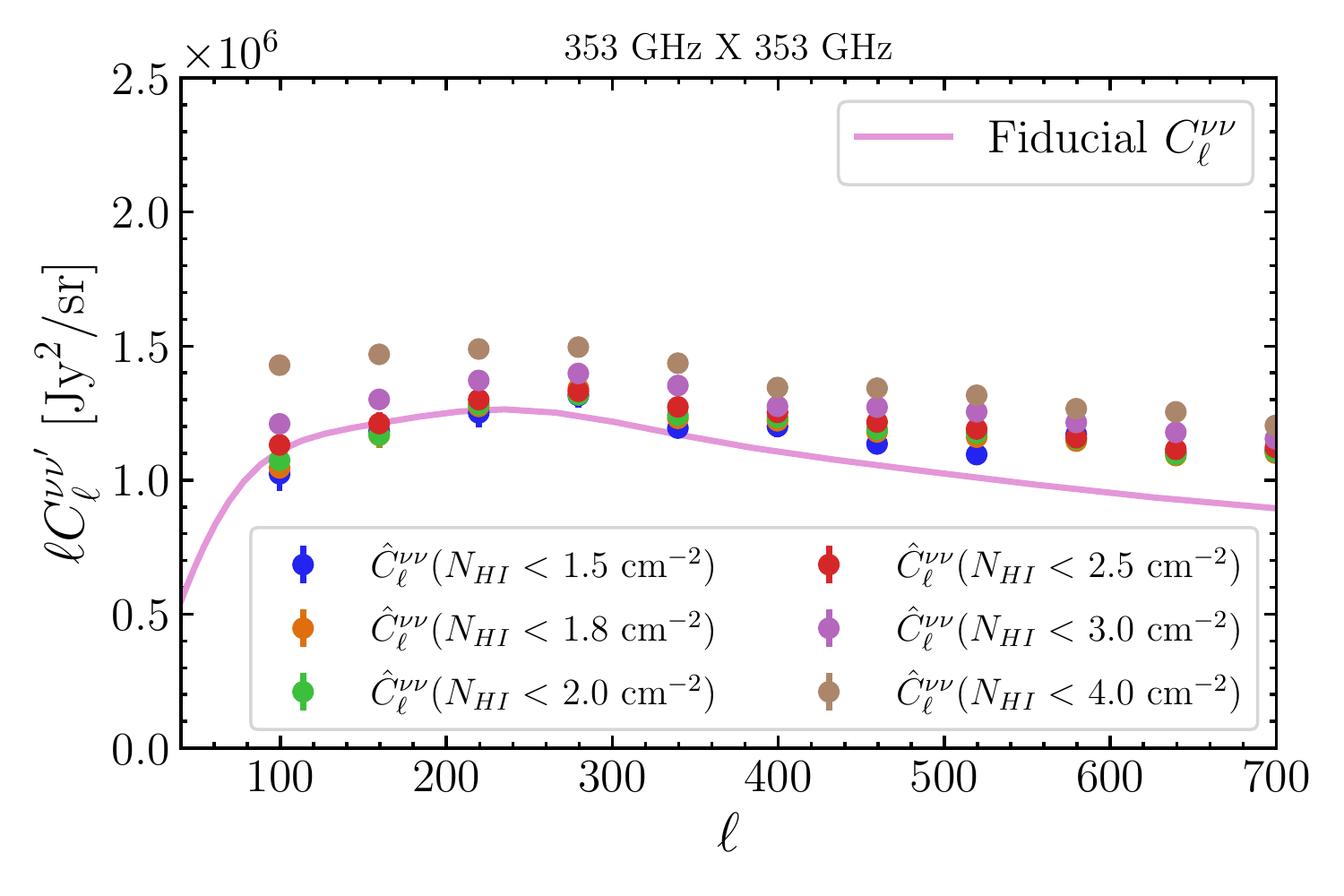}
\includegraphics[width=\columnwidth]{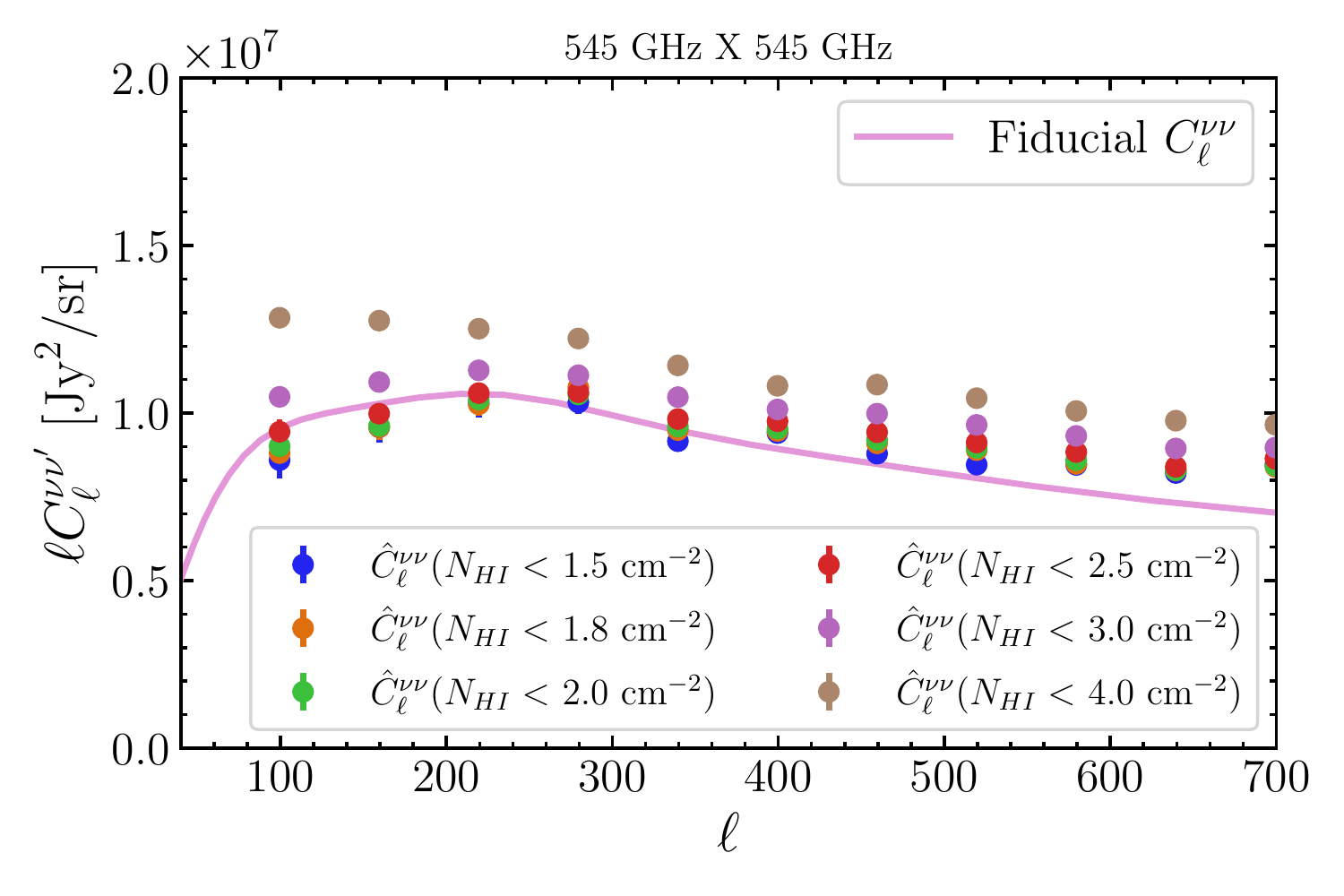}
\includegraphics[width=\columnwidth]{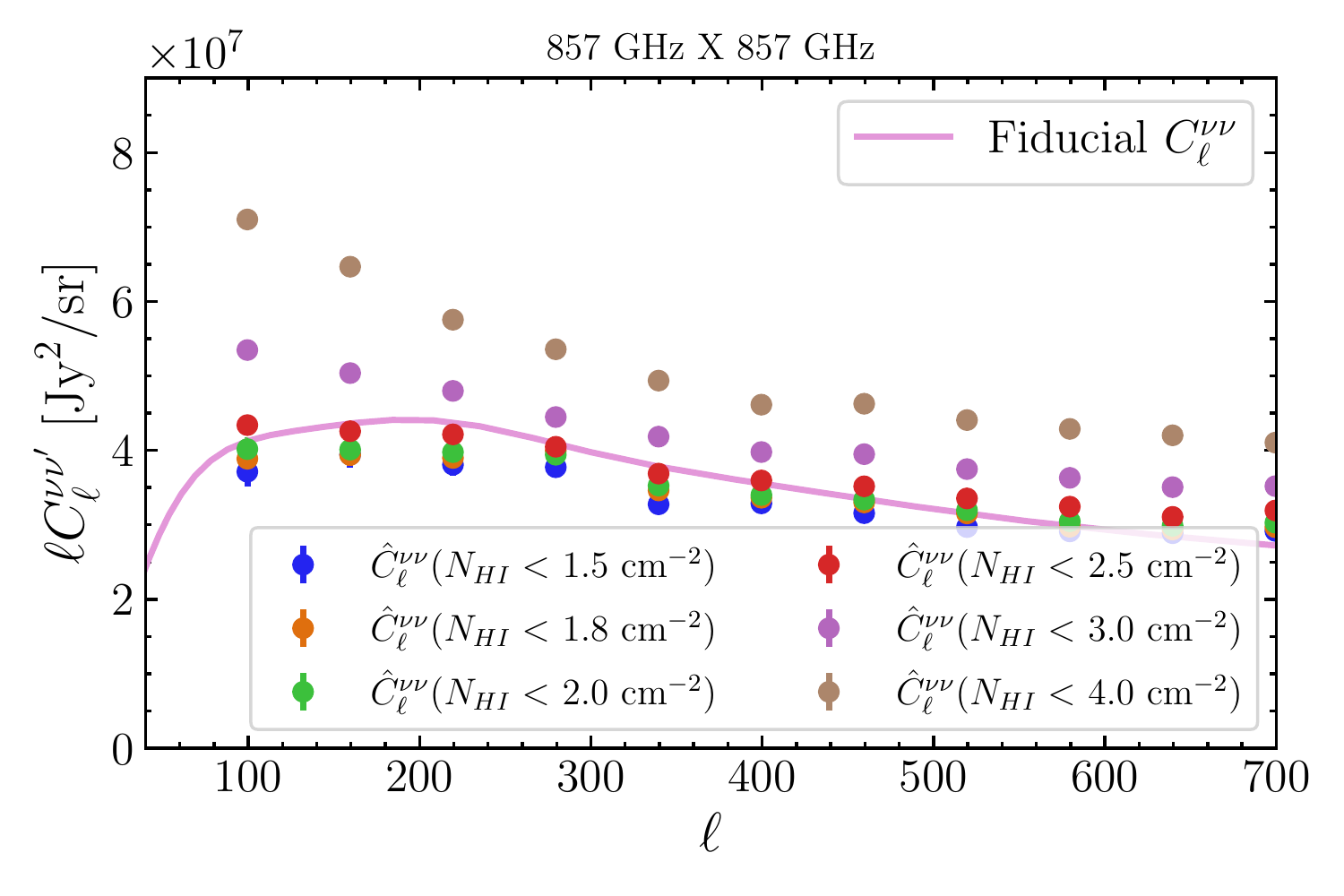}
\caption{The  fiducial auto power and noise, and the measured power spectrum for various sky areas. For CMB lensing (top left), the fiducial models for $C_\ell$ and $N_\ell$ give an appropriate estimation of the measured power in the maps, and thus are appropriate to use in the theoretical covariance matrix. However, it is clear that in the CIB maps the fiducial component is insufficient, especially for the large sky areas (corresponding to dustier maps); this is also a problem for the $\nu\ne\nu^\prime$ power spectra. To avoid this dust bias, we only perform analysis on the maps with $N_{HI}<2.5\,\mathrm{cm}^{-2}$ and below. }\label{fig:Cell_error}
\end{figure*}

$\mathbb{C}(\hat C_\ell^{\alpha\beta},\hat  C_\ell^{\gamma\delta})$ can be converted directly into uncertainties on the measurement of the $\hat C_\ell$s. In Fig~\ref{fig:Cell_error}, we show the fiducial power ($C_\ell+N_\ell$) for the auto-spectra along with the measured power. We can see that the measured CMB lensing power is captured appropriately by the fiducial model; for the more aggressive cleaning thresholds ($N_{HI}<2.5\,\mathrm{cm}^{-2}$ and lower), the large-scale auto CIB power is also captured by the model. However, for less aggressive thresholds, namely $N_{HI}=\{3.0,4.0\}\,\mathrm{cm}^{-2}$, there is some remaining power due to Galactic dust. Neglecting this power would lead to under-estimation of the uncertainties on $\hat C_\ell^{\nu\kappa}$; this could be incorporated by including some Galactic dust power in the covariance matrix. However, we choose to restrict our analysis to the maps with $N_{HI}<2.5\,\mathrm{cm}^{-2}$.

Note that, as we have not accounted for Galactic dust in the uncertainties in Figure~\ref{fig:Cell_error}, the uncertainties for $N_{HI}=\{3.0,4.0\}\,\mathrm{cm}^{-2}$ are underestimated; however, as noted above, we do not use these data in our analysis. We also ignore any possible non-Gaussian contributions to the noise as these are expected to be small since we use relatively clean parts of the sky with our HI thresholds. Our covariance naturally includes the effects of mask decoupling since this is performed on the simulations as well.

\subsection{Priors}

We include three priors in our analysis: 
\begin{itemize}
\item a prior on the CIB-mean;
\item a prior on the calibration parameters; 
\item and a prior on the CIB bias at $z=0$. 
\end{itemize}
For these priors, we follow~\cite{2018A&A...614A..39M}. All priors are Gaussian; the details are given in Table~\ref{tab:cibmeanprior}. The CIB-mean prior comes from measurements of the CIB mean~\cite{2006A&A...451..417D,bethermin_numbercounts}; for further details and references we refer to~\cite{2018A&A...614A..39M}.  The calibration parameters allow for uncertainty in the \textit{Planck} calibration and are implemented as multiplicative factors multiplying the power spectra at the appropriate frequency. We marginalize over these,  with three independent Gaussian priors centered on 1 with a width of 5\%. The prior on the CIB bias at $z=0$ is $b_0=0.83\pm0.11$~\cite{2018A&A...614A..39M}.

\begin{table*}
\begin{tabular}{|c||c||c|c|}
\hline
Frequency [GHz]&   $\bar{\nu I_\nu }[\mathrm{n W m^{-2}sr^{-1}}]$&$\sigma^+ [\mathrm{n W m^{-2}sr^{-1}}]$&$\sigma^- [\mathrm{n W m^{-2}sr^{-1}}]$\\\hline
353 & 0.46 & 0.04&0.05\\\hline
600 & 2.8& 0.93 & 0.81\\\hline
857 & 6.6&  1.70& 1.60\\\hline
1200 & 10.2& 2.6 & 2.3\\\hline
1875 & 13.63 & 3.53 & 0.85\\\hline
3000 & 12.61& 8.31& 1.74\\\hline
\end{tabular}
\caption{The priors on the CIB mean; for more details and references see Table 2 of~\cite{2018A&A...614A..39M}.  This prior is implemented as a Gaussian prior on $\nu I_\nu$ as calculated by Equation~\eqref{cibemiss}, with mean $\bar{\nu I_\nu} $ and standard deviation $\sigma^+$ for $\nu I_\nu$  higher than the mean, and standard deviation $\sigma^-$ for  $\nu I_\nu$  lower than the mean.}\label{tab:cibmeanprior}
\end{table*}
 
\subsection{Sky area and masks}
In~\cite{1905.00426}, the cleaning process allowed for subtraction of differing amounts of Galactic dust by defining different thresholds for the amount of HI in the maps; the cleanest maps, with $N_{HI}<1.5\, \mathrm{cm}^{-2}$, are on $\sim 10\%$ of the sky, with the largest maps, at $N_{HI}<4.0 \,\mathrm{cm}^{-2}$, on $\sim34\%$ of the sky. Each threshold has a different sky mask provided. We perform the analysis separately on the four cleanest maps: $N_{HI}\le\{1.5,1.8,2.0,2.5\}\, \mathrm{cm}^{-2}$. In every case, we multiply the appropriate $353$, $545$, $857$ GHz Boolean masks with each other and with the mask used for the \textit{Planck} CMB lensing reconstruction, such that our analysis is done on one common area of sky for each $N_{HI}$;  the resulting sky areas are given in Table~\ref{tab:skyareas}. Following~\cite{1905.00426}, we apodize the maps with a kernel with a full width at half maximum (FWHM) of 15' before estimating the auto- and cross-power spectra on the remaining sky. Ref.~\cite{1905.00426} also includes maps with $N_{HI}\le\{3.0,4.0\}\,\mathrm{cm}^{-2}$; we also measure the power spectra of these maps but we do not use them in our analysis as they contain significant amounts of dust on large scales.

\begin{table}[h!]
\begin{tabular}{|c||c|c|c||c||c|}
\hline
HI threshold & $f_{\rm{sky}}^{353}[\%] $ & $f_{\rm{sky}}^{545}[\%]$ & $f_{\rm{sky}}^{858}[\%]$ & $f_{\rm{sky}}^{\rm CIB}[\%]$ & $f^{\rm{CIB},\kappa}_{\rm{sky}}[\%]$\\\hline\hline
$1.5\,\mathrm{cm}^{-2}$ & 10.56 & 10.52 & 10.41 & 10.37 & 10.20\\\hline
$1.8\,\mathrm{cm}^{-2}$ & 14.63 & 14.57 & 14.42 & 14.36 & 14.06\\\hline
$2.0\,\mathrm{cm}^{-2}$ & 16.38 & 16.31 & 16.15 & 16.08 & 15.73\\\hline
$2.5\,\mathrm{cm}^{-2}$ & 18.7 & 18.62 & 18.42 & 18.34 & 17.95\\\hline
$3.0\,\mathrm{cm}^{-2}$ & 27.57 & 27.44 & 27.15 & 27.03 & 26.46\\\hline
$4.0\,\mathrm{cm}^{-2}$ & 34.42 & 34.23 & 33.83 & 33.67 & 32.99\\\hline
\end{tabular}

\caption{The sky-areas (in percentage of the full sky) of the 3 CIB maps at each HI threshold $f_{\rm{sky}}^{\nu}$, their overlap area $f_{\rm{sky}}^{\rm CIB}$, and their overlap area with the CMB lensing reconstruction $f^{\rm{CIB},\kappa}_{\rm{sky}}$. We calculate $f_{\rm{sky}}^{\rm CIB}$ by calculating the area of the mask defined by the product of the  binary masks for each of the three CIB frequencies. We calculate $f^{\rm{CIB},\kappa}_{\rm{sky}}$ by multiplying this mask with the \textit{Planck} lensing reconstruction mask (which itself has a total sky area of 67.06\%). As we only concentrate on regions of the sky where all the fields can be measured simultaneously, $f^{\rm{CIB},\kappa}_{\rm{sky}}$ is the relevant sky fraction for us; we see that the cleanest maps are on 10.20\% of the sky, with areas as large as 33\% possible at the cost of higher dust contamination.  }\label{tab:skyareas}
\end{table}

\subsection{Power spectrum measurement}\label{sec:powerspecmeas}

We measure the cross-power spectrum of the CMB lensing map with the CIB maps at frequencies (353, 545, 857 GHz) using \texttt{NaMaster}~\cite{1809.09603}. We bin the $C_\ell$ in bins of equal width $\Delta \ell=60$, between $\ell=70$ and $\ell=610$; we have checked robustness of our results for different bin-widths. We deconvolve the instrument beam with the effective window functions provided by~\cite{1905.00426}. In total, we have 45 data points from the $C_\ell$;  6 priors from the CIB mean measurement; and 4 external priors, to constrain 17 parameters.

To avoid noise bias in the auto power spectra, we use the half-mission splits provided by~\cite{1905.00426} to measure $C_\ell^{\nu\nu}$; for $C_\ell^{\nu\nu^\prime}$ with $\nu\ne\nu^\prime$ and for $C_\ell^{\nu\kappa}$ we use the full mission maps. 

\begin{figure}
\includegraphics[width=\columnwidth]{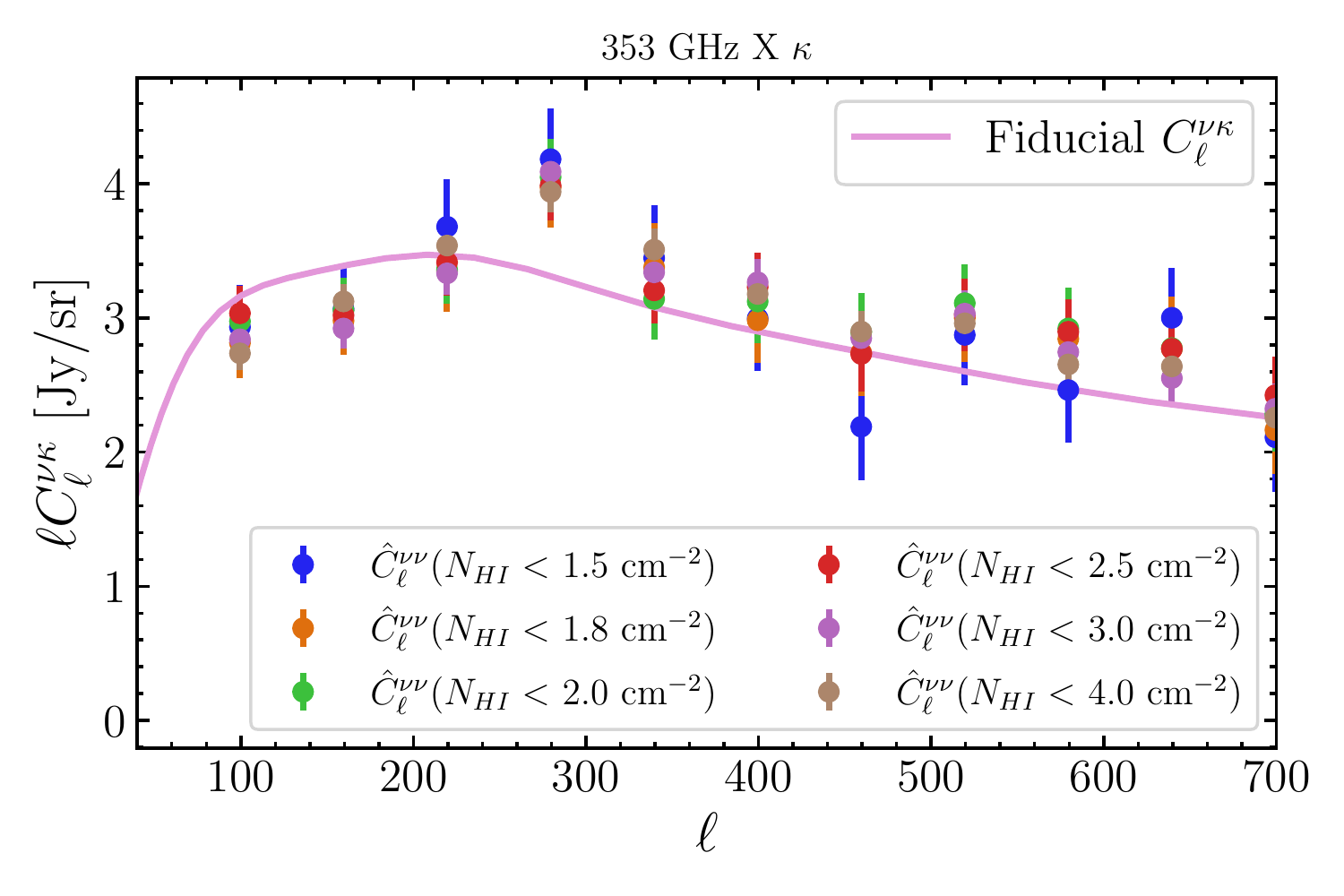}
\includegraphics[width=\columnwidth]{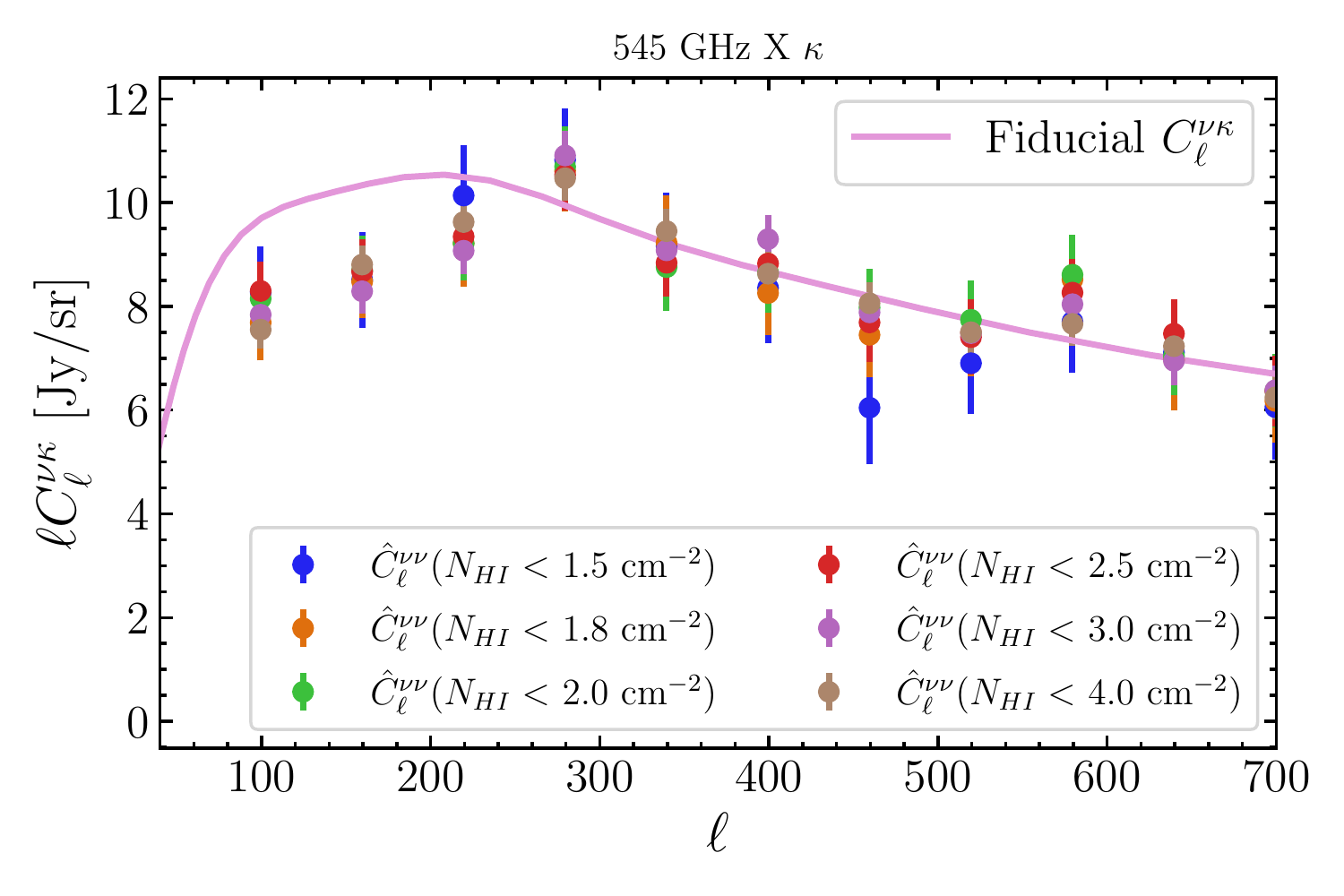}
\includegraphics[width=\columnwidth]{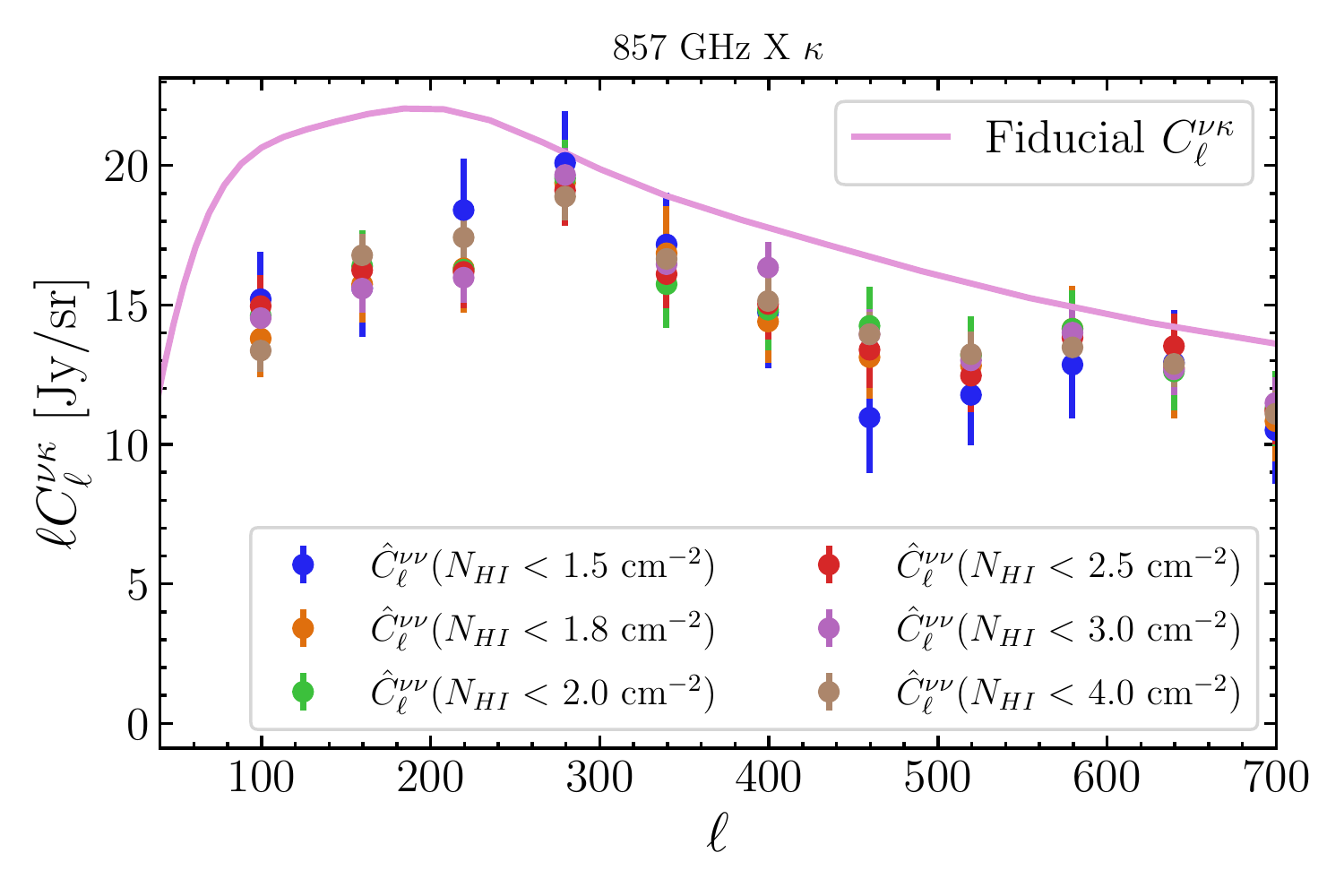}
\caption{The measured cross-power spectra of the CIB with the CMB lensing potential, along with the fiducial model. Note that, in contrast to the auto power (Figure~\ref{fig:Cell_error}), we see no Galactic dust contamination by eye at low $\ell$ regardless of the $N_{HI}$ threshold used for cleaning. The error-bars are calculated by taking the square root of the covariance matrix used in our MCMC analysis, which is calculated as described in Section~\ref{sec:covmat} (note that the $N_{HI}<3.0\, \mathrm{cm}^{-2}$ and $N_{HI}<4.0 \,\mathrm{cm}^{-2}$ uncertainties are therefore underestimated, as they do not include the significant variance contribution from Galactic dust at these thresholds; however, we do not use these thresholds in our analysis).}\label{fig:crosspower_data}
\end{figure}

We show in Figure~\ref{fig:crosspower_data} the measured cross-power spectra, for various values of $N_{HI}$ thresholds. In contrast to the auto power spectra, we see that there is no large-scale bias visible by eye in the dustier maps.

\subsection{$\fnl$ extraction and uncertainty calculation}

To extract  the best-fit $\fnl$, we  maximize the likelihood~\eqref{likelihood}. To calculate the uncertainties, we apply our pipeline to 200 Gaussian simulations. We histogram the best-fit values of $\fnl$, and fit a Gaussian to this histogram. We verify that the mean of the histogram is close to 0 (which ensures that our pipeline is unbiased). The standard deviation of this Gaussian is our $1\sigma$ uncertainty.

We also explore the posterior by using ~\texttt{cobaya}~\cite{2019ascl.soft10019T,Torrado:2020dgo} to perform Markov Chain Monte-Carlo (MCMC) sampling. We do this for the four different cleaning thresholds $N_{HI}=\{1.5,1.8,2.0,2.5\}\,\mathrm{cm}^{-2}$ separately, although note that the data are not independent as the smaller sky areas are subsets of the larger ones, meaning that the constraints are not independent. We run our chains until they are converged with a Gelman-Rubin convergence criterion~\cite{1992StaSc...7..457G} of $R-1<0.01$
  
 We find our best-fit ``measured'' $\fnl$ by minimizing our $\chi^2$ directly, using the above pipeline on the measured data.

\section{Results}\label{sec:results}

\begin{figure*}
\includegraphics[width=0.49\textwidth]{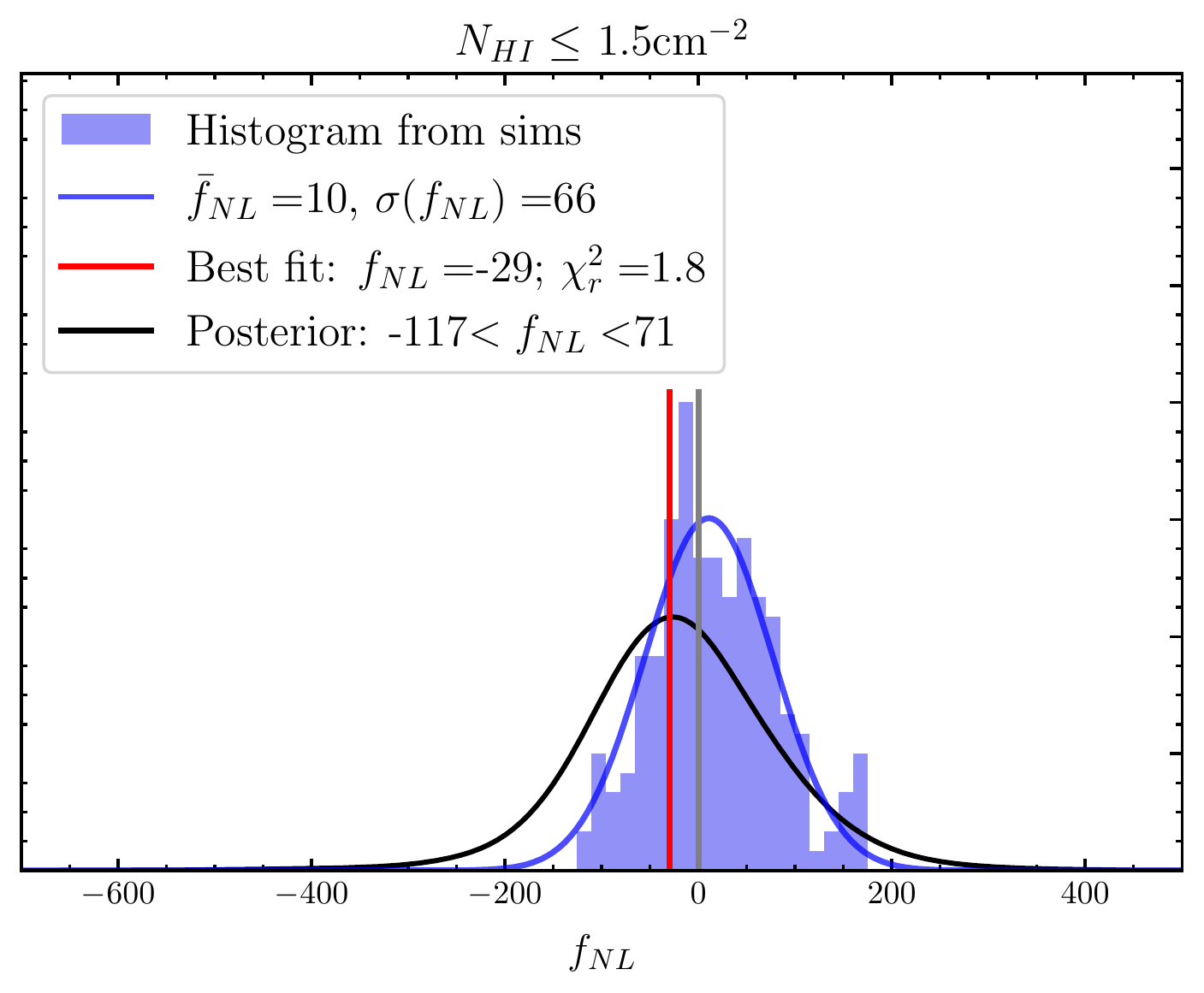}
\includegraphics[width=0.49\textwidth]{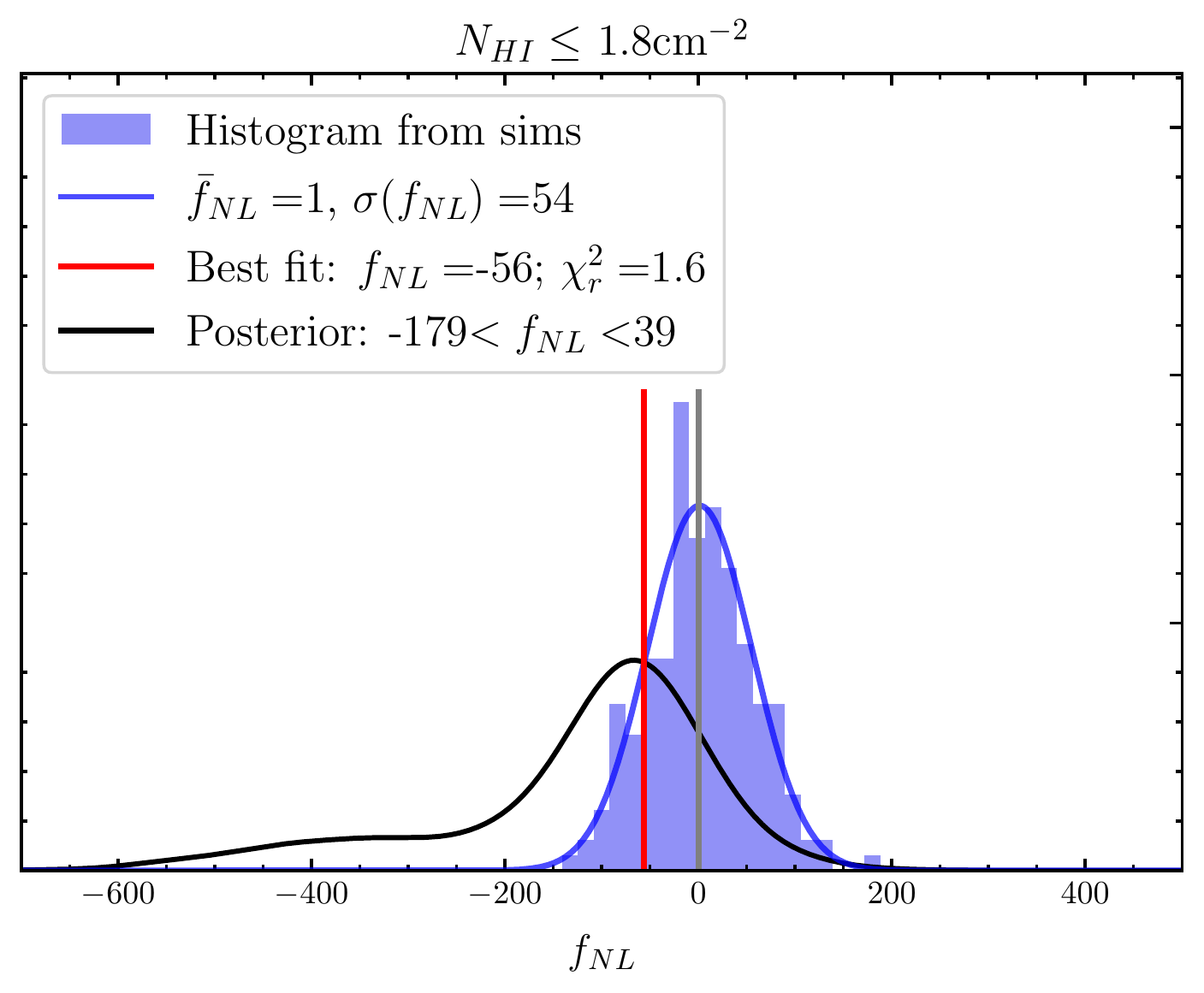}
\includegraphics[width=0.49\textwidth]{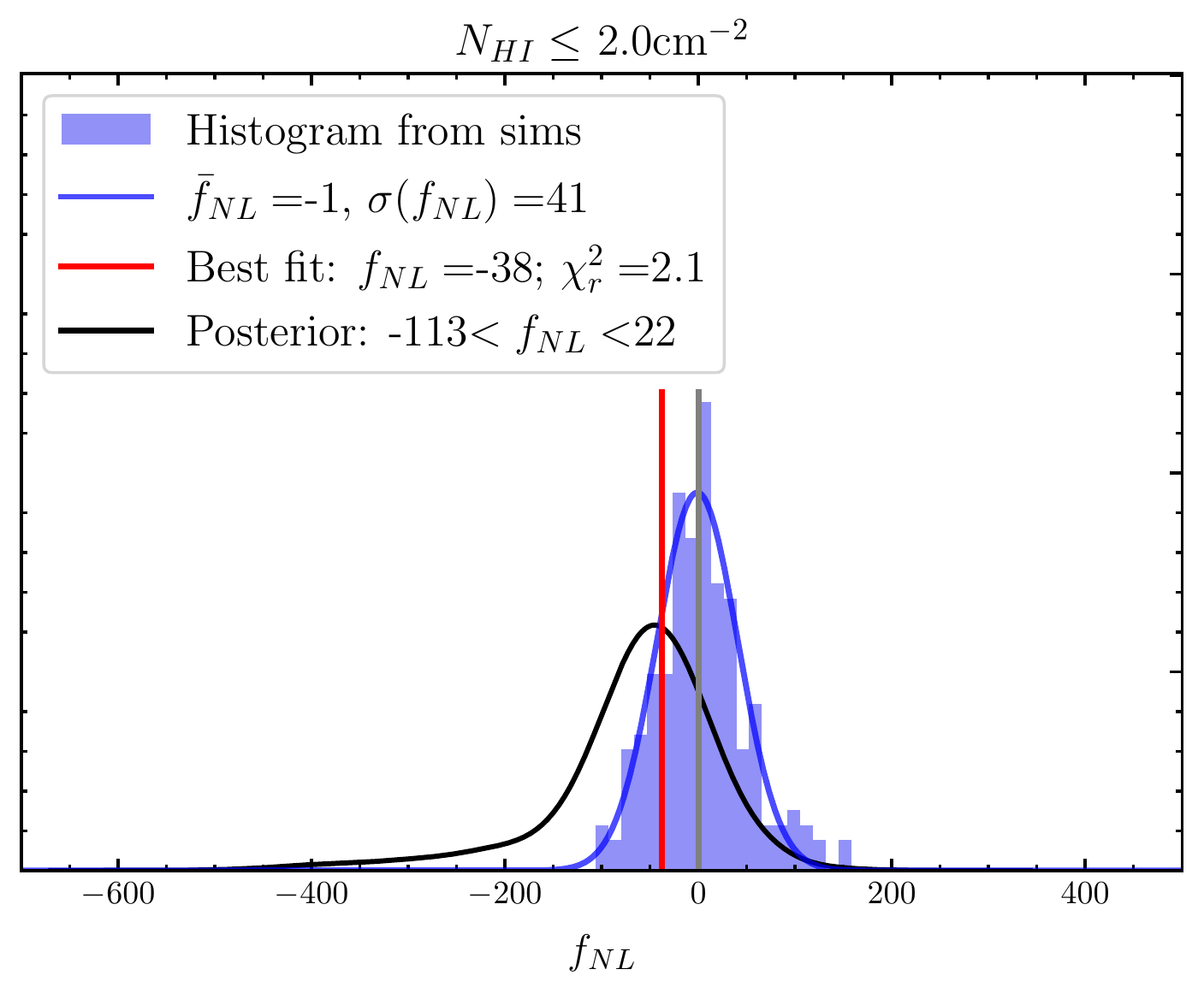}
\includegraphics[width=0.49\textwidth]{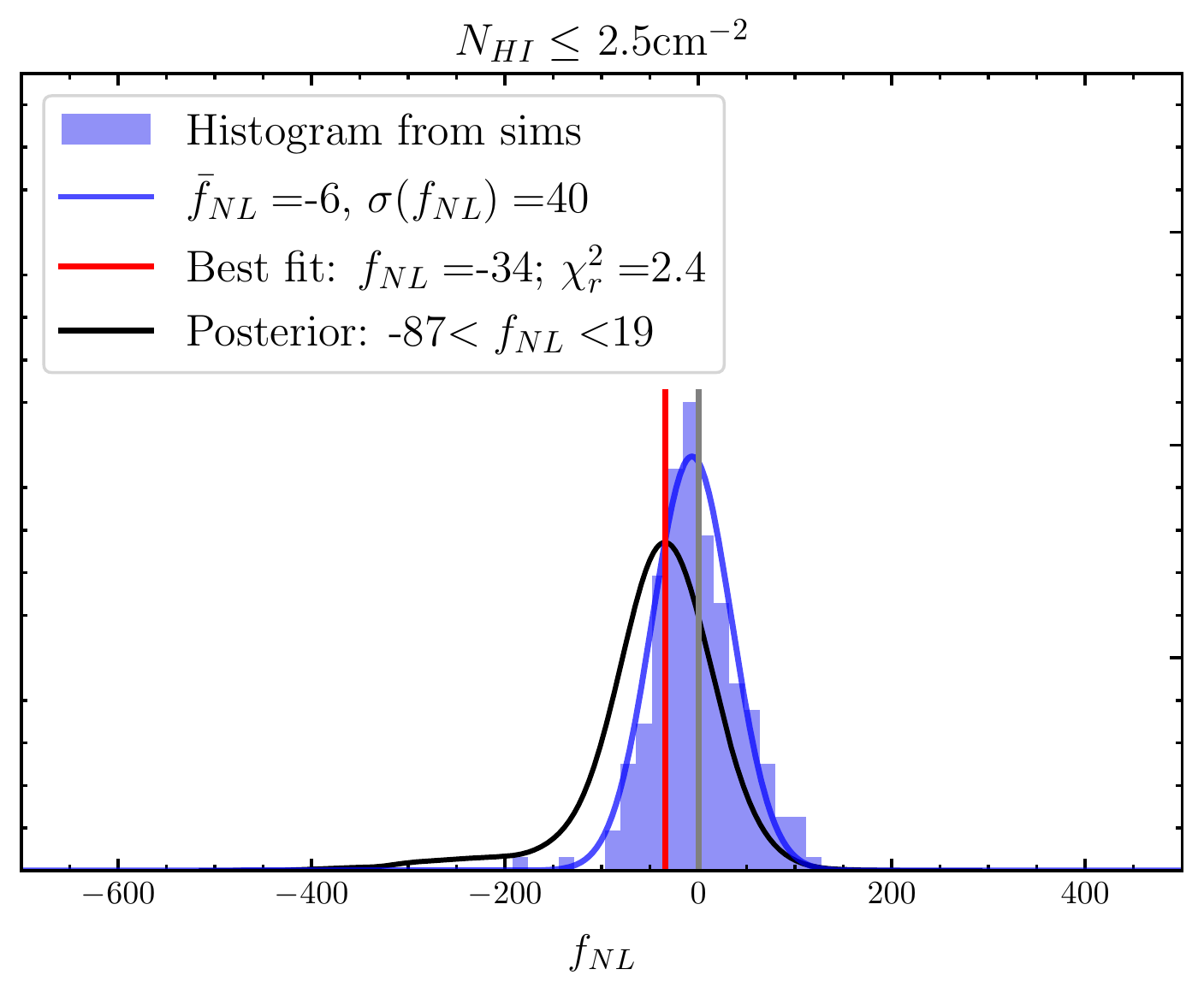}
\caption{Constraints on $\fnl$, for CIB maps with different HI thresholds.  All measurements are consistent with $\fnl=0$ corresponding to Gaussian initial conditions. In every plot we show the histogram of the recovered best-fit $\fnl$ from 200 simulations, and a Gaussian fit to this histogram; we indicate in the legend the mean and standard deviation of these Gaussians. We also show the posterior on $\fnl$ from our MCMC analyses, and indicate the $68\%$ confidence interval in the legends. We also show the best-fit $\fnl$ from the data with a red vertical line, and indicate the reduced $\chi^2$ ($\chi^2_r\equiv\chi^2/dof$)  in the legend.}\label{fig:skyareas_constraints}
\end{figure*}

Our results, for different sky-areas, are presented in Figure~\ref{fig:skyareas_constraints}. From our baseline $N_{HI}<2.5\,\mathrm{cm}^{-2}$ configuration we get a best-fit value of $\fnl=-34\pm40$. 

\begin{figure}[h!]
\includegraphics[width=\columnwidth]{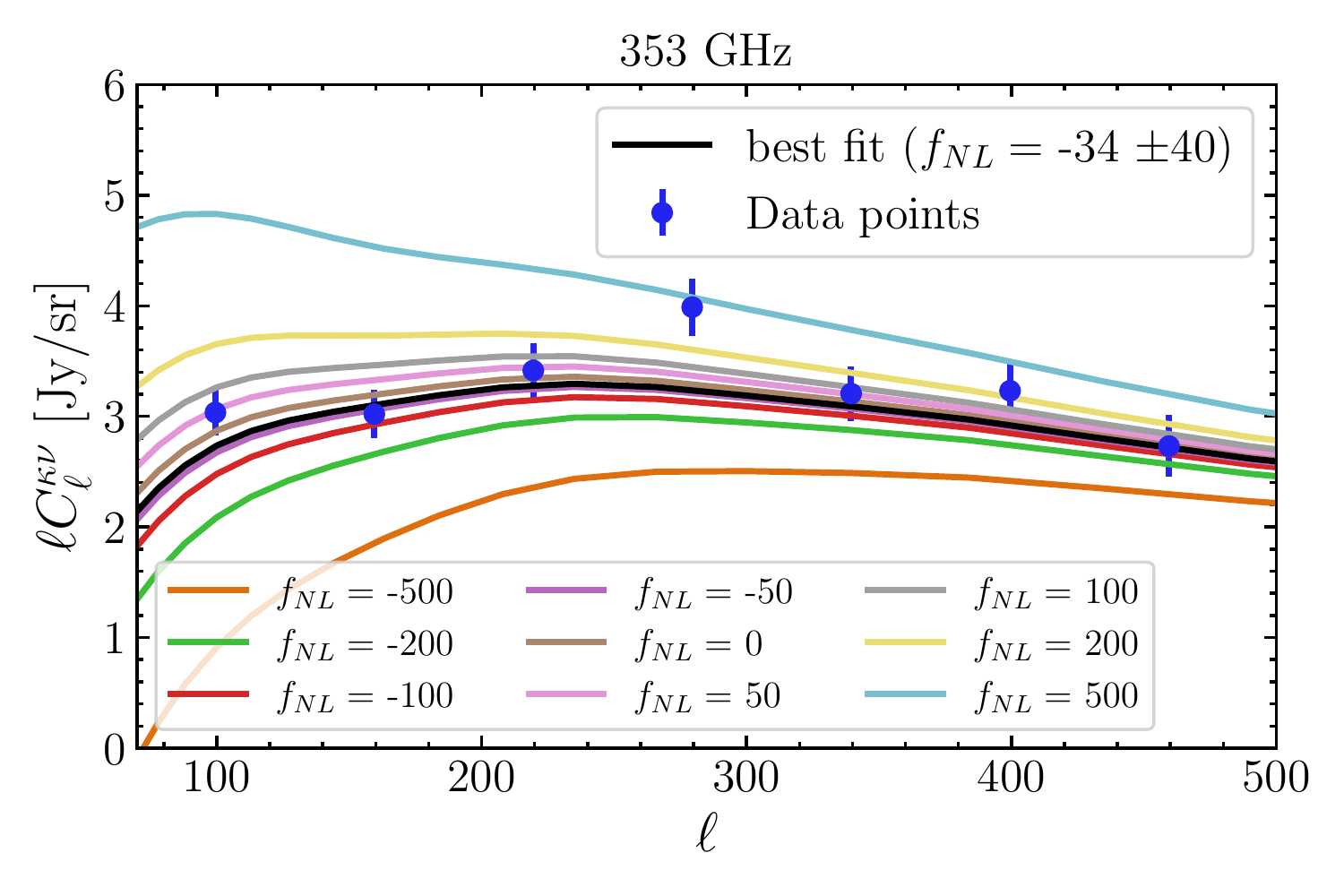}
\includegraphics[width=\columnwidth]{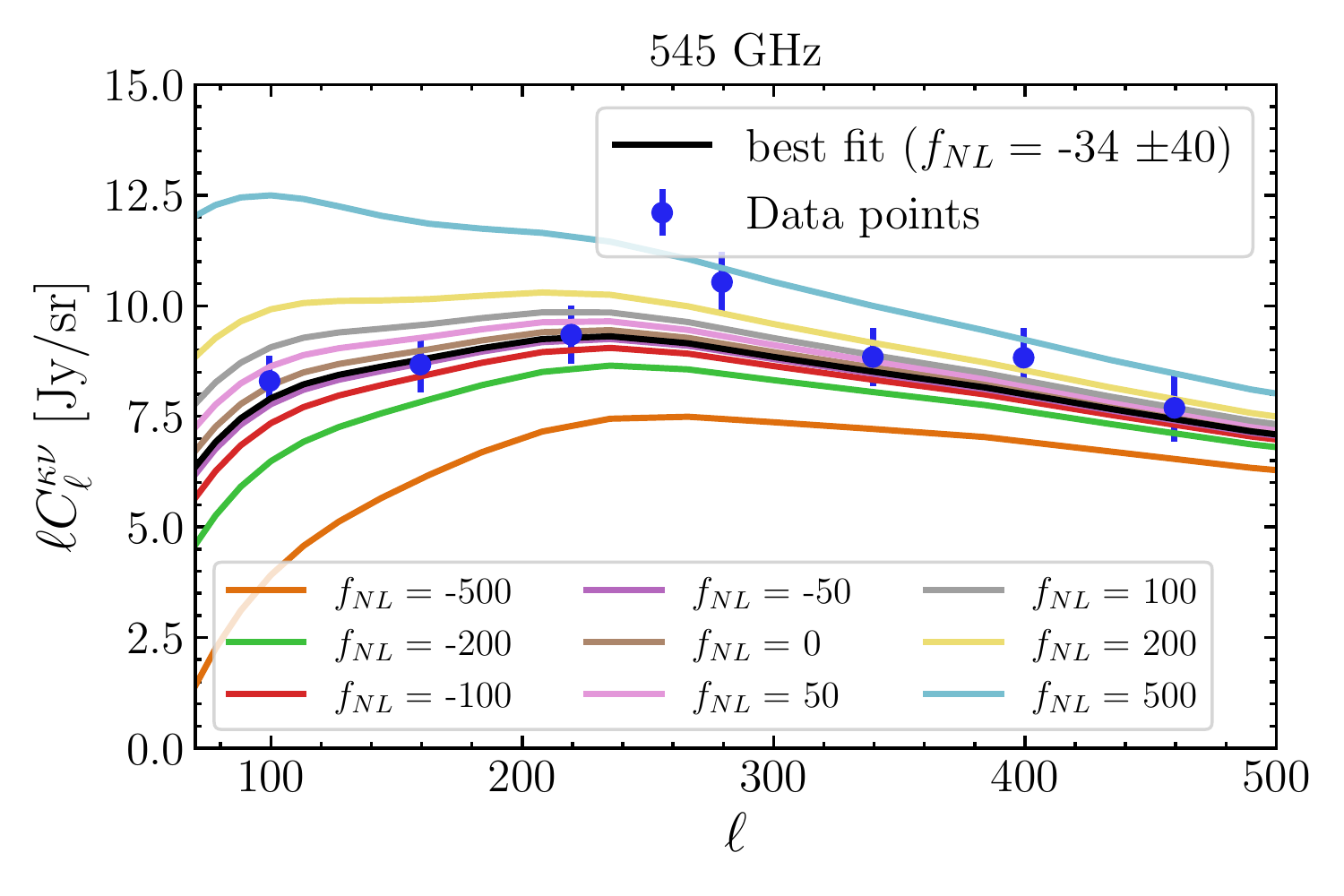}
\includegraphics[width=\columnwidth]{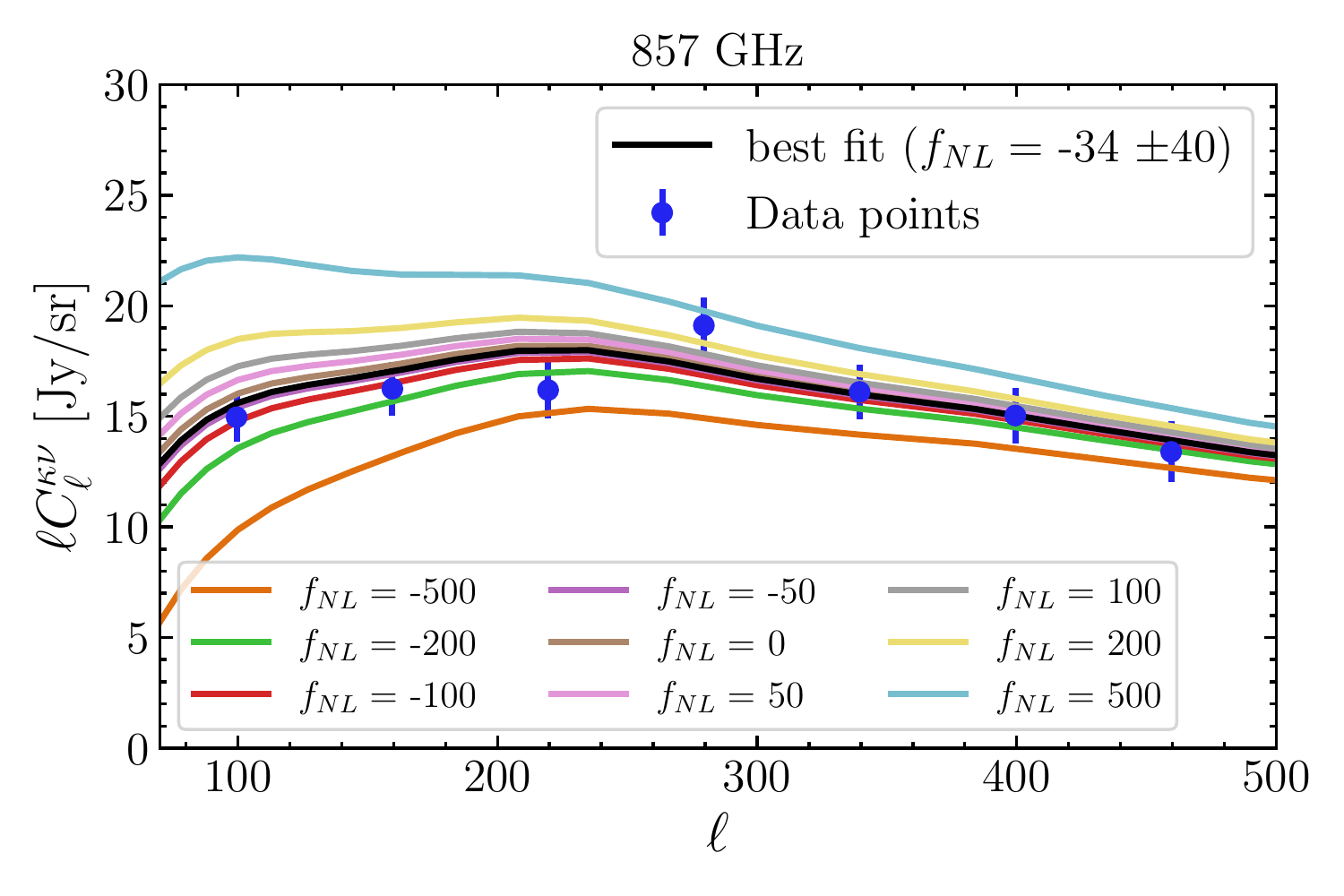}
\caption{The best-fit curves, and the data points, with different values of $\fnl$ (but the remaining parameters the same as the best-fit parameters, except $\fnl$) indicated.}\label{fig:bestfit}
\end{figure}

We show in Figure~\ref{fig:bestfit} the best-fit theory curves for $N_{HI}<2.5\,\mathrm{cm}^{-2}$. We also show other plots with varying values of $\fnl$, with the remaining parameters fixed to their best-fit values.  

We list the values of the best-fit $\fnl$, quantify the posteriors, and the histograms of the best-fit $\fnl$ from the simulations in Table~\ref{tab:results_fnl}. Our tightest constraint on $\fnl$, from the $N_{HI}<2.5\,\mathrm{cm}^{-2}$ maps, is \fnlconfidenceinterval; the reduced $\chi^2$ at the best-fit point is 2.3, with a Gaussian standard deviation of 41. We get a better fit, although a degraded constraint, from the smaller maps, as indicated in Table~\ref{tab:results_fnl}.

\begin{table}[h!]
\begin{tabular}{|c||c|c|c|}\hline
$N_{HI}$ & $67\%$ confidence limit & $\sigma(\fnl^{sim})$  & $\chi^2_r$\\\hline\hline
1.5cm$^{-2}$   &   -117$<\fnl<$71  &   66  &   1.8\\\hline
1.8cm$^{-2}$   &   -179$<\fnl<$39  &   54  &   1.6\\\hline
2.0cm$^{-2}$   &   -114$<\fnl<$22  &   41  &   2.1\\\hline
2.5cm$^{-2}$   &   -87$<\fnl<$19  &   40  &   2.4\\\hline
\end{tabular}
\caption{A summary of our $\fnl$ constraints, with the $67\%$ confidence interval from our $\fnl$ posteriors; the standard deviation of the recovered best-fit $\fnl$ from 200 simulations; and the reduced $\chi^2$ at the best-fit point in each case.}\label{tab:results_fnl}
\end{table}

\section{Future constraints with this method}\label{sec:future_constraints}

\subsection{Improvements from future CMB lensing experiments}

\begin{figure}
\includegraphics[width=\columnwidth]{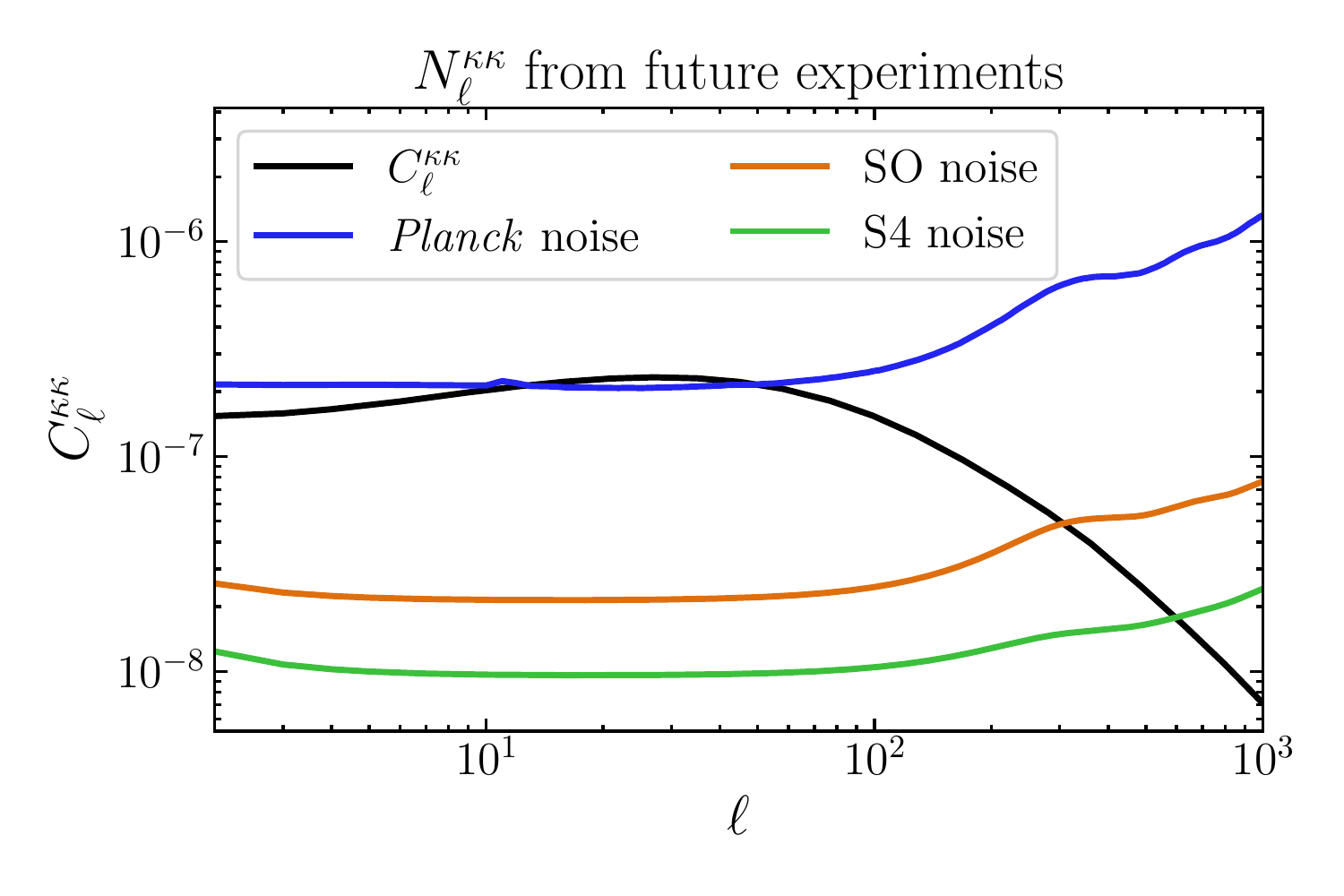}
\caption{The CMB lensing convergence signal $C_\ell^{\kappa\kappa}$ and the noise from \textit{Planck} and the future experiments SO and CMB-S4 (forecasts). The noise for \Planck could be further reduced by using the `GMV' CMB lensing quadratic estimator from \cite{Maniyar_2021} as done in \cite{2022JCAP...09..039C}; we will explore using the improved \Planck lensing map in future work.} \label{fig:lensingnoise_future}
\end{figure}

In the coming years, experiments such as ACT, SPT, the Simons Observatory (SO), and CMB-S4 will produce CMB lensing maps with far lower noise; see Figure~\ref{fig:lensingnoise_future} where we plot the signal and forecast noise from SO \cite{Ade:2018sbj} and CMB-S4 \cite{CMBS4DSR}. This will directly result in lower uncertainties in the measured CIB-$\kappa$ cross-correlation, and improved uncertainties on $\fnl$.

\begin{figure*}[t]
\includegraphics[width=\columnwidth]{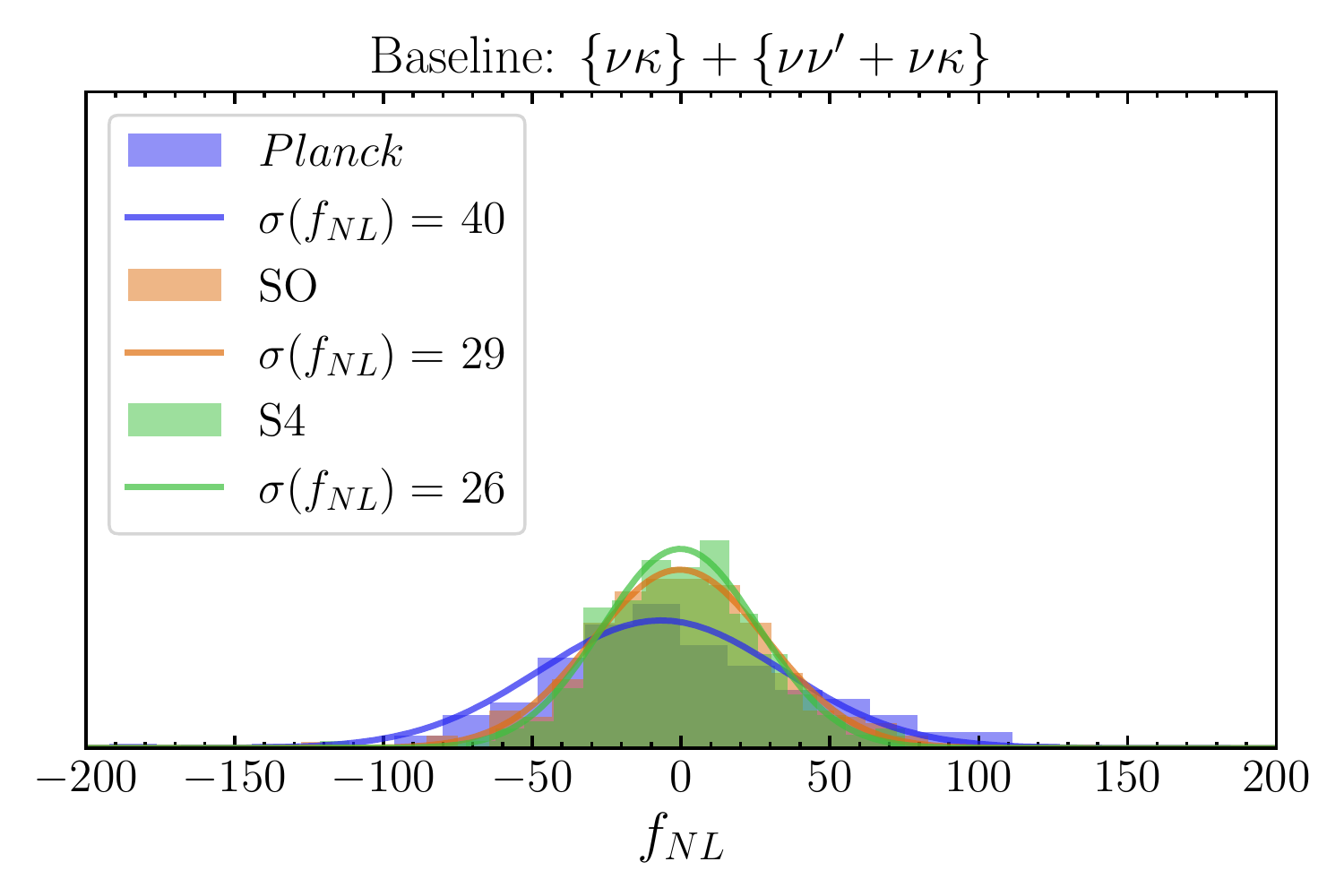}
\includegraphics[width=\columnwidth]{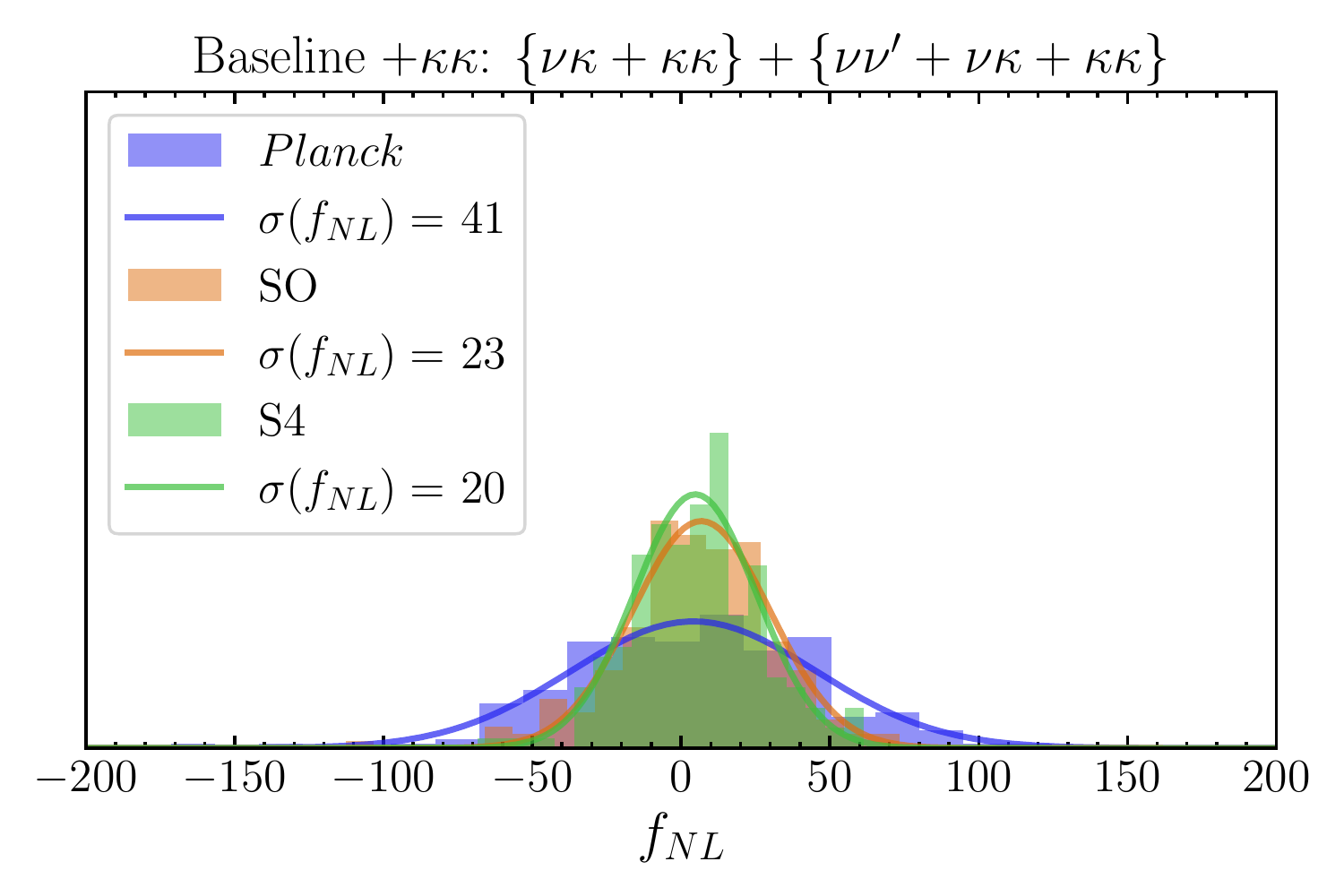}

\caption{Forecast constraints on $\fnl$ with this method, from future CMB experiments. We replace the \textit{Planck} noise curve in the CMB lensing data of our simulations with one appropriate for an SO-like experiment and a CMBS4-like experiment. We find that, for a ``baseline'' analysis exactly like the one we used in this work, there is potential for the uncertainties to decrease by a factor of $\sim30\%$ with SO; CMB-S4 can improve slightly further on this. However, if the lensing auto power spectrum is included (as on the right), there is room for further improvement via sample variance cancellation in the future experiments; however, for current (\textit{Planck}) data the noise on the lensing power spectrum is too high. 
For these comparisons we used the sky area corresponding to the  $N_{HI}<2.5\,\mathrm{cm}^{-2}$ maps.}\label{fig:future_constraints}
\end{figure*}

In  Figure~\ref{fig:future_constraints}, we show the $1\sigma$ forecast constraint from future experiments, calculated by simulating 200 datasets in each case and histogramming the recovered $\fnl$. We find that a similar analysis to ours but with an experiment like SO for the CMB lensing data will improve on our uncertainties by a factor of about 1.4, a significant improvement; however, at that point the uncertainties will saturate and there will be only slightly further improvement from an S4-like experiment. For future experiments we can exploit sample variance cancellation by including the $C^{\kappa\kappa}_{\ell}$ auto-power spectrum, and achieve with an SO-like experiment or an S4-like experiment $\sigma(\fnl)$ of 23 and 20 respectively; however, the higher noise levels of the \textit{Planck} $\kappa$ measurement prevent us from gaining significantly from including $C^{\kappa\kappa}_{\ell}$ in our current analysis (sample variance cancellation gains are possible typically when the fields are highly signal-dominated).

\begin{figure*}[t]
\includegraphics[width=0.3\textwidth]{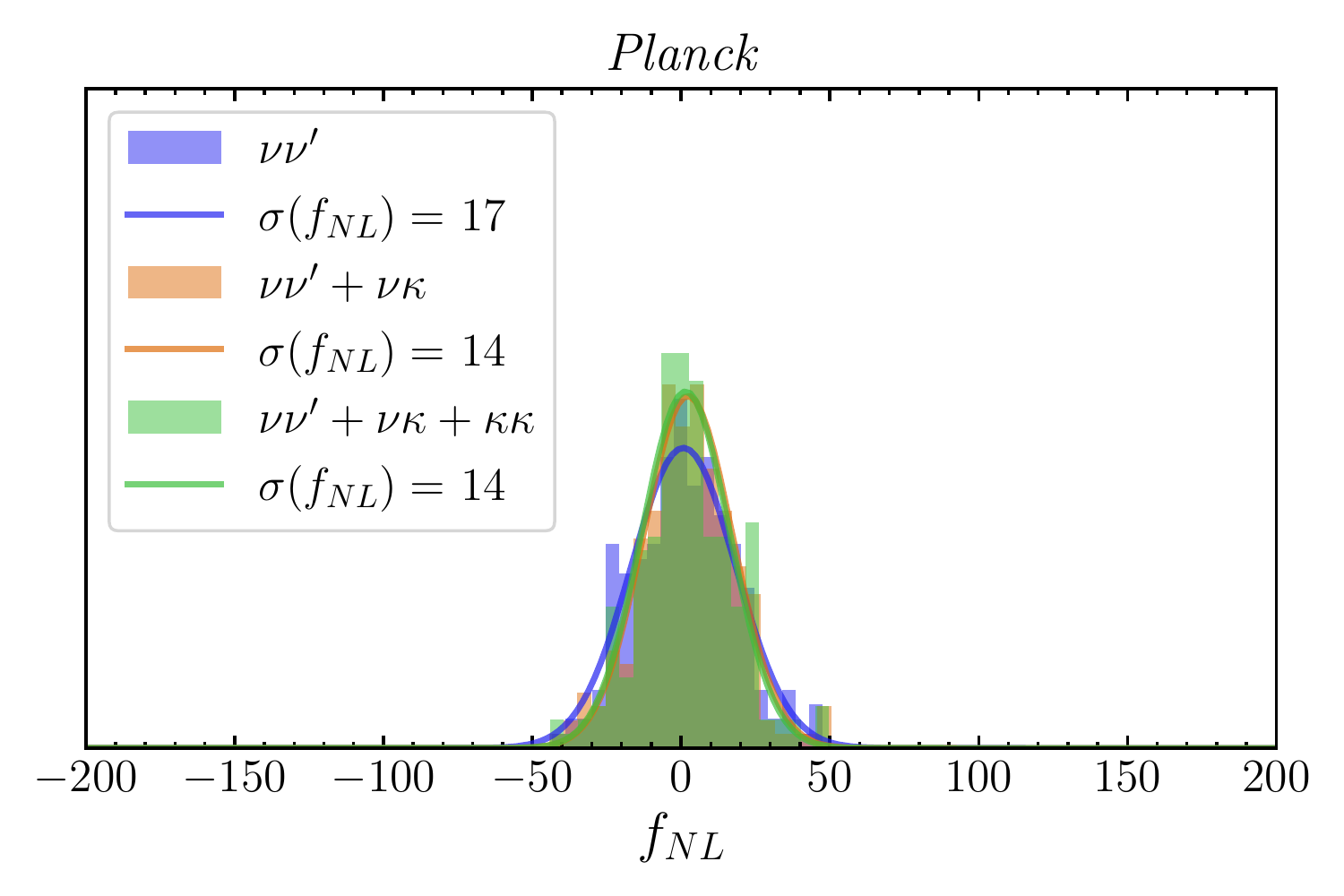}
\includegraphics[width=0.3\textwidth]{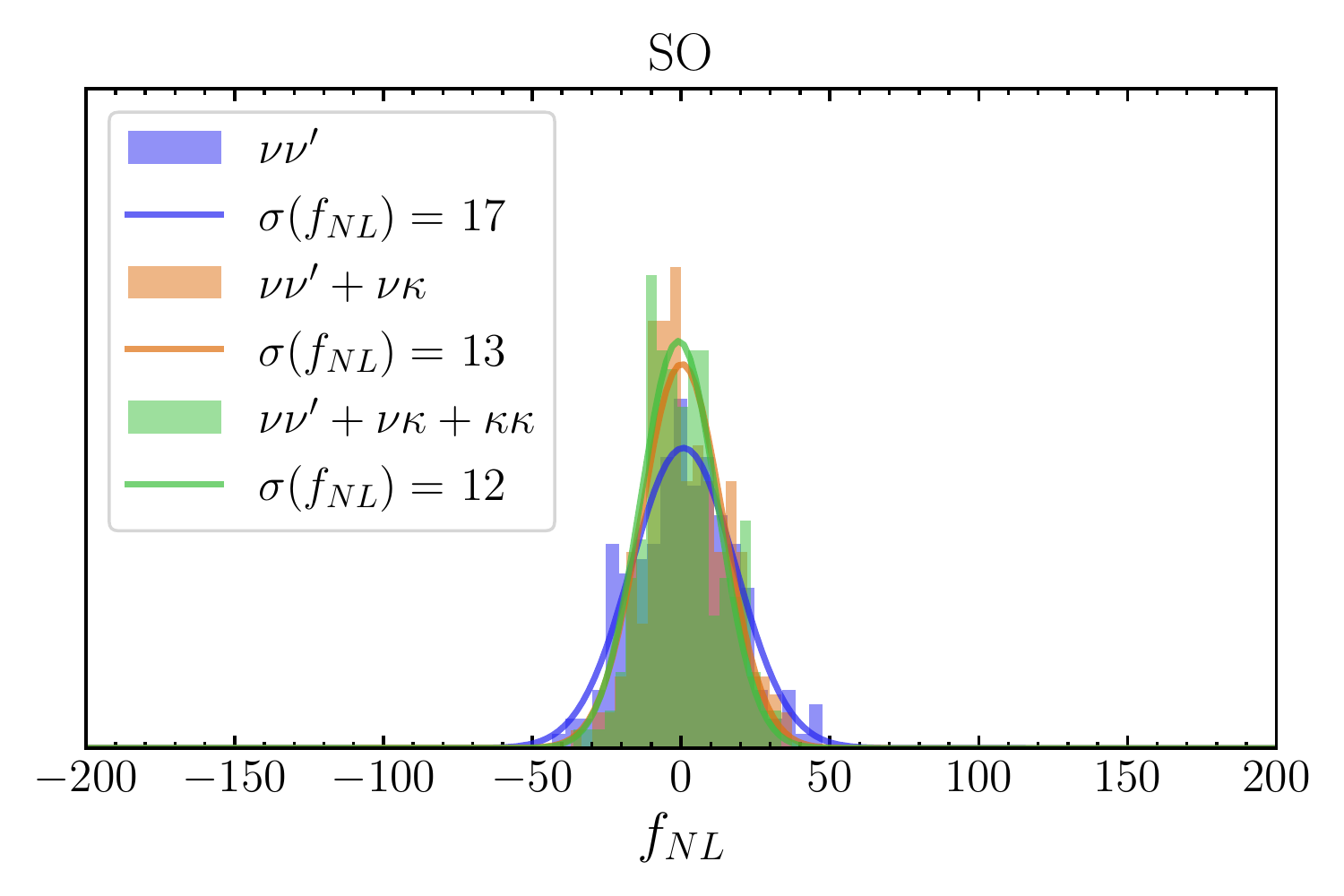}
\includegraphics[width=0.3\textwidth]{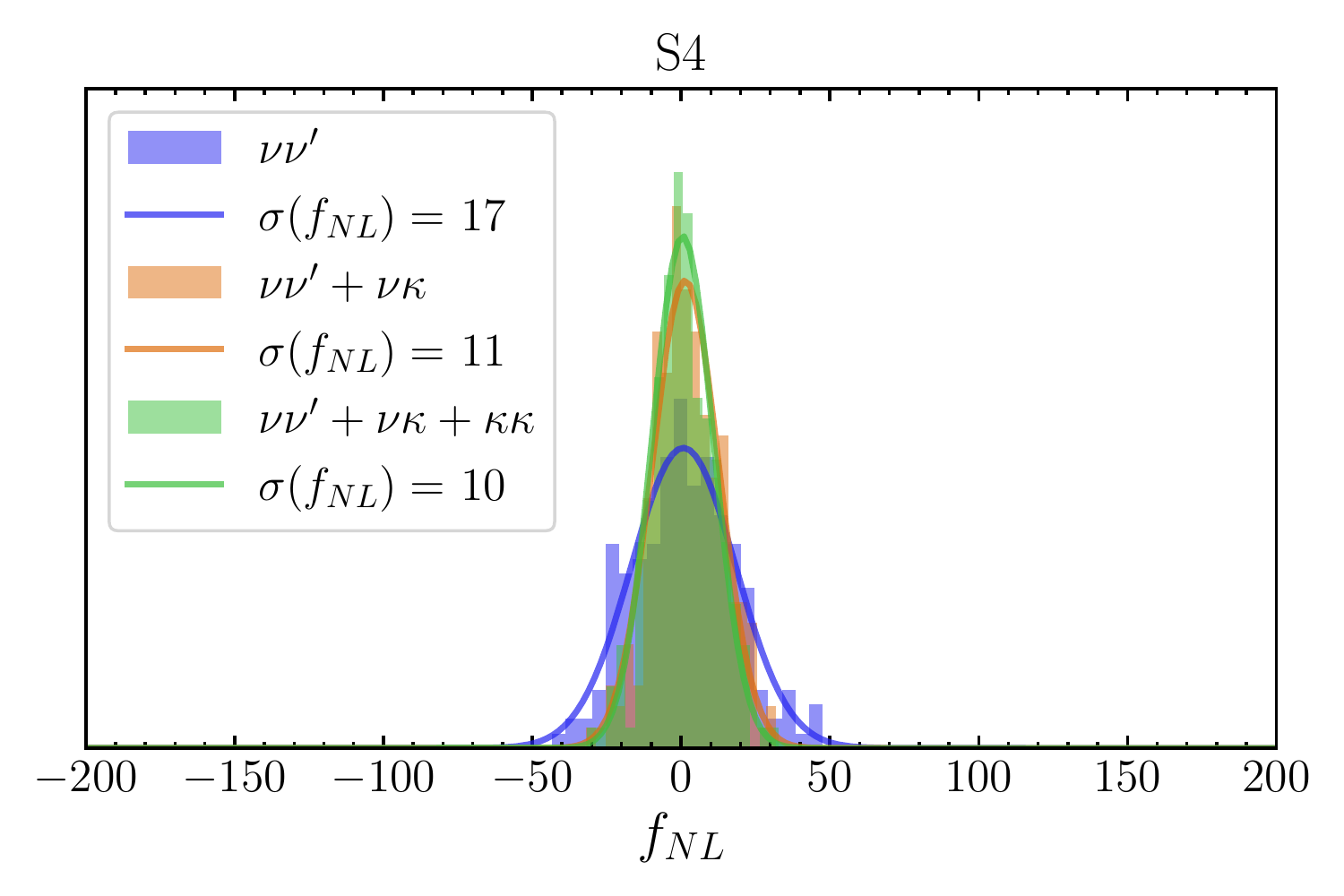}

\caption{Forecast constraints on $\fnl$ for the  CIB maps combined with various lensing experiments, including the $C_\ell^{\nu\nu}$ at all scales. We continue to use a minimum multipole of 70. We see that if we were to include the auto power spectrum at all scales, we could constrain $\fnl$ to $\sim14$ with the data we have at hand. The sample variance cancellation improvements from including the $\kappa\nu$ cross power spectrum and the $\kappa\kappa$ auto-power are also indicated.
}\label{fig:svcancellation}
\end{figure*}

In this analysis, we avoided using $C_\ell^{\nu\nu^\prime}$ at $\ell<430$ to avoid bias from Galactic dust. In Figure~\ref{fig:svcancellation} we show that the forecast uncertainty on a $C_\ell^{\nu\nu^\prime}$-alone analysis of the $N_{HI}<2.5\,\mathrm{cm}^{-2}$ field would achieve an impressive $\sigma(\fnl)=17$, with our baseline minimum multipole of $\ell_{\rm min}=70$. This could be improved with the \textit{Planck} lensing measurements to $\sigma(\fnl)=14$.

\subsection{Dependence on $\ell_{\mathrm{min}}$}

\begin{figure}
    \includegraphics[width=\columnwidth]{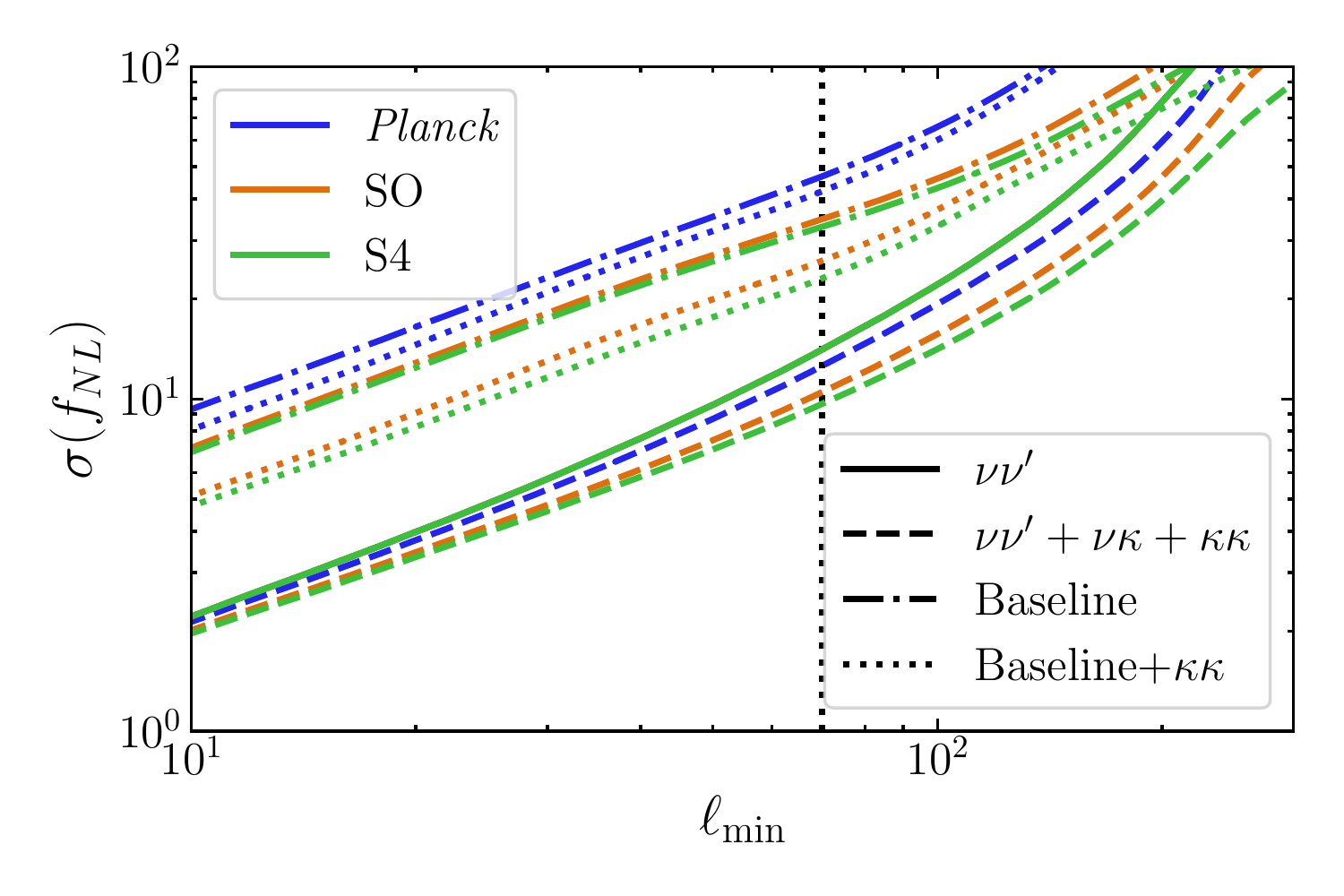}
    \caption{The behavior of the forecast uncertainty on $\fnl$ with $\ell_{\mathrm{min}}$. Note that we calculate this analytically, and do not include dust variance in the $\nu\nu$ covariance matrix, and so this is optimistic given current methods for Galactic dust subtraction (e.g \cite{Lenz:2019ugy}). We show for comparison the forecast constraints from $\nu\nu$ and from the full analysis $\nu\nu+\nu\kappa+\kappa\kappa$; we also include our one ``baseline'' configuration, i.e. $\nu\kappa$ for $\ell<430$ and $\nu\nu+\nu\kappa$ for $\ell>430$, along with the ``baseline$+\kappa\kappa$'' configuration, i.e. $\nu\kappa+\kappa\kappa$ for $\ell<430$ and $\nu\nu+\nu\kappa+\kappa\kappa$ for $\ell>430$. In all cases we take a maximum multipole $\ell_{\mathrm{max}}=610$. Note that the $\ell_{\mathrm{min}}$ we used in our analysis is $\ell_{\mathrm{min}}=70$, which we have indicated on the plot with a vertical dotted line. }
    \label{fig:lmin_forecast}
\end{figure}

In this analysis, we have restricted ourselves to a minimum multipole $\ell_{\mathrm{min}}=70$, as the CIB maps of~\cite{1905.00426} are only unbiased above $\ell_{\mathrm{min}}\sim70$. Since scale-dependent bias from $\fnl$ has a $1/k^2$ dependence, the loss of information at low $\ell$ is a severe hindrance. Maintaining optimism that additional external data or new analysis techniques could help clean Galactic dust at lower multipoles in the future, we explore how $\fnl$ constraints could improve if future CIB maps were reliable on larger scales than used in our analysis. We show in Figure~\ref{fig:lmin_forecast} the constraints we would expect to get if we could decrease $\ell_{\mathrm{min}}$. We calculate these forecast uncertainties on $\fnl$ with a Fisher matrix for the parameters, calculated according to
\be
F_{ij}(\Pi)= \sum_\ell\frac{\partial C_\ell(\Pi)^{T}}{\partial \Pi^i} \mathbb{C}_\ell^{-1}\frac{\partial C_\ell(\Pi)}{\partial \Pi^j}
\ee
where $C_\ell(\Pi)$ is the theoretical data vector which depends on the parameter vector $\Pi$, and $\mathbb{C}_\ell$ is the covariance matrix defined in Equation~\eqref{covm_defn} (note that we use the analytical covariance matrix in this forecast, not a covariance matrix from simulations as we did in our analysis). We take the sky area to be $f_{\mathrm{sky}}=0.1795$, corresponding to the $N_{HI}<2.5\,\mathrm{cm}^{-2}$ threshold. The priors on the parameters $b_0$ and $f_\nu$ are included according to
\be
F = F(\Pi)+\sum_iC^{-1}_{\mathrm{prior}_i}
\ee
where $C^{-1}_{\mathrm{prior}_i}$ is a matrix of zeros with $C^{-1}_{i,i}=1/\sigma_{\mathrm{prior}_i}^2$. For simplicity, we do not include the prior on the mean value of the CIB in the Fisher forecast.

The marginalized forecast parameter constraints are calculated from the diagonal of the inverse Fisher matrix according to
\be
\sigma_{\Pi^i} = \sqrt{(F^{-1})_{ii}},
\ee
so $\sigma_{\fnl} = \sqrt{(F^{-1})_{\fnl\fnl}}$.

The resulting forecast constraints are shown in Figure~\ref{fig:lmin_forecast}. Although we have avoided the CIB auto power spectrum in our analysis, we show the constraints for $\nu\nu^\prime$ along with the $\nu\nu^\prime+\nu\kappa+\kappa\kappa$ constraints which can take full advantage of sample variance as the noise on the CMB lensing reconstruction is reduced. We also show our ``baseline'' constraints, which agree well with the calculated constraint from simulations of $\sigma(\fnl)=40$ for \textit{Planck} with $\ell_{\mathrm{min}}=70.$

Notably, for $\ell_{\mathrm{min}}=10$ a constraint with $\sigma(\fnl)\sim4$---better than the existing constraints from the primary CMB bispectrum---can be obtained through cross-correlation alone. Including CIB auto-spectra allows constraints stronger than $\sigma(\fnl)\sim 2$, beginning to probe multi-field inflation. These forecasts are optimistic (they also neglect dust variance on the $\nu\kappa$ cross correlation), but serve to show what can be achieved with CIB maps cleaned to the extent of~\cite{1905.00426} to lower multipoles and provide a guide for the full ``Fisher information'' in the CIB field.

\section{Discussion}\label{sec:discussion}

In this work, we have presented the  strongest large-scale structure (LSS) constraint on local primordial non-Gaussianity that utilizes cross-correlations alone; using the CIB as a halo tracer and CMB lensing as a probe of the underlying matter distribution, we constrain scale-dependent halo bias avoiding potential systematics associated with auto-spectra.  In particular, we exploited the independent large-scale systematics of CMB lensing and the CIB emission to achieve an estimate of $\fnl$ without Galactic dust bias. With future CMB experiments, the reconstruction noise will decrease to such a level that there will be potential improvements on the constraint from including the CMB lensing auto power spectrum, which reduces the impact of sample variance in the measurement. 

Our baseline constraint on $\fnl$ from the largest maps we used is $-87<\fnl<19$ (68\% confidence interval from an MCMC sample); this compares to a Gaussian uncertainty of $\sigma(\fnl)=40$, with a reduced $\chi^2$ of $\chi^2_r=2.4$. We also analysed smaller maps, which in principle have less Galactic dust power, and found in our most conservative case that $-117<\fnl<72$, comparing to a Gaussian uncertainty of  $\sigma(\fnl)=66$, with an improved reduced $\chi^2$ of $\chi^2_r=1.8$ (note that any extra Galactic dust in the larger maps adds variance although not bias on large scales). Thus we found no evidence for an $\fnl$ signal in any data that we considered. This is consistent with the independent constraint from the CMB bispectrum of $\fnl=-0.9\pm5.1$, and from the scale dependent bias of BOSS quasars~\cite{2021arXiv210613725M} of  $-12\pm21$.

There is a large amount of constraining power on $\fnl$ in the CIB auto power spectrum, which we have conservatively avoided by including the CIB auto power spectrum only at multipoles $\ge430$, which were required to constrain the CIB bias and star formation rate. We have explicitly calculated the constraining power on $\fnl$ in the CIB auto power spectrum, if we were to obtain a dust-free measurement (or indeed, to do a less conservative analysis on the data at-hand). We find that even with an $\ell_{\mathrm{min}}$ of 70, the CIB auto power spectrum alone could constrain $\fnl$ to $\sigma(\fnl)=14$ when combined with CMB lensing. Future CMB lensing data, in particular those of CMBS4, could improve this to $\sigma(\fnl)=10$. Even remaining conservative and including no CIB auto power spectrum below $\ell_{\mathrm{min}}=430$ (as we have done in this analysis), we find that Simons Observatory and CMBS4 can constrain $\fnl$ to $\sigma(\fnl)=23$ and $\sigma(\fnl)=20$ respectively.

A key limiting factor in our analysis is that we only use multipoles $\ell>70$ in our analysis, since the CIB maps from \cite{1905.00426} contain a multiplicative transfer function below those scales that would bias a cross-correlation. This bias arises from the way a template of the Galactic dust is constructed after splitting the sky into HEALPIX~\cite{2005ApJ...622..759G} superpixel patches; a linear model is fit against the neutral hydrogen and observed \Planck far-infrared data in these superpixels, but the finite size of the patches effectively induces a high-pass filter that is significant below around $\ell \sim 70$. We will explore in future work whether alternative analysis techniques (including obtaining the multiplicative transfer function from simulations) can overcome this limitation thus allowing us to vastly improve the $\fnl$ constraint from existing data. We have shown how a dust-free CIB map down to $\ell=10$ can provide $\sigma(\fnl)=10$ with currently available \Planck data, which would be the tightest $\fnl$ constraint from the late-universe imprint of primordial non-Gaussianity in large-scale structure. With future CMB lensing data from SO or S4, such a CIB map could provide a better $\fnl$ constraint --- $\sigma(\fnl)$ below 4 --- than any existing measurement, improving that from the primary CMB early-universe bispectrum measured by \Planck. If the CIB auto-spectrum could be reliable in such a map, then a $\sigma(\fnl)$ below 2 could be achieved with just the \Planck CIB auto-spectrum.

While we aimed to be conservative in this work by allowing for a redshift dependent CIB bias and marginalizing over three associated bias parameters, we have made an assumption regarding the universality of the halo mass function, i.e., we assume that the relation in Eq. \ref{fnlbias} is exact. There has been compelling recent work that shows that this relation does not hold universally and exactly \cite{2006.09368,2009.06622,2107.06887,2022arXiv220505673B,2209.07251}, introducing dependences degenerate with $\fnl$ not only through the kinds of galaxies or halos used but even their formation history or assembly bias \cite{1004.1637}. In this picture, our constraints can be thought of as a constraint on $b_\phi \fnl$, with a fiducial value of $b_\phi = 2\delta_c(b^{G}-1)$ (obtained from universality) that may differ from the $b_\phi$ expected for dusty galaxies constituting the CIB. It is important to note that despite this, a detection of scale-dependent bias on large scales still constitutes evidence for primordial non-Gaussianity; however, the interpretation of such a detection in terms of multi-field inflation models then becomes more challenging. Nevertheless, the line of work pursued in e.g.\cite{2022arXiv220505673B} strongly motivates exploring this further with simulations which may deliver strong priors on $b_\phi$.  We leave further investigation of this issue to future work.

\begin{acknowledgements}
 We thank Anthony Challinor, Neal Dalal, Blake Sherwin and Alexander van Engelen for helpful discussions. We thank the Scientific Computing Core staff at the Flatiron Institute for computational support.  The Flatiron Institute is supported by the Simons Foundation.  Research at the Perimeter Institute is supported in part by the Government of Canada through the Department of Innovation, Science and Economic Development, and by the Province of Ontario through the Ministry of Colleges and Universities. 
\end{acknowledgements}

\clearpage

\bibliography{references}

\begin{thebibliography}{83}
\expandafter\ifx\csname natexlab\endcsname\relax\def\natexlab#1{#1}\fi
\expandafter\ifx\csname bibnamefont\endcsname\relax
  \def\bibnamefont#1{#1}\fi
\expandafter\ifx\csname bibfnamefont\endcsname\relax
  \def\bibfnamefont#1{#1}\fi
\expandafter\ifx\csname citenamefont\endcsname\relax
  \def\citenamefont#1{#1}\fi
\expandafter\ifx\csname url\endcsname\relax
  \def\url#1{\texttt{#1}}\fi
\expandafter\ifx\csname urlprefix\endcsname\relax\def\urlprefix{URL }\fi
\providecommand{\bibinfo}[2]{#2}
\providecommand{\eprint}[2][]{\url{#2}}

\bibitem[{\citenamefont{{de Putter} et~al.}(2017)\citenamefont{{de Putter},
  {Gleyzes}, and {Dor{\'e}}}}]{2017PhRvD..95l3507D}
\bibinfo{author}{\bibfnamefont{R.}~\bibnamefont{{de Putter}}},
  \bibinfo{author}{\bibfnamefont{J.}~\bibnamefont{{Gleyzes}}},
  \bibnamefont{and}
  \bibinfo{author}{\bibfnamefont{O.}~\bibnamefont{{Dor{\'e}}}},
  \bibinfo{journal}{\prd} \textbf{\bibinfo{volume}{95}}, \bibinfo{eid}{123507}
  (\bibinfo{year}{2017}), \eprint{1612.05248}.

\bibitem[{\citenamefont{{Planck Collaboration}
  et~al.}(2020{\natexlab{a}})\citenamefont{{Planck Collaboration}, {Akrami},
  {Arroja}, {Ashdown}, {Aumont}, {Baccigalupi}, {Ballardini}, {Banday},
  {Barreiro}, {Bartolo} et~al.}}]{2020A&A...641A...9P}
\bibinfo{author}{\bibnamefont{{Planck Collaboration}}},
  \bibinfo{author}{\bibfnamefont{Y.}~\bibnamefont{{Akrami}}},
  \bibinfo{author}{\bibfnamefont{F.}~\bibnamefont{{Arroja}}},
  \bibinfo{author}{\bibfnamefont{M.}~\bibnamefont{{Ashdown}}},
  \bibinfo{author}{\bibfnamefont{J.}~\bibnamefont{{Aumont}}},
  \bibinfo{author}{\bibfnamefont{C.}~\bibnamefont{{Baccigalupi}}},
  \bibinfo{author}{\bibfnamefont{M.}~\bibnamefont{{Ballardini}}},
  \bibinfo{author}{\bibfnamefont{A.~J.} \bibnamefont{{Banday}}},
  \bibinfo{author}{\bibfnamefont{R.~B.} \bibnamefont{{Barreiro}}},
  \bibinfo{author}{\bibfnamefont{N.}~\bibnamefont{{Bartolo}}},
  \bibnamefont{et~al.}, \bibinfo{journal}{\aap} \textbf{\bibinfo{volume}{641}},
  \bibinfo{eid}{A9} (\bibinfo{year}{2020}{\natexlab{a}}), \eprint{1905.05697}.

\bibitem[{\citenamefont{{Ade} et~al.}(2019)\citenamefont{{Ade}, {Aguirre},
  {Ahmed}, {Aiola}, {Ali}, {Alonso}, {Alvarez}, {Arnold}, {Ashton},
  {Austermann} et~al.}}]{Ade:2018sbj}
\bibinfo{author}{\bibfnamefont{P.}~\bibnamefont{{Ade}}},
  \bibinfo{author}{\bibfnamefont{J.}~\bibnamefont{{Aguirre}}},
  \bibinfo{author}{\bibfnamefont{Z.}~\bibnamefont{{Ahmed}}},
  \bibinfo{author}{\bibfnamefont{S.}~\bibnamefont{{Aiola}}},
  \bibinfo{author}{\bibfnamefont{A.}~\bibnamefont{{Ali}}},
  \bibinfo{author}{\bibfnamefont{D.}~\bibnamefont{{Alonso}}},
  \bibinfo{author}{\bibfnamefont{M.~A.} \bibnamefont{{Alvarez}}},
  \bibinfo{author}{\bibfnamefont{K.}~\bibnamefont{{Arnold}}},
  \bibinfo{author}{\bibfnamefont{P.}~\bibnamefont{{Ashton}}},
  \bibinfo{author}{\bibfnamefont{J.}~\bibnamefont{{Austermann}}},
  \bibnamefont{et~al.}, \bibinfo{journal}{\jcap}
  \textbf{\bibinfo{volume}{2019}}, \bibinfo{eid}{056} (\bibinfo{year}{2019}),
  \eprint{1808.07445}.

\bibitem[{\citenamefont{{Baldauf} et~al.}(2011)\citenamefont{{Baldauf},
  {Seljak}, and {Senatore}}}]{1011.1513}
\bibinfo{author}{\bibfnamefont{T.}~\bibnamefont{{Baldauf}}},
  \bibinfo{author}{\bibfnamefont{U.}~\bibnamefont{{Seljak}}}, \bibnamefont{and}
  \bibinfo{author}{\bibfnamefont{L.}~\bibnamefont{{Senatore}}},
  \bibinfo{journal}{\jcap} \textbf{\bibinfo{volume}{2011}}, \bibinfo{eid}{006}
  (\bibinfo{year}{2011}), \eprint{1011.1513}.

\bibitem[{\citenamefont{{Shirasaki} et~al.}(2021)\citenamefont{{Shirasaki},
  {Sugiyama}, {Takahashi}, and {Kitaura}}}]{2010.04567}
\bibinfo{author}{\bibfnamefont{M.}~\bibnamefont{{Shirasaki}}},
  \bibinfo{author}{\bibfnamefont{N.~S.} \bibnamefont{{Sugiyama}}},
  \bibinfo{author}{\bibfnamefont{R.}~\bibnamefont{{Takahashi}}},
  \bibnamefont{and} \bibinfo{author}{\bibfnamefont{F.-S.}
  \bibnamefont{{Kitaura}}}, \bibinfo{journal}{\prd}
  \textbf{\bibinfo{volume}{103}}, \bibinfo{eid}{023506} (\bibinfo{year}{2021}),
  \eprint{2010.04567}.

\bibitem[{\citenamefont{{Coulton}
  et~al.}(2022{\natexlab{a}})\citenamefont{{Coulton}, {Villaescusa-Navarro},
  {Jamieson}, {Baldi}, {Jung}, {Karagiannis}, {Liguori}, {Verde}, and
  {Wandelt}}}]{2206.01619}
\bibinfo{author}{\bibfnamefont{W.~R.} \bibnamefont{{Coulton}}},
  \bibinfo{author}{\bibfnamefont{F.}~\bibnamefont{{Villaescusa-Navarro}}},
  \bibinfo{author}{\bibfnamefont{D.}~\bibnamefont{{Jamieson}}},
  \bibinfo{author}{\bibfnamefont{M.}~\bibnamefont{{Baldi}}},
  \bibinfo{author}{\bibfnamefont{G.}~\bibnamefont{{Jung}}},
  \bibinfo{author}{\bibfnamefont{D.}~\bibnamefont{{Karagiannis}}},
  \bibinfo{author}{\bibfnamefont{M.}~\bibnamefont{{Liguori}}},
  \bibinfo{author}{\bibfnamefont{L.}~\bibnamefont{{Verde}}}, \bibnamefont{and}
  \bibinfo{author}{\bibfnamefont{B.~D.} \bibnamefont{{Wandelt}}},
  \bibinfo{journal}{arXiv e-prints} \bibinfo{eid}{arXiv:2206.01619}
  (\bibinfo{year}{2022}{\natexlab{a}}), \eprint{2206.01619}.

\bibitem[{\citenamefont{{Coulton}
  et~al.}(2022{\natexlab{b}})\citenamefont{{Coulton}, {Villaescusa-Navarro},
  {Jamieson}, {Baldi}, {Jung}, {Karagiannis}, {Liguori}, {Verde}, and
  {Wandelt}}}]{2206.15450}
\bibinfo{author}{\bibfnamefont{W.~R.} \bibnamefont{{Coulton}}},
  \bibinfo{author}{\bibfnamefont{F.}~\bibnamefont{{Villaescusa-Navarro}}},
  \bibinfo{author}{\bibfnamefont{D.}~\bibnamefont{{Jamieson}}},
  \bibinfo{author}{\bibfnamefont{M.}~\bibnamefont{{Baldi}}},
  \bibinfo{author}{\bibfnamefont{G.}~\bibnamefont{{Jung}}},
  \bibinfo{author}{\bibfnamefont{D.}~\bibnamefont{{Karagiannis}}},
  \bibinfo{author}{\bibfnamefont{M.}~\bibnamefont{{Liguori}}},
  \bibinfo{author}{\bibfnamefont{L.}~\bibnamefont{{Verde}}}, \bibnamefont{and}
  \bibinfo{author}{\bibfnamefont{B.~D.} \bibnamefont{{Wandelt}}},
  \bibinfo{journal}{arXiv e-prints} \bibinfo{eid}{arXiv:2206.15450}
  (\bibinfo{year}{2022}{\natexlab{b}}), \eprint{2206.15450}.

\bibitem[{\citenamefont{{Dalal} et~al.}(2008)\citenamefont{{Dalal}, {Dor{\'e}},
  {Huterer}, and {Shirokov}}}]{0710.4560}
\bibinfo{author}{\bibfnamefont{N.}~\bibnamefont{{Dalal}}},
  \bibinfo{author}{\bibfnamefont{O.}~\bibnamefont{{Dor{\'e}}}},
  \bibinfo{author}{\bibfnamefont{D.}~\bibnamefont{{Huterer}}},
  \bibnamefont{and}
  \bibinfo{author}{\bibfnamefont{A.}~\bibnamefont{{Shirokov}}},
  \bibinfo{journal}{\prd} \textbf{\bibinfo{volume}{77}}, \bibinfo{eid}{123514}
  (\bibinfo{year}{2008}), \eprint{0710.4560}.

\bibitem[{\citenamefont{{Slosar} et~al.}(2008)\citenamefont{{Slosar}, {Hirata},
  {Seljak}, {Ho}, and {Padmanabhan}}}]{2008JCAP...08..031S}
\bibinfo{author}{\bibfnamefont{A.}~\bibnamefont{{Slosar}}},
  \bibinfo{author}{\bibfnamefont{C.}~\bibnamefont{{Hirata}}},
  \bibinfo{author}{\bibfnamefont{U.}~\bibnamefont{{Seljak}}},
  \bibinfo{author}{\bibfnamefont{S.}~\bibnamefont{{Ho}}}, \bibnamefont{and}
  \bibinfo{author}{\bibfnamefont{N.}~\bibnamefont{{Padmanabhan}}},
  \bibinfo{journal}{\jcap} \textbf{\bibinfo{volume}{2008}}, \bibinfo{eid}{031}
  (\bibinfo{year}{2008}), \eprint{0805.3580}.

\bibitem[{\citenamefont{{Ross} et~al.}(2013)\citenamefont{{Ross}, {Percival},
  {Carnero}, {Zhao}, {Manera}, {Raccanelli}, {Aubourg}, {Bizyaev},
  {Brewington}, {Brinkmann} et~al.}}]{2013MNRAS.428.1116R}
\bibinfo{author}{\bibfnamefont{A.~J.} \bibnamefont{{Ross}}},
  \bibinfo{author}{\bibfnamefont{W.~J.} \bibnamefont{{Percival}}},
  \bibinfo{author}{\bibfnamefont{A.}~\bibnamefont{{Carnero}}},
  \bibinfo{author}{\bibfnamefont{G.-b.} \bibnamefont{{Zhao}}},
  \bibinfo{author}{\bibfnamefont{M.}~\bibnamefont{{Manera}}},
  \bibinfo{author}{\bibfnamefont{A.}~\bibnamefont{{Raccanelli}}},
  \bibinfo{author}{\bibfnamefont{E.}~\bibnamefont{{Aubourg}}},
  \bibinfo{author}{\bibfnamefont{D.}~\bibnamefont{{Bizyaev}}},
  \bibinfo{author}{\bibfnamefont{H.}~\bibnamefont{{Brewington}}},
  \bibinfo{author}{\bibfnamefont{J.}~\bibnamefont{{Brinkmann}}},
  \bibnamefont{et~al.}, \bibinfo{journal}{\mnras}
  \textbf{\bibinfo{volume}{428}}, \bibinfo{pages}{1116} (\bibinfo{year}{2013}),
  \eprint{1208.1491}.

\bibitem[{\citenamefont{{Giannantonio} and
  {Percival}}(2014)}]{2014MNRAS.441L..16G}
\bibinfo{author}{\bibfnamefont{T.}~\bibnamefont{{Giannantonio}}}
  \bibnamefont{and} \bibinfo{author}{\bibfnamefont{W.~J.}
  \bibnamefont{{Percival}}}, \bibinfo{journal}{\mnras}
  \textbf{\bibinfo{volume}{441}}, \bibinfo{pages}{L16} (\bibinfo{year}{2014}),
  \eprint{1312.5154}.

\bibitem[{\citenamefont{{Giannantonio}
  et~al.}(2014)\citenamefont{{Giannantonio}, {Ross}, {Percival}, {Crittenden},
  {Bacher}, {Kilbinger}, {Nichol}, and {Weller}}}]{2014PhRvD..89b3511G}
\bibinfo{author}{\bibfnamefont{T.}~\bibnamefont{{Giannantonio}}},
  \bibinfo{author}{\bibfnamefont{A.~J.} \bibnamefont{{Ross}}},
  \bibinfo{author}{\bibfnamefont{W.~J.} \bibnamefont{{Percival}}},
  \bibinfo{author}{\bibfnamefont{R.}~\bibnamefont{{Crittenden}}},
  \bibinfo{author}{\bibfnamefont{D.}~\bibnamefont{{Bacher}}},
  \bibinfo{author}{\bibfnamefont{M.}~\bibnamefont{{Kilbinger}}},
  \bibinfo{author}{\bibfnamefont{R.}~\bibnamefont{{Nichol}}}, \bibnamefont{and}
  \bibinfo{author}{\bibfnamefont{J.}~\bibnamefont{{Weller}}},
  \bibinfo{journal}{\prd} \textbf{\bibinfo{volume}{89}}, \bibinfo{eid}{023511}
  (\bibinfo{year}{2014}), \eprint{1303.1349}.

\bibitem[{\citenamefont{{Leistedt} et~al.}(2014)\citenamefont{{Leistedt},
  {Peiris}, and {Roth}}}]{2014PhRvL.113v1301L}
\bibinfo{author}{\bibfnamefont{B.}~\bibnamefont{{Leistedt}}},
  \bibinfo{author}{\bibfnamefont{H.~V.} \bibnamefont{{Peiris}}},
  \bibnamefont{and} \bibinfo{author}{\bibfnamefont{N.}~\bibnamefont{{Roth}}},
  \bibinfo{journal}{\prl} \textbf{\bibinfo{volume}{113}}, \bibinfo{eid}{221301}
  (\bibinfo{year}{2014}), \eprint{1405.4315}.

\bibitem[{\citenamefont{{Mueller} et~al.}(2022)\citenamefont{{Mueller},
  {Rezaie}, {Percival}, {Ross}, {Ruggeri}, {Seo}, {Gil-Mar{\'\i}n}, {Bautista},
  {Brownstein}, {Dawson} et~al.}}]{2022MNRAS.514.3396M}
\bibinfo{author}{\bibfnamefont{E.-M.} \bibnamefont{{Mueller}}},
  \bibinfo{author}{\bibfnamefont{M.}~\bibnamefont{{Rezaie}}},
  \bibinfo{author}{\bibfnamefont{W.~J.} \bibnamefont{{Percival}}},
  \bibinfo{author}{\bibfnamefont{A.~J.} \bibnamefont{{Ross}}},
  \bibinfo{author}{\bibfnamefont{R.}~\bibnamefont{{Ruggeri}}},
  \bibinfo{author}{\bibfnamefont{H.-J.} \bibnamefont{{Seo}}},
  \bibinfo{author}{\bibfnamefont{H.}~\bibnamefont{{Gil-Mar{\'\i}n}}},
  \bibinfo{author}{\bibfnamefont{J.}~\bibnamefont{{Bautista}}},
  \bibinfo{author}{\bibfnamefont{J.~R.} \bibnamefont{{Brownstein}}},
  \bibinfo{author}{\bibfnamefont{K.}~\bibnamefont{{Dawson}}},
  \bibnamefont{et~al.}, \bibinfo{journal}{\mnras}
  \textbf{\bibinfo{volume}{514}}, \bibinfo{pages}{3396} (\bibinfo{year}{2022}).

\bibitem[{\citenamefont{{D'Amico} et~al.}(2022)\citenamefont{{D'Amico},
  {Lewandowski}, {Senatore}, and {Zhang}}}]{2022arXiv220111518D}
\bibinfo{author}{\bibfnamefont{G.}~\bibnamefont{{D'Amico}}},
  \bibinfo{author}{\bibfnamefont{M.}~\bibnamefont{{Lewandowski}}},
  \bibinfo{author}{\bibfnamefont{L.}~\bibnamefont{{Senatore}}},
  \bibnamefont{and} \bibinfo{author}{\bibfnamefont{P.}~\bibnamefont{{Zhang}}},
  \bibinfo{journal}{arXiv e-prints} \bibinfo{eid}{arXiv:2201.11518}
  (\bibinfo{year}{2022}), \eprint{2201.11518}.

\bibitem[{\citenamefont{{Cabass} et~al.}(2022)\citenamefont{{Cabass}, {Ivanov},
  {Philcox}, {Simonovi{\'c}}, and {Zaldarriaga}}}]{2022PhRvD.106d3506C}
\bibinfo{author}{\bibfnamefont{G.}~\bibnamefont{{Cabass}}},
  \bibinfo{author}{\bibfnamefont{M.~M.} \bibnamefont{{Ivanov}}},
  \bibinfo{author}{\bibfnamefont{O.~H.~E.} \bibnamefont{{Philcox}}},
  \bibinfo{author}{\bibfnamefont{M.}~\bibnamefont{{Simonovi{\'c}}}},
  \bibnamefont{and}
  \bibinfo{author}{\bibfnamefont{M.}~\bibnamefont{{Zaldarriaga}}},
  \bibinfo{journal}{\prd} \textbf{\bibinfo{volume}{106}}, \bibinfo{eid}{043506}
  (\bibinfo{year}{2022}), \eprint{2204.01781}.

\bibitem[{\citenamefont{{LSST Science Collaboration}
  et~al.}(2009)\citenamefont{{LSST Science Collaboration}, {Abell}, {Allison},
  {Anderson}, {Andrew}, {Angel}, {Armus}, {Arnett}, {Asztalos}, {Axelrod}
  et~al.}}]{lsst}
\bibinfo{author}{\bibnamefont{{LSST Science Collaboration}}},
  \bibinfo{author}{\bibfnamefont{P.~A.} \bibnamefont{{Abell}}},
  \bibinfo{author}{\bibfnamefont{J.}~\bibnamefont{{Allison}}},
  \bibinfo{author}{\bibfnamefont{S.~F.} \bibnamefont{{Anderson}}},
  \bibinfo{author}{\bibfnamefont{J.~R.} \bibnamefont{{Andrew}}},
  \bibinfo{author}{\bibfnamefont{J.~R.~P.} \bibnamefont{{Angel}}},
  \bibinfo{author}{\bibfnamefont{L.}~\bibnamefont{{Armus}}},
  \bibinfo{author}{\bibfnamefont{D.}~\bibnamefont{{Arnett}}},
  \bibinfo{author}{\bibfnamefont{S.~J.} \bibnamefont{{Asztalos}}},
  \bibinfo{author}{\bibfnamefont{T.~S.} \bibnamefont{{Axelrod}}},
  \bibnamefont{et~al.}, \bibinfo{journal}{arXiv e-prints}
  \bibinfo{eid}{arXiv:0912.0201} (\bibinfo{year}{2009}), \eprint{0912.0201}.

\bibitem[{\citenamefont{{Dor{\'e}} et~al.}(2014)\citenamefont{{Dor{\'e}},
  {Bock}, {Ashby}, {Capak}, {Cooray}, {de Putter}, {Eifler}, {Flagey}, {Gong},
  {Habib} et~al.}}]{2014arXiv1412.4872D}
\bibinfo{author}{\bibfnamefont{O.}~\bibnamefont{{Dor{\'e}}}},
  \bibinfo{author}{\bibfnamefont{J.}~\bibnamefont{{Bock}}},
  \bibinfo{author}{\bibfnamefont{M.}~\bibnamefont{{Ashby}}},
  \bibinfo{author}{\bibfnamefont{P.}~\bibnamefont{{Capak}}},
  \bibinfo{author}{\bibfnamefont{A.}~\bibnamefont{{Cooray}}},
  \bibinfo{author}{\bibfnamefont{R.}~\bibnamefont{{de Putter}}},
  \bibinfo{author}{\bibfnamefont{T.}~\bibnamefont{{Eifler}}},
  \bibinfo{author}{\bibfnamefont{N.}~\bibnamefont{{Flagey}}},
  \bibinfo{author}{\bibfnamefont{Y.}~\bibnamefont{{Gong}}},
  \bibinfo{author}{\bibfnamefont{S.}~\bibnamefont{{Habib}}},
  \bibnamefont{et~al.}, \bibinfo{journal}{arXiv e-prints}
  \bibinfo{eid}{arXiv:1412.4872} (\bibinfo{year}{2014}), \eprint{1412.4872}.

\bibitem[{\citenamefont{{Seljak}}(2009)}]{Seljak:2008xr}
\bibinfo{author}{\bibfnamefont{U.}~\bibnamefont{{Seljak}}},
  \bibinfo{journal}{\prl} \textbf{\bibinfo{volume}{102}}, \bibinfo{eid}{021302}
  (\bibinfo{year}{2009}), \eprint{0807.1770}.

\bibitem[{\citenamefont{{Schmittfull} and
  {Seljak}}(2018)}]{Schmittfull:2017ffw}
\bibinfo{author}{\bibfnamefont{M.}~\bibnamefont{{Schmittfull}}}
  \bibnamefont{and} \bibinfo{author}{\bibfnamefont{U.}~\bibnamefont{{Seljak}}},
  \bibinfo{journal}{\prd} \textbf{\bibinfo{volume}{97}}, \bibinfo{eid}{123540}
  (\bibinfo{year}{2018}), \eprint{1710.09465}.

\bibitem[{\citenamefont{{M{\"u}nchmeyer}
  et~al.}(2019)\citenamefont{{M{\"u}nchmeyer}, {Madhavacheril}, {Ferraro},
  {Johnson}, and {Smith}}}]{meyer:2018eey}
\bibinfo{author}{\bibfnamefont{M.}~\bibnamefont{{M{\"u}nchmeyer}}},
  \bibinfo{author}{\bibfnamefont{M.~S.} \bibnamefont{{Madhavacheril}}},
  \bibinfo{author}{\bibfnamefont{S.}~\bibnamefont{{Ferraro}}},
  \bibinfo{author}{\bibfnamefont{M.~C.} \bibnamefont{{Johnson}}},
  \bibnamefont{and} \bibinfo{author}{\bibfnamefont{K.~M.}
  \bibnamefont{{Smith}}}, \bibinfo{journal}{\prd}
  \textbf{\bibinfo{volume}{100}}, \bibinfo{eid}{083508} (\bibinfo{year}{2019}),
  \eprint{1810.13424}.

\bibitem[{\citenamefont{{Madau} and {Dickinson}}(2014)}]{2014ARA&A..52..415M}
\bibinfo{author}{\bibfnamefont{P.}~\bibnamefont{{Madau}}} \bibnamefont{and}
  \bibinfo{author}{\bibfnamefont{M.}~\bibnamefont{{Dickinson}}},
  \bibinfo{journal}{\araa} \textbf{\bibinfo{volume}{52}}, \bibinfo{pages}{415}
  (\bibinfo{year}{2014}), \eprint{1403.0007}.

\bibitem[{\citenamefont{{Knox} et~al.}(2001)\citenamefont{{Knox}, {Cooray},
  {Eisenstein}, and {Haiman}}}]{2001ApJ...550....7K}
\bibinfo{author}{\bibfnamefont{L.}~\bibnamefont{{Knox}}},
  \bibinfo{author}{\bibfnamefont{A.}~\bibnamefont{{Cooray}}},
  \bibinfo{author}{\bibfnamefont{D.}~\bibnamefont{{Eisenstein}}},
  \bibnamefont{and} \bibinfo{author}{\bibfnamefont{Z.}~\bibnamefont{{Haiman}}},
  \bibinfo{journal}{\apj} \textbf{\bibinfo{volume}{550}}, \bibinfo{pages}{7}
  (\bibinfo{year}{2001}), \eprint{astro-ph/0009151}.

\bibitem[{\citenamefont{{Blanchard} and
  {Schneider}}(1987)}]{1987A&A...184....1B}
\bibinfo{author}{\bibfnamefont{A.}~\bibnamefont{{Blanchard}}} \bibnamefont{and}
  \bibinfo{author}{\bibfnamefont{J.}~\bibnamefont{{Schneider}}},
  \bibinfo{journal}{\aap} \textbf{\bibinfo{volume}{184}}, \bibinfo{pages}{1}
  (\bibinfo{year}{1987}).

\bibitem[{\citenamefont{{Lewis} and {Challinor}}(2006)}]{Lewis:2006fu}
\bibinfo{author}{\bibfnamefont{A.}~\bibnamefont{{Lewis}}} \bibnamefont{and}
  \bibinfo{author}{\bibfnamefont{A.}~\bibnamefont{{Challinor}}},
  \bibinfo{journal}{\physrep} \textbf{\bibinfo{volume}{429}},
  \bibinfo{pages}{1} (\bibinfo{year}{2006}), \eprint{astro-ph/0601594}.

\bibitem[{\citenamefont{{Planck Collaboration}
  et~al.}(2014{\natexlab{a}})\citenamefont{{Planck Collaboration}, {Ade},
  {Aghanim}, {Armitage-Caplan}, {Arnaud}, {Ashdown}, {Atrio-Barandela},
  {Aumont}, {Baccigalupi}, {Banday} et~al.}}]{1303.5077}
\bibinfo{author}{\bibnamefont{{Planck Collaboration}}},
  \bibinfo{author}{\bibfnamefont{P.~A.~R.} \bibnamefont{{Ade}}},
  \bibinfo{author}{\bibfnamefont{N.}~\bibnamefont{{Aghanim}}},
  \bibinfo{author}{\bibfnamefont{C.}~\bibnamefont{{Armitage-Caplan}}},
  \bibinfo{author}{\bibfnamefont{M.}~\bibnamefont{{Arnaud}}},
  \bibinfo{author}{\bibfnamefont{M.}~\bibnamefont{{Ashdown}}},
  \bibinfo{author}{\bibfnamefont{F.}~\bibnamefont{{Atrio-Barandela}}},
  \bibinfo{author}{\bibfnamefont{J.}~\bibnamefont{{Aumont}}},
  \bibinfo{author}{\bibfnamefont{C.}~\bibnamefont{{Baccigalupi}}},
  \bibinfo{author}{\bibfnamefont{A.~J.} \bibnamefont{{Banday}}},
  \bibnamefont{et~al.}, \bibinfo{journal}{\aap} \textbf{\bibinfo{volume}{571}},
  \bibinfo{eid}{A17} (\bibinfo{year}{2014}{\natexlab{a}}), \eprint{1303.5077}.

\bibitem[{\citenamefont{{Planck Collaboration}
  et~al.}(2016)\citenamefont{{Planck Collaboration}, {Ade}, {Aghanim},
  {Arnaud}, {Ashdown}, {Aumont}, {Baccigalupi}, {Banday}, {Barreiro},
  {Bartlett} et~al.}}]{1502.01591}
\bibinfo{author}{\bibnamefont{{Planck Collaboration}}},
  \bibinfo{author}{\bibfnamefont{P.~A.~R.} \bibnamefont{{Ade}}},
  \bibinfo{author}{\bibfnamefont{N.}~\bibnamefont{{Aghanim}}},
  \bibinfo{author}{\bibfnamefont{M.}~\bibnamefont{{Arnaud}}},
  \bibinfo{author}{\bibfnamefont{M.}~\bibnamefont{{Ashdown}}},
  \bibinfo{author}{\bibfnamefont{J.}~\bibnamefont{{Aumont}}},
  \bibinfo{author}{\bibfnamefont{C.}~\bibnamefont{{Baccigalupi}}},
  \bibinfo{author}{\bibfnamefont{A.~J.} \bibnamefont{{Banday}}},
  \bibinfo{author}{\bibfnamefont{R.~B.} \bibnamefont{{Barreiro}}},
  \bibinfo{author}{\bibfnamefont{J.~G.} \bibnamefont{{Bartlett}}},
  \bibnamefont{et~al.}, \bibinfo{journal}{\aap} \textbf{\bibinfo{volume}{594}},
  \bibinfo{eid}{A15} (\bibinfo{year}{2016}), \eprint{1502.01591}.

\bibitem[{\citenamefont{{Planck Collaboration}
  et~al.}(2020{\natexlab{b}})\citenamefont{{Planck Collaboration}, {Aghanim},
  {Akrami}, {Ashdown}, {Aumont}, {Baccigalupi}, {Ballardini}, {Banday},
  {Barreiro}, {Bartolo} et~al.}}]{1807.06210}
\bibinfo{author}{\bibnamefont{{Planck Collaboration}}},
  \bibinfo{author}{\bibfnamefont{N.}~\bibnamefont{{Aghanim}}},
  \bibinfo{author}{\bibfnamefont{Y.}~\bibnamefont{{Akrami}}},
  \bibinfo{author}{\bibfnamefont{M.}~\bibnamefont{{Ashdown}}},
  \bibinfo{author}{\bibfnamefont{J.}~\bibnamefont{{Aumont}}},
  \bibinfo{author}{\bibfnamefont{C.}~\bibnamefont{{Baccigalupi}}},
  \bibinfo{author}{\bibfnamefont{M.}~\bibnamefont{{Ballardini}}},
  \bibinfo{author}{\bibfnamefont{A.~J.} \bibnamefont{{Banday}}},
  \bibinfo{author}{\bibfnamefont{R.~B.} \bibnamefont{{Barreiro}}},
  \bibinfo{author}{\bibfnamefont{N.}~\bibnamefont{{Bartolo}}},
  \bibnamefont{et~al.}, \bibinfo{journal}{\aap} \textbf{\bibinfo{volume}{641}},
  \bibinfo{eid}{A8} (\bibinfo{year}{2020}{\natexlab{b}}), \eprint{1807.06210}.

\bibitem[{\citenamefont{{Carron}
  et~al.}(2022{\natexlab{a}})\citenamefont{{Carron}, {Mirmelstein}, and
  {Lewis}}}]{2206.07773}
\bibinfo{author}{\bibfnamefont{J.}~\bibnamefont{{Carron}}},
  \bibinfo{author}{\bibfnamefont{M.}~\bibnamefont{{Mirmelstein}}},
  \bibnamefont{and} \bibinfo{author}{\bibfnamefont{A.}~\bibnamefont{{Lewis}}},
  \bibinfo{journal}{\jcap} \textbf{\bibinfo{volume}{2022}}, \bibinfo{eid}{039}
  (\bibinfo{year}{2022}{\natexlab{a}}), \eprint{2206.07773}.

\bibitem[{\citenamefont{{Darwish} et~al.}(2021)\citenamefont{{Darwish},
  {Madhavacheril}, {Sherwin}, {Aiola}, {Battaglia}, {Beall}, {Becker}, {Bond},
  {Calabrese}, {Choi} et~al.}}]{Darwish:2020fwf}
\bibinfo{author}{\bibfnamefont{O.}~\bibnamefont{{Darwish}}},
  \bibinfo{author}{\bibfnamefont{M.~S.} \bibnamefont{{Madhavacheril}}},
  \bibinfo{author}{\bibfnamefont{B.~D.} \bibnamefont{{Sherwin}}},
  \bibinfo{author}{\bibfnamefont{S.}~\bibnamefont{{Aiola}}},
  \bibinfo{author}{\bibfnamefont{N.}~\bibnamefont{{Battaglia}}},
  \bibinfo{author}{\bibfnamefont{J.~A.} \bibnamefont{{Beall}}},
  \bibinfo{author}{\bibfnamefont{D.~T.} \bibnamefont{{Becker}}},
  \bibinfo{author}{\bibfnamefont{J.~R.} \bibnamefont{{Bond}}},
  \bibinfo{author}{\bibfnamefont{E.}~\bibnamefont{{Calabrese}}},
  \bibinfo{author}{\bibfnamefont{S.~K.} \bibnamefont{{Choi}}},
  \bibnamefont{et~al.}, \bibinfo{journal}{\mnras}
  \textbf{\bibinfo{volume}{500}}, \bibinfo{pages}{2250} (\bibinfo{year}{2021}),
  \eprint{2004.01139}.

\bibitem[{\citenamefont{{Sherwin} et~al.}(2017)\citenamefont{{Sherwin}, {van
  Engelen}, {Sehgal}, {Madhavacheril}, {Addison}, {Aiola}, {Allison},
  {Battaglia}, {Becker}, {Beall} et~al.}}]{Sherwin:2016tyf}
\bibinfo{author}{\bibfnamefont{B.~D.} \bibnamefont{{Sherwin}}},
  \bibinfo{author}{\bibfnamefont{A.}~\bibnamefont{{van Engelen}}},
  \bibinfo{author}{\bibfnamefont{N.}~\bibnamefont{{Sehgal}}},
  \bibinfo{author}{\bibfnamefont{M.}~\bibnamefont{{Madhavacheril}}},
  \bibinfo{author}{\bibfnamefont{G.~E.} \bibnamefont{{Addison}}},
  \bibinfo{author}{\bibfnamefont{S.}~\bibnamefont{{Aiola}}},
  \bibinfo{author}{\bibfnamefont{R.}~\bibnamefont{{Allison}}},
  \bibinfo{author}{\bibfnamefont{N.}~\bibnamefont{{Battaglia}}},
  \bibinfo{author}{\bibfnamefont{D.~T.} \bibnamefont{{Becker}}},
  \bibinfo{author}{\bibfnamefont{J.~A.} \bibnamefont{{Beall}}},
  \bibnamefont{et~al.}, \bibinfo{journal}{\prd} \textbf{\bibinfo{volume}{95}},
  \bibinfo{eid}{123529} (\bibinfo{year}{2017}), \eprint{1611.09753}.

\bibitem[{\citenamefont{{Das} et~al.}(2011)\citenamefont{{Das}, {Sherwin},
  {Aguirre}, {Appel}, {Bond}, {Carvalho}, {Devlin}, {Dunkley}, {D{\"u}nner},
  {Essinger-Hileman} et~al.}}]{2011PhRvL.107b1301D}
\bibinfo{author}{\bibfnamefont{S.}~\bibnamefont{{Das}}},
  \bibinfo{author}{\bibfnamefont{B.~D.} \bibnamefont{{Sherwin}}},
  \bibinfo{author}{\bibfnamefont{P.}~\bibnamefont{{Aguirre}}},
  \bibinfo{author}{\bibfnamefont{J.~W.} \bibnamefont{{Appel}}},
  \bibinfo{author}{\bibfnamefont{J.~R.} \bibnamefont{{Bond}}},
  \bibinfo{author}{\bibfnamefont{C.~S.} \bibnamefont{{Carvalho}}},
  \bibinfo{author}{\bibfnamefont{M.~J.} \bibnamefont{{Devlin}}},
  \bibinfo{author}{\bibfnamefont{J.}~\bibnamefont{{Dunkley}}},
  \bibinfo{author}{\bibfnamefont{R.}~\bibnamefont{{D{\"u}nner}}},
  \bibinfo{author}{\bibfnamefont{T.}~\bibnamefont{{Essinger-Hileman}}},
  \bibnamefont{et~al.}, \bibinfo{journal}{\prl} \textbf{\bibinfo{volume}{107}},
  \bibinfo{eid}{021301} (\bibinfo{year}{2011}), \eprint{1103.2124}.

\bibitem[{\citenamefont{{Millea} et~al.}(2021)\citenamefont{{Millea}, {Daley},
  {Chou}, {Anderes}, {Ade}, {Anderson}, {Austermann}, {Avva}, {Beall}, {Bender}
  et~al.}}]{2021ApJ...922..259M}
\bibinfo{author}{\bibfnamefont{M.}~\bibnamefont{{Millea}}},
  \bibinfo{author}{\bibfnamefont{C.~M.} \bibnamefont{{Daley}}},
  \bibinfo{author}{\bibfnamefont{T.~L.} \bibnamefont{{Chou}}},
  \bibinfo{author}{\bibfnamefont{E.}~\bibnamefont{{Anderes}}},
  \bibinfo{author}{\bibfnamefont{P.~A.~R.} \bibnamefont{{Ade}}},
  \bibinfo{author}{\bibfnamefont{A.~J.} \bibnamefont{{Anderson}}},
  \bibinfo{author}{\bibfnamefont{J.~E.} \bibnamefont{{Austermann}}},
  \bibinfo{author}{\bibfnamefont{J.~S.} \bibnamefont{{Avva}}},
  \bibinfo{author}{\bibfnamefont{J.~A.} \bibnamefont{{Beall}}},
  \bibinfo{author}{\bibfnamefont{A.~N.} \bibnamefont{{Bender}}},
  \bibnamefont{et~al.}, \bibinfo{journal}{\apj} \textbf{\bibinfo{volume}{922}},
  \bibinfo{eid}{259} (\bibinfo{year}{2021}), \eprint{2012.01709}.

\bibitem[{\citenamefont{{Wu} et~al.}(2019)\citenamefont{{Wu}, {Mocanu}, {Ade},
  {Anderson}, {Austermann}, {Avva}, {Beall}, {Bender}, {Benson}, {Bianchini}
  et~al.}}]{2019ApJ...884...70W}
\bibinfo{author}{\bibfnamefont{W.~L.~K.} \bibnamefont{{Wu}}},
  \bibinfo{author}{\bibfnamefont{L.~M.} \bibnamefont{{Mocanu}}},
  \bibinfo{author}{\bibfnamefont{P.~A.~R.} \bibnamefont{{Ade}}},
  \bibinfo{author}{\bibfnamefont{A.~J.} \bibnamefont{{Anderson}}},
  \bibinfo{author}{\bibfnamefont{J.~E.} \bibnamefont{{Austermann}}},
  \bibinfo{author}{\bibfnamefont{J.~S.} \bibnamefont{{Avva}}},
  \bibinfo{author}{\bibfnamefont{J.~A.} \bibnamefont{{Beall}}},
  \bibinfo{author}{\bibfnamefont{A.~N.} \bibnamefont{{Bender}}},
  \bibinfo{author}{\bibfnamefont{B.~A.} \bibnamefont{{Benson}}},
  \bibinfo{author}{\bibfnamefont{F.}~\bibnamefont{{Bianchini}}},
  \bibnamefont{et~al.}, \bibinfo{journal}{\apj} \textbf{\bibinfo{volume}{884}},
  \bibinfo{eid}{70} (\bibinfo{year}{2019}), \eprint{1905.05777}.

\bibitem[{\citenamefont{{Omori} et~al.}(2017)\citenamefont{{Omori}, {Chown},
  {Simard}, {Story}, {Aylor}, {Baxter}, {Benson}, {Bleem}, {Carlstrom}, {Chang}
  et~al.}}]{2017ApJ...849..124O}
\bibinfo{author}{\bibfnamefont{Y.}~\bibnamefont{{Omori}}},
  \bibinfo{author}{\bibfnamefont{R.}~\bibnamefont{{Chown}}},
  \bibinfo{author}{\bibfnamefont{G.}~\bibnamefont{{Simard}}},
  \bibinfo{author}{\bibfnamefont{K.~T.} \bibnamefont{{Story}}},
  \bibinfo{author}{\bibfnamefont{K.}~\bibnamefont{{Aylor}}},
  \bibinfo{author}{\bibfnamefont{E.~J.} \bibnamefont{{Baxter}}},
  \bibinfo{author}{\bibfnamefont{B.~A.} \bibnamefont{{Benson}}},
  \bibinfo{author}{\bibfnamefont{L.~E.} \bibnamefont{{Bleem}}},
  \bibinfo{author}{\bibfnamefont{J.~E.} \bibnamefont{{Carlstrom}}},
  \bibinfo{author}{\bibfnamefont{C.~L.} \bibnamefont{{Chang}}},
  \bibnamefont{et~al.}, \bibinfo{journal}{\apj} \textbf{\bibinfo{volume}{849}},
  \bibinfo{eid}{124} (\bibinfo{year}{2017}), \eprint{1705.00743}.

\bibitem[{\citenamefont{{Story} et~al.}(2015)\citenamefont{{Story}, {Hanson},
  {Ade}, {Aird}, {Austermann}, {Beall}, {Bender}, {Benson}, {Bleem},
  {Carlstrom} et~al.}}]{2015ApJ...810...50S}
\bibinfo{author}{\bibfnamefont{K.~T.} \bibnamefont{{Story}}},
  \bibinfo{author}{\bibfnamefont{D.}~\bibnamefont{{Hanson}}},
  \bibinfo{author}{\bibfnamefont{P.~A.~R.} \bibnamefont{{Ade}}},
  \bibinfo{author}{\bibfnamefont{K.~A.} \bibnamefont{{Aird}}},
  \bibinfo{author}{\bibfnamefont{J.~E.} \bibnamefont{{Austermann}}},
  \bibinfo{author}{\bibfnamefont{J.~A.} \bibnamefont{{Beall}}},
  \bibinfo{author}{\bibfnamefont{A.~N.} \bibnamefont{{Bender}}},
  \bibinfo{author}{\bibfnamefont{B.~A.} \bibnamefont{{Benson}}},
  \bibinfo{author}{\bibfnamefont{L.~E.} \bibnamefont{{Bleem}}},
  \bibinfo{author}{\bibfnamefont{J.~E.} \bibnamefont{{Carlstrom}}},
  \bibnamefont{et~al.}, \bibinfo{journal}{\apj} \textbf{\bibinfo{volume}{810}},
  \bibinfo{eid}{50} (\bibinfo{year}{2015}), \eprint{1412.4760}.

\bibitem[{\citenamefont{{van Engelen} et~al.}(2012)\citenamefont{{van Engelen},
  {Keisler}, {Zahn}, {Aird}, {Benson}, {Bleem}, {Carlstrom}, {Chang}, {Cho},
  {Crawford} et~al.}}]{2012ApJ...756..142V}
\bibinfo{author}{\bibfnamefont{A.}~\bibnamefont{{van Engelen}}},
  \bibinfo{author}{\bibfnamefont{R.}~\bibnamefont{{Keisler}}},
  \bibinfo{author}{\bibfnamefont{O.}~\bibnamefont{{Zahn}}},
  \bibinfo{author}{\bibfnamefont{K.~A.} \bibnamefont{{Aird}}},
  \bibinfo{author}{\bibfnamefont{B.~A.} \bibnamefont{{Benson}}},
  \bibinfo{author}{\bibfnamefont{L.~E.} \bibnamefont{{Bleem}}},
  \bibinfo{author}{\bibfnamefont{J.~E.} \bibnamefont{{Carlstrom}}},
  \bibinfo{author}{\bibfnamefont{C.~L.} \bibnamefont{{Chang}}},
  \bibinfo{author}{\bibfnamefont{H.~M.} \bibnamefont{{Cho}}},
  \bibinfo{author}{\bibfnamefont{T.~M.} \bibnamefont{{Crawford}}},
  \bibnamefont{et~al.}, \bibinfo{journal}{\apj} \textbf{\bibinfo{volume}{756}},
  \bibinfo{eid}{142} (\bibinfo{year}{2012}), \eprint{1202.0546}.

\bibitem[{\citenamefont{{Tucci} et~al.}(2016)\citenamefont{{Tucci},
  {Desjacques}, and {Kunz}}}]{1606.02323}
\bibinfo{author}{\bibfnamefont{M.}~\bibnamefont{{Tucci}}},
  \bibinfo{author}{\bibfnamefont{V.}~\bibnamefont{{Desjacques}}},
  \bibnamefont{and} \bibinfo{author}{\bibfnamefont{M.}~\bibnamefont{{Kunz}}},
  \bibinfo{journal}{\mnras} \textbf{\bibinfo{volume}{463}},
  \bibinfo{pages}{2046} (\bibinfo{year}{2016}), \eprint{1606.02323}.

\bibitem[{\citenamefont{{Lenz} et~al.}(2019{\natexlab{a}})\citenamefont{{Lenz},
  {Dor{\'e}}, and {Lagache}}}]{1905.00426}
\bibinfo{author}{\bibfnamefont{D.}~\bibnamefont{{Lenz}}},
  \bibinfo{author}{\bibfnamefont{O.}~\bibnamefont{{Dor{\'e}}}},
  \bibnamefont{and}
  \bibinfo{author}{\bibfnamefont{G.}~\bibnamefont{{Lagache}}},
  \bibinfo{journal}{\apj} \textbf{\bibinfo{volume}{883}}, \bibinfo{eid}{75}
  (\bibinfo{year}{2019}{\natexlab{a}}), \eprint{1905.00426}.

\bibitem[{\citenamefont{{Planck Collaboration}
  et~al.}(2014{\natexlab{b}})\citenamefont{{Planck Collaboration}, {Ade},
  {Aghanim}, {Armitage-Caplan}, {Arnaud}, {Ashdown}, {Atrio-Barand ela},
  {Aumont}, {Baccigalupi}, {Banday} et~al.}}]{Ade:2013zuv}
\bibinfo{author}{\bibnamefont{{Planck Collaboration}}},
  \bibinfo{author}{\bibfnamefont{P.~A.~R.} \bibnamefont{{Ade}}},
  \bibinfo{author}{\bibfnamefont{N.}~\bibnamefont{{Aghanim}}},
  \bibinfo{author}{\bibfnamefont{C.}~\bibnamefont{{Armitage-Caplan}}},
  \bibinfo{author}{\bibfnamefont{M.}~\bibnamefont{{Arnaud}}},
  \bibinfo{author}{\bibfnamefont{M.}~\bibnamefont{{Ashdown}}},
  \bibinfo{author}{\bibfnamefont{F.}~\bibnamefont{{Atrio-Barand ela}}},
  \bibinfo{author}{\bibfnamefont{J.}~\bibnamefont{{Aumont}}},
  \bibinfo{author}{\bibfnamefont{C.}~\bibnamefont{{Baccigalupi}}},
  \bibinfo{author}{\bibfnamefont{A.~J.} \bibnamefont{{Banday}}},
  \bibnamefont{et~al.}, \bibinfo{journal}{\aap} \textbf{\bibinfo{volume}{571}},
  \bibinfo{eid}{A16} (\bibinfo{year}{2014}{\natexlab{b}}), \eprint{1303.5076}.

\bibitem[{\citenamefont{{Lewis} et~al.}(2000)\citenamefont{{Lewis},
  {Challinor}, and {Lasenby}}}]{2000ApJ...538..473L}
\bibinfo{author}{\bibfnamefont{A.}~\bibnamefont{{Lewis}}},
  \bibinfo{author}{\bibfnamefont{A.}~\bibnamefont{{Challinor}}},
  \bibnamefont{and}
  \bibinfo{author}{\bibfnamefont{A.}~\bibnamefont{{Lasenby}}},
  \bibinfo{journal}{\apj} \textbf{\bibinfo{volume}{538}}, \bibinfo{pages}{473}
  (\bibinfo{year}{2000}), \eprint{astro-ph/9911177}.

\bibitem[{\citenamefont{{Shang} et~al.}(2012)\citenamefont{{Shang}, {Haiman},
  {Knox}, and {Oh}}}]{2012MNRAS.421.2832S}
\bibinfo{author}{\bibfnamefont{C.}~\bibnamefont{{Shang}}},
  \bibinfo{author}{\bibfnamefont{Z.}~\bibnamefont{{Haiman}}},
  \bibinfo{author}{\bibfnamefont{L.}~\bibnamefont{{Knox}}}, \bibnamefont{and}
  \bibinfo{author}{\bibfnamefont{S.~P.} \bibnamefont{{Oh}}},
  \bibinfo{journal}{\mnras} \textbf{\bibinfo{volume}{421}},
  \bibinfo{pages}{2832} (\bibinfo{year}{2012}), \eprint{1109.1522}.

\bibitem[{\citenamefont{{Wu} and {Dor{\'e}}}(2017)}]{Wu:2016vpb}
\bibinfo{author}{\bibfnamefont{H.-Y.} \bibnamefont{{Wu}}} \bibnamefont{and}
  \bibinfo{author}{\bibfnamefont{O.}~\bibnamefont{{Dor{\'e}}}},
  \bibinfo{journal}{\mnras} \textbf{\bibinfo{volume}{466}},
  \bibinfo{pages}{4651} (\bibinfo{year}{2017}), \eprint{1611.04517}.

\bibitem[{\citenamefont{{Maniyar} et~al.}(2019)\citenamefont{{Maniyar},
  {Lagache}, {B{\'e}thermin}, and {Ili{\'c}}}}]{Maniyar:2018hfp}
\bibinfo{author}{\bibfnamefont{A.}~\bibnamefont{{Maniyar}}},
  \bibinfo{author}{\bibfnamefont{G.}~\bibnamefont{{Lagache}}},
  \bibinfo{author}{\bibfnamefont{M.}~\bibnamefont{{B{\'e}thermin}}},
  \bibnamefont{and}
  \bibinfo{author}{\bibfnamefont{S.}~\bibnamefont{{Ili{\'c}}}},
  \bibinfo{journal}{\aap} \textbf{\bibinfo{volume}{621}}, \bibinfo{eid}{A32}
  (\bibinfo{year}{2019}), \eprint{1809.04551}.

\bibitem[{\citenamefont{{Maniyar}
  et~al.}(2021{\natexlab{a}})\citenamefont{{Maniyar}, {B{\'e}thermin}, and
  {Lagache}}}]{2021A&A...645A..40M}
\bibinfo{author}{\bibfnamefont{A.}~\bibnamefont{{Maniyar}}},
  \bibinfo{author}{\bibfnamefont{M.}~\bibnamefont{{B{\'e}thermin}}},
  \bibnamefont{and}
  \bibinfo{author}{\bibfnamefont{G.}~\bibnamefont{{Lagache}}},
  \bibinfo{journal}{\aap} \textbf{\bibinfo{volume}{645}}, \bibinfo{eid}{A40}
  (\bibinfo{year}{2021}{\natexlab{a}}), \eprint{2006.16329}.

\bibitem[{\citenamefont{{Limber}}(1953)}]{1953ApJ...117..134L}
\bibinfo{author}{\bibfnamefont{D.~N.} \bibnamefont{{Limber}}},
  \bibinfo{journal}{\apj} \textbf{\bibinfo{volume}{117}}, \bibinfo{pages}{134}
  (\bibinfo{year}{1953}).

\bibitem[{\citenamefont{{Maniyar} et~al.}(2018)\citenamefont{{Maniyar},
  {B{\'e}thermin}, and {Lagache}}}]{2018A&A...614A..39M}
\bibinfo{author}{\bibfnamefont{A.~S.} \bibnamefont{{Maniyar}}},
  \bibinfo{author}{\bibfnamefont{M.}~\bibnamefont{{B{\'e}thermin}}},
  \bibnamefont{and}
  \bibinfo{author}{\bibfnamefont{G.}~\bibnamefont{{Lagache}}},
  \bibinfo{journal}{\aap} \textbf{\bibinfo{volume}{614}}, \bibinfo{eid}{A39}
  (\bibinfo{year}{2018}), \eprint{1801.10146}.

\bibitem[{\citenamefont{{Kennicutt}}(1998)}]{1998ARA&A..36..189K}
\bibinfo{author}{\bibfnamefont{J.}~\bibnamefont{{Kennicutt}},
  \bibfnamefont{Robert~C.}}, \bibinfo{journal}{ARAA}
  \textbf{\bibinfo{volume}{36}}, \bibinfo{pages}{189} (\bibinfo{year}{1998}),
  \eprint{astro-ph/9807187}.

\bibitem[{\citenamefont{{B{\'e}thermin}
  et~al.}(2013)\citenamefont{{B{\'e}thermin}, {Wang}, {Dor{\'e}}, {Lagache},
  {Sargent}, {Daddi}, {Cousin}, and {Aussel}}}]{Bethermin_SFR}
\bibinfo{author}{\bibfnamefont{M.}~\bibnamefont{{B{\'e}thermin}}},
  \bibinfo{author}{\bibfnamefont{L.}~\bibnamefont{{Wang}}},
  \bibinfo{author}{\bibfnamefont{O.}~\bibnamefont{{Dor{\'e}}}},
  \bibinfo{author}{\bibfnamefont{G.}~\bibnamefont{{Lagache}}},
  \bibinfo{author}{\bibfnamefont{M.}~\bibnamefont{{Sargent}}},
  \bibinfo{author}{\bibfnamefont{E.}~\bibnamefont{{Daddi}}},
  \bibinfo{author}{\bibfnamefont{M.}~\bibnamefont{{Cousin}}}, \bibnamefont{and}
  \bibinfo{author}{\bibfnamefont{H.}~\bibnamefont{{Aussel}}},
  \bibinfo{journal}{\aap} \textbf{\bibinfo{volume}{557}}, \bibinfo{eid}{A66}
  (\bibinfo{year}{2013}), \eprint{1304.3936}.

\bibitem[{\citenamefont{{B{\'e}thermin}
  et~al.}(2015)\citenamefont{{B{\'e}thermin}, {Daddi}, {Magdis}, {Lagos},
  {Sargent}, {Albrecht}, {Aussel}, {Bertoldi}, {Buat}, {Galametz}
  et~al.}}]{Bethermin_SEDs}
\bibinfo{author}{\bibfnamefont{M.}~\bibnamefont{{B{\'e}thermin}}},
  \bibinfo{author}{\bibfnamefont{E.}~\bibnamefont{{Daddi}}},
  \bibinfo{author}{\bibfnamefont{G.}~\bibnamefont{{Magdis}}},
  \bibinfo{author}{\bibfnamefont{C.}~\bibnamefont{{Lagos}}},
  \bibinfo{author}{\bibfnamefont{M.}~\bibnamefont{{Sargent}}},
  \bibinfo{author}{\bibfnamefont{M.}~\bibnamefont{{Albrecht}}},
  \bibinfo{author}{\bibfnamefont{H.}~\bibnamefont{{Aussel}}},
  \bibinfo{author}{\bibfnamefont{F.}~\bibnamefont{{Bertoldi}}},
  \bibinfo{author}{\bibfnamefont{V.}~\bibnamefont{{Buat}}},
  \bibinfo{author}{\bibfnamefont{M.}~\bibnamefont{{Galametz}}},
  \bibnamefont{et~al.}, \bibinfo{journal}{\aap} \textbf{\bibinfo{volume}{573}},
  \bibinfo{eid}{A113} (\bibinfo{year}{2015}), \eprint{1409.5796}.

\bibitem[{\citenamefont{{B{\'e}thermin}
  et~al.}(2017)\citenamefont{{B{\'e}thermin}, {Wu}, {Lagache}, {Davidzon},
  {Ponthieu}, {Cousin}, {Wang}, {Dor{\'e}}, {Daddi}, and
  {Lapi}}}]{2017A&A...607A..89B}
\bibinfo{author}{\bibfnamefont{M.}~\bibnamefont{{B{\'e}thermin}}},
  \bibinfo{author}{\bibfnamefont{H.-Y.} \bibnamefont{{Wu}}},
  \bibinfo{author}{\bibfnamefont{G.}~\bibnamefont{{Lagache}}},
  \bibinfo{author}{\bibfnamefont{I.}~\bibnamefont{{Davidzon}}},
  \bibinfo{author}{\bibfnamefont{N.}~\bibnamefont{{Ponthieu}}},
  \bibinfo{author}{\bibfnamefont{M.}~\bibnamefont{{Cousin}}},
  \bibinfo{author}{\bibfnamefont{L.}~\bibnamefont{{Wang}}},
  \bibinfo{author}{\bibfnamefont{O.}~\bibnamefont{{Dor{\'e}}}},
  \bibinfo{author}{\bibfnamefont{E.}~\bibnamefont{{Daddi}}}, \bibnamefont{and}
  \bibinfo{author}{\bibfnamefont{A.}~\bibnamefont{{Lapi}}},
  \bibinfo{journal}{\aap} \textbf{\bibinfo{volume}{607}}, \bibinfo{eid}{A89}
  (\bibinfo{year}{2017}), \eprint{1703.08795}.

\bibitem[{\citenamefont{{Hu} and {Okamoto}}(2002)}]{astro-ph/0111606}
\bibinfo{author}{\bibfnamefont{W.}~\bibnamefont{{Hu}}} \bibnamefont{and}
  \bibinfo{author}{\bibfnamefont{T.}~\bibnamefont{{Okamoto}}},
  \bibinfo{journal}{\apj} \textbf{\bibinfo{volume}{574}}, \bibinfo{pages}{566}
  (\bibinfo{year}{2002}), \eprint{astro-ph/0111606}.

\bibitem[{\citenamefont{{Planck Collaboration}
  et~al.}(2014{\natexlab{c}})\citenamefont{{Planck Collaboration}, {Ade},
  {Aghanim}, {Armitage-Caplan}, {Arnaud}, {Ashdown}, {Atrio-Barand ela},
  {Aumont}, {Baccigalupi}, {Banday} et~al.}}]{Ade:2013aro}
\bibinfo{author}{\bibnamefont{{Planck Collaboration}}},
  \bibinfo{author}{\bibfnamefont{P.~A.~R.} \bibnamefont{{Ade}}},
  \bibinfo{author}{\bibfnamefont{N.}~\bibnamefont{{Aghanim}}},
  \bibinfo{author}{\bibfnamefont{C.}~\bibnamefont{{Armitage-Caplan}}},
  \bibinfo{author}{\bibfnamefont{M.}~\bibnamefont{{Arnaud}}},
  \bibinfo{author}{\bibfnamefont{M.}~\bibnamefont{{Ashdown}}},
  \bibinfo{author}{\bibfnamefont{F.}~\bibnamefont{{Atrio-Barand ela}}},
  \bibinfo{author}{\bibfnamefont{J.}~\bibnamefont{{Aumont}}},
  \bibinfo{author}{\bibfnamefont{C.}~\bibnamefont{{Baccigalupi}}},
  \bibinfo{author}{\bibfnamefont{A.~J.} \bibnamefont{{Banday}}},
  \bibnamefont{et~al.}, \bibinfo{journal}{\aap} \textbf{\bibinfo{volume}{571}},
  \bibinfo{eid}{A18} (\bibinfo{year}{2014}{\natexlab{c}}), \eprint{1303.5078}.

\bibitem[{\citenamefont{{Holder} et~al.}(2013)\citenamefont{{Holder}, {Viero},
  {Zahn}, {Aird}, {Benson}, {Bhattacharya}, {Bleem}, {Bock}, {Brodwin},
  {Carlstrom} et~al.}}]{1303.5048}
\bibinfo{author}{\bibfnamefont{G.~P.} \bibnamefont{{Holder}}},
  \bibinfo{author}{\bibfnamefont{M.~P.} \bibnamefont{{Viero}}},
  \bibinfo{author}{\bibfnamefont{O.}~\bibnamefont{{Zahn}}},
  \bibinfo{author}{\bibfnamefont{K.~A.} \bibnamefont{{Aird}}},
  \bibinfo{author}{\bibfnamefont{B.~A.} \bibnamefont{{Benson}}},
  \bibinfo{author}{\bibfnamefont{S.}~\bibnamefont{{Bhattacharya}}},
  \bibinfo{author}{\bibfnamefont{L.~E.} \bibnamefont{{Bleem}}},
  \bibinfo{author}{\bibfnamefont{J.}~\bibnamefont{{Bock}}},
  \bibinfo{author}{\bibfnamefont{M.}~\bibnamefont{{Brodwin}}},
  \bibinfo{author}{\bibfnamefont{J.~E.} \bibnamefont{{Carlstrom}}},
  \bibnamefont{et~al.}, \bibinfo{journal}{\apjl}
  \textbf{\bibinfo{volume}{771}}, \bibinfo{eid}{L16} (\bibinfo{year}{2013}),
  \eprint{1303.5048}.

\bibitem[{\citenamefont{{van Engelen} et~al.}(2015)\citenamefont{{van Engelen},
  {Sherwin}, {Sehgal}, {Addison}, {Allison}, {Battaglia}, {de Bernardis},
  {Bond}, {Calabrese}, {Coughlin} et~al.}}]{1412.0626}
\bibinfo{author}{\bibfnamefont{A.}~\bibnamefont{{van Engelen}}},
  \bibinfo{author}{\bibfnamefont{B.~D.} \bibnamefont{{Sherwin}}},
  \bibinfo{author}{\bibfnamefont{N.}~\bibnamefont{{Sehgal}}},
  \bibinfo{author}{\bibfnamefont{G.~E.} \bibnamefont{{Addison}}},
  \bibinfo{author}{\bibfnamefont{R.}~\bibnamefont{{Allison}}},
  \bibinfo{author}{\bibfnamefont{N.}~\bibnamefont{{Battaglia}}},
  \bibinfo{author}{\bibfnamefont{F.}~\bibnamefont{{de Bernardis}}},
  \bibinfo{author}{\bibfnamefont{J.~R.} \bibnamefont{{Bond}}},
  \bibinfo{author}{\bibfnamefont{E.}~\bibnamefont{{Calabrese}}},
  \bibinfo{author}{\bibfnamefont{K.}~\bibnamefont{{Coughlin}}},
  \bibnamefont{et~al.}, \bibinfo{journal}{\apj} \textbf{\bibinfo{volume}{808}},
  \bibinfo{eid}{7} (\bibinfo{year}{2015}), \eprint{1412.0626}.

\bibitem[{\citenamefont{{Winkel} et~al.}(2010)\citenamefont{{Winkel}, {Kerp},
  {Kalberla}, and {Ben Bekhti}}}]{2010ASPC..438..381W}
\bibinfo{author}{\bibfnamefont{B.}~\bibnamefont{{Winkel}}},
  \bibinfo{author}{\bibfnamefont{J.}~\bibnamefont{{Kerp}}},
  \bibinfo{author}{\bibfnamefont{P.~M.~W.} \bibnamefont{{Kalberla}}},
  \bibnamefont{and} \bibinfo{author}{\bibfnamefont{N.}~\bibnamefont{{Ben
  Bekhti}}}, in \emph{\bibinfo{booktitle}{The Dynamic Interstellar Medium: A
  Celebration of the Canadian Galactic Plane Survey}}, edited by
  \bibinfo{editor}{\bibfnamefont{R.}~\bibnamefont{{Kothes}}},
  \bibinfo{editor}{\bibfnamefont{T.~L.} \bibnamefont{{Landecker}}},
  \bibnamefont{and} \bibinfo{editor}{\bibfnamefont{A.~G.}
  \bibnamefont{{Willis}}} (\bibinfo{year}{2010}), vol. \bibinfo{volume}{438} of
  \emph{\bibinfo{series}{Astronomical Society of the Pacific Conference
  Series}}, p. \bibinfo{pages}{381}, \eprint{1007.3363}.

\bibitem[{\citenamefont{{Kerp} et~al.}(2011)\citenamefont{{Kerp}, {Winkel},
  {Ben Bekhti}, {Fl{\"o}er}, and {Kalberla}}}]{2011AN....332..637K}
\bibinfo{author}{\bibfnamefont{J.}~\bibnamefont{{Kerp}}},
  \bibinfo{author}{\bibfnamefont{B.}~\bibnamefont{{Winkel}}},
  \bibinfo{author}{\bibfnamefont{N.}~\bibnamefont{{Ben Bekhti}}},
  \bibinfo{author}{\bibfnamefont{L.}~\bibnamefont{{Fl{\"o}er}}},
  \bibnamefont{and} \bibinfo{author}{\bibfnamefont{P.~M.~W.}
  \bibnamefont{{Kalberla}}}, \bibinfo{journal}{Astronomische Nachrichten}
  \textbf{\bibinfo{volume}{332}}, \bibinfo{pages}{637} (\bibinfo{year}{2011}),
  \eprint{1104.1185}.

\bibitem[{\citenamefont{{Winkel} et~al.}(2016)\citenamefont{{Winkel}, {Kerp},
  {Fl{\"o}er}, {Kalberla}, {Ben Bekhti}, {Keller}, and
  {Lenz}}}]{2016A&A...585A..41W}
\bibinfo{author}{\bibfnamefont{B.}~\bibnamefont{{Winkel}}},
  \bibinfo{author}{\bibfnamefont{J.}~\bibnamefont{{Kerp}}},
  \bibinfo{author}{\bibfnamefont{L.}~\bibnamefont{{Fl{\"o}er}}},
  \bibinfo{author}{\bibfnamefont{P.~M.~W.} \bibnamefont{{Kalberla}}},
  \bibinfo{author}{\bibfnamefont{N.}~\bibnamefont{{Ben Bekhti}}},
  \bibinfo{author}{\bibfnamefont{R.}~\bibnamefont{{Keller}}}, \bibnamefont{and}
  \bibinfo{author}{\bibfnamefont{D.}~\bibnamefont{{Lenz}}},
  \bibinfo{journal}{\aap} \textbf{\bibinfo{volume}{585}}, \bibinfo{eid}{A41}
  (\bibinfo{year}{2016}), \eprint{1512.05348}.

\bibitem[{\citenamefont{{McClure-Griffiths}
  et~al.}(2009)\citenamefont{{McClure-Griffiths}, {Pisano}, {Calabretta},
  {Ford}, {Lockman}, {Staveley-Smith}, {Kalberla}, {Bailin}, {Dedes},
  {Janowiecki} et~al.}}]{2009ApJS..181..398M}
\bibinfo{author}{\bibfnamefont{N.~M.} \bibnamefont{{McClure-Griffiths}}},
  \bibinfo{author}{\bibfnamefont{D.~J.} \bibnamefont{{Pisano}}},
  \bibinfo{author}{\bibfnamefont{M.~R.} \bibnamefont{{Calabretta}}},
  \bibinfo{author}{\bibfnamefont{H.~A.} \bibnamefont{{Ford}}},
  \bibinfo{author}{\bibfnamefont{F.~J.} \bibnamefont{{Lockman}}},
  \bibinfo{author}{\bibfnamefont{L.}~\bibnamefont{{Staveley-Smith}}},
  \bibinfo{author}{\bibfnamefont{P.~M.~W.} \bibnamefont{{Kalberla}}},
  \bibinfo{author}{\bibfnamefont{J.}~\bibnamefont{{Bailin}}},
  \bibinfo{author}{\bibfnamefont{L.}~\bibnamefont{{Dedes}}},
  \bibinfo{author}{\bibfnamefont{S.}~\bibnamefont{{Janowiecki}}},
  \bibnamefont{et~al.}, \bibinfo{journal}{\apjs}
  \textbf{\bibinfo{volume}{181}}, \bibinfo{pages}{398} (\bibinfo{year}{2009}),
  \eprint{0901.1159}.

\bibitem[{\citenamefont{{Kalberla} et~al.}(2010)\citenamefont{{Kalberla},
  {McClure-Griffiths}, {Pisano}, {Calabretta}, {Ford}, {Lockman},
  {Staveley-Smith}, {Kerp}, {Winkel}, {Murphy} et~al.}}]{2010A&A...521A..17K}
\bibinfo{author}{\bibfnamefont{P.~M.~W.} \bibnamefont{{Kalberla}}},
  \bibinfo{author}{\bibfnamefont{N.~M.} \bibnamefont{{McClure-Griffiths}}},
  \bibinfo{author}{\bibfnamefont{D.~J.} \bibnamefont{{Pisano}}},
  \bibinfo{author}{\bibfnamefont{M.~R.} \bibnamefont{{Calabretta}}},
  \bibinfo{author}{\bibfnamefont{H.~A.} \bibnamefont{{Ford}}},
  \bibinfo{author}{\bibfnamefont{F.~J.} \bibnamefont{{Lockman}}},
  \bibinfo{author}{\bibfnamefont{L.}~\bibnamefont{{Staveley-Smith}}},
  \bibinfo{author}{\bibfnamefont{J.}~\bibnamefont{{Kerp}}},
  \bibinfo{author}{\bibfnamefont{B.}~\bibnamefont{{Winkel}}},
  \bibinfo{author}{\bibfnamefont{T.}~\bibnamefont{{Murphy}}},
  \bibnamefont{et~al.}, \bibinfo{journal}{\aap} \textbf{\bibinfo{volume}{521}},
  \bibinfo{eid}{A17} (\bibinfo{year}{2010}), \eprint{1007.0686}.

\bibitem[{\citenamefont{{Kalberla} and {Haud}}(2015)}]{2015A&A...578A..78K}
\bibinfo{author}{\bibfnamefont{P.~M.~W.} \bibnamefont{{Kalberla}}}
  \bibnamefont{and} \bibinfo{author}{\bibfnamefont{U.}~\bibnamefont{{Haud}}},
  \bibinfo{journal}{\aap} \textbf{\bibinfo{volume}{578}}, \bibinfo{eid}{A78}
  (\bibinfo{year}{2015}), \eprint{1505.01011}.

\bibitem[{\citenamefont{{HI4PI Collaboration} et~al.}(2016)\citenamefont{{HI4PI
  Collaboration}, {Ben Bekhti}, {Fl{\"o}er}, {Keller}, {Kerp}, {Lenz},
  {Winkel}, {Bailin}, {Calabretta}, {Dedes} et~al.}}]{1610.06175}
\bibinfo{author}{\bibnamefont{{HI4PI Collaboration}}},
  \bibinfo{author}{\bibfnamefont{N.}~\bibnamefont{{Ben Bekhti}}},
  \bibinfo{author}{\bibfnamefont{L.}~\bibnamefont{{Fl{\"o}er}}},
  \bibinfo{author}{\bibfnamefont{R.}~\bibnamefont{{Keller}}},
  \bibinfo{author}{\bibfnamefont{J.}~\bibnamefont{{Kerp}}},
  \bibinfo{author}{\bibfnamefont{D.}~\bibnamefont{{Lenz}}},
  \bibinfo{author}{\bibfnamefont{B.}~\bibnamefont{{Winkel}}},
  \bibinfo{author}{\bibfnamefont{J.}~\bibnamefont{{Bailin}}},
  \bibinfo{author}{\bibfnamefont{M.~R.} \bibnamefont{{Calabretta}}},
  \bibinfo{author}{\bibfnamefont{L.}~\bibnamefont{{Dedes}}},
  \bibnamefont{et~al.}, \bibinfo{journal}{\aap} \textbf{\bibinfo{volume}{594}},
  \bibinfo{eid}{A116} (\bibinfo{year}{2016}), \eprint{1610.06175}.

\bibitem[{\citenamefont{{McCarthy} and {Madhavacheril}}(2021)}]{2010.16405}
\bibinfo{author}{\bibfnamefont{F.}~\bibnamefont{{McCarthy}}} \bibnamefont{and}
  \bibinfo{author}{\bibfnamefont{M.~S.} \bibnamefont{{Madhavacheril}}},
  \bibinfo{journal}{\prd} \textbf{\bibinfo{volume}{103}}, \bibinfo{eid}{103515}
  (\bibinfo{year}{2021}), \eprint{2010.16405}.

\bibitem[{\citenamefont{{Zonca} et~al.}(2019)\citenamefont{{Zonca}, {Singer},
  {Lenz}, {Reinecke}, {Rosset}, {Hivon}, and {Gorski}}}]{Zonca2019}
\bibinfo{author}{\bibfnamefont{A.}~\bibnamefont{{Zonca}}},
  \bibinfo{author}{\bibfnamefont{L.}~\bibnamefont{{Singer}}},
  \bibinfo{author}{\bibfnamefont{D.}~\bibnamefont{{Lenz}}},
  \bibinfo{author}{\bibfnamefont{M.}~\bibnamefont{{Reinecke}}},
  \bibinfo{author}{\bibfnamefont{C.}~\bibnamefont{{Rosset}}},
  \bibinfo{author}{\bibfnamefont{E.}~\bibnamefont{{Hivon}}}, \bibnamefont{and}
  \bibinfo{author}{\bibfnamefont{K.}~\bibnamefont{{Gorski}}},
  \bibinfo{journal}{The Journal of Open Source Software}
  \textbf{\bibinfo{volume}{4}}, \bibinfo{eid}{1298} (\bibinfo{year}{2019}).

\bibitem[{\citenamefont{{G{\'o}rski} et~al.}(2005)\citenamefont{{G{\'o}rski},
  {Hivon}, {Banday}, {Wandelt}, {Hansen}, {Reinecke}, and
  {Bartelmann}}}]{2005ApJ...622..759G}
\bibinfo{author}{\bibfnamefont{K.~M.} \bibnamefont{{G{\'o}rski}}},
  \bibinfo{author}{\bibfnamefont{E.}~\bibnamefont{{Hivon}}},
  \bibinfo{author}{\bibfnamefont{A.~J.} \bibnamefont{{Banday}}},
  \bibinfo{author}{\bibfnamefont{B.~D.} \bibnamefont{{Wandelt}}},
  \bibinfo{author}{\bibfnamefont{F.~K.} \bibnamefont{{Hansen}}},
  \bibinfo{author}{\bibfnamefont{M.}~\bibnamefont{{Reinecke}}},
  \bibnamefont{and}
  \bibinfo{author}{\bibfnamefont{M.}~\bibnamefont{{Bartelmann}}},
  \bibinfo{journal}{\apj} \textbf{\bibinfo{volume}{622}}, \bibinfo{pages}{759}
  (\bibinfo{year}{2005}), \eprint{arXiv:astro-ph/0409513}.

\bibitem[{\citenamefont{{Planck Collaboration}
  et~al.}(2014{\natexlab{d}})\citenamefont{{Planck Collaboration}, {Ade},
  {Aghanim}, {Armitage-Caplan}, {Arnaud}, {Ashdown}, {Atrio-Barandela},
  {Aumont}, {Baccigalupi}, {Banday} et~al.}}]{2014A&A...571A...6P}
\bibinfo{author}{\bibnamefont{{Planck Collaboration}}},
  \bibinfo{author}{\bibfnamefont{P.~A.~R.} \bibnamefont{{Ade}}},
  \bibinfo{author}{\bibfnamefont{N.}~\bibnamefont{{Aghanim}}},
  \bibinfo{author}{\bibfnamefont{C.}~\bibnamefont{{Armitage-Caplan}}},
  \bibinfo{author}{\bibfnamefont{M.}~\bibnamefont{{Arnaud}}},
  \bibinfo{author}{\bibfnamefont{M.}~\bibnamefont{{Ashdown}}},
  \bibinfo{author}{\bibfnamefont{F.}~\bibnamefont{{Atrio-Barandela}}},
  \bibinfo{author}{\bibfnamefont{J.}~\bibnamefont{{Aumont}}},
  \bibinfo{author}{\bibfnamefont{C.}~\bibnamefont{{Baccigalupi}}},
  \bibinfo{author}{\bibfnamefont{A.~J.} \bibnamefont{{Banday}}},
  \bibnamefont{et~al.}, \bibinfo{journal}{\aap} \textbf{\bibinfo{volume}{571}},
  \bibinfo{eid}{A6} (\bibinfo{year}{2014}{\natexlab{d}}), \eprint{1303.5067}.

\bibitem[{\citenamefont{{Dole} et~al.}(2006)\citenamefont{{Dole}, {Lagache},
  {Puget}, {Caputi}, {Fern{\'a}ndez-Conde}, {Le Floc'h}, {Papovich},
  {P{\'e}rez-Gonz{\'a}lez}, {Rieke}, and {Blaylock}}}]{2006A&A...451..417D}
\bibinfo{author}{\bibfnamefont{H.}~\bibnamefont{{Dole}}},
  \bibinfo{author}{\bibfnamefont{G.}~\bibnamefont{{Lagache}}},
  \bibinfo{author}{\bibfnamefont{J.~L.} \bibnamefont{{Puget}}},
  \bibinfo{author}{\bibfnamefont{K.~I.} \bibnamefont{{Caputi}}},
  \bibinfo{author}{\bibfnamefont{N.}~\bibnamefont{{Fern{\'a}ndez-Conde}}},
  \bibinfo{author}{\bibfnamefont{E.}~\bibnamefont{{Le Floc'h}}},
  \bibinfo{author}{\bibfnamefont{C.}~\bibnamefont{{Papovich}}},
  \bibinfo{author}{\bibfnamefont{P.~G.}
  \bibnamefont{{P{\'e}rez-Gonz{\'a}lez}}},
  \bibinfo{author}{\bibfnamefont{G.~H.} \bibnamefont{{Rieke}}},
  \bibnamefont{and}
  \bibinfo{author}{\bibfnamefont{M.}~\bibnamefont{{Blaylock}}},
  \bibinfo{journal}{\aap} \textbf{\bibinfo{volume}{451}}, \bibinfo{pages}{417}
  (\bibinfo{year}{2006}), \eprint{astro-ph/0603208}.

\bibitem[{\citenamefont{{B{\'e}thermin}
  et~al.}(2012)\citenamefont{{B{\'e}thermin}, {Le Floc'h}, {Ilbert}, {Conley},
  {Lagache}, {Amblard}, {Arumugam}, {Aussel}, {Berta}, {Bock}
  et~al.}}]{bethermin_numbercounts}
\bibinfo{author}{\bibfnamefont{M.}~\bibnamefont{{B{\'e}thermin}}},
  \bibinfo{author}{\bibfnamefont{E.}~\bibnamefont{{Le Floc'h}}},
  \bibinfo{author}{\bibfnamefont{O.}~\bibnamefont{{Ilbert}}},
  \bibinfo{author}{\bibfnamefont{A.}~\bibnamefont{{Conley}}},
  \bibinfo{author}{\bibfnamefont{G.}~\bibnamefont{{Lagache}}},
  \bibinfo{author}{\bibfnamefont{A.}~\bibnamefont{{Amblard}}},
  \bibinfo{author}{\bibfnamefont{V.}~\bibnamefont{{Arumugam}}},
  \bibinfo{author}{\bibfnamefont{H.}~\bibnamefont{{Aussel}}},
  \bibinfo{author}{\bibfnamefont{S.}~\bibnamefont{{Berta}}},
  \bibinfo{author}{\bibfnamefont{J.}~\bibnamefont{{Bock}}},
  \bibnamefont{et~al.}, \bibinfo{journal}{\aap} \textbf{\bibinfo{volume}{542}},
  \bibinfo{eid}{A58} (\bibinfo{year}{2012}), \eprint{1203.1925}.

\bibitem[{\citenamefont{{Alonso} et~al.}(2019)\citenamefont{{Alonso},
  {Sanchez}, {Slosar}, and {LSST Dark Energy Science
  Collaboration}}}]{1809.09603}
\bibinfo{author}{\bibfnamefont{D.}~\bibnamefont{{Alonso}}},
  \bibinfo{author}{\bibfnamefont{J.}~\bibnamefont{{Sanchez}}},
  \bibinfo{author}{\bibfnamefont{A.}~\bibnamefont{{Slosar}}}, \bibnamefont{and}
  \bibinfo{author}{\bibnamefont{{LSST Dark Energy Science Collaboration}}},
  \bibinfo{journal}{\mnras} \textbf{\bibinfo{volume}{484}},
  \bibinfo{pages}{4127} (\bibinfo{year}{2019}), \eprint{1809.09603}.

\bibitem[{\citenamefont{{Torrado} and {Lewis}}(2019)}]{2019ascl.soft10019T}
\bibinfo{author}{\bibfnamefont{J.}~\bibnamefont{{Torrado}}} \bibnamefont{and}
  \bibinfo{author}{\bibfnamefont{A.}~\bibnamefont{{Lewis}}},
  \emph{\bibinfo{title}{{Cobaya: Bayesian analysis in cosmology}}}
  (\bibinfo{year}{2019}), \eprint{1910.019}.

\bibitem[{\citenamefont{{Torrado} and {Lewis}}(2021)}]{Torrado:2020dgo}
\bibinfo{author}{\bibfnamefont{J.}~\bibnamefont{{Torrado}}} \bibnamefont{and}
  \bibinfo{author}{\bibfnamefont{A.}~\bibnamefont{{Lewis}}},
  \bibinfo{journal}{\jcap} \textbf{\bibinfo{volume}{2021}}, \bibinfo{eid}{057}
  (\bibinfo{year}{2021}), \eprint{2005.05290}.

\bibitem[{\citenamefont{{Gelman} and {Rubin}}(1992)}]{1992StaSc...7..457G}
\bibinfo{author}{\bibfnamefont{A.}~\bibnamefont{{Gelman}}} \bibnamefont{and}
  \bibinfo{author}{\bibfnamefont{D.~B.} \bibnamefont{{Rubin}}},
  \bibinfo{journal}{Statistical Science} \textbf{\bibinfo{volume}{7}},
  \bibinfo{pages}{457} (\bibinfo{year}{1992}).

\bibitem[{\citenamefont{{Abazajian} et~al.}(2019)\citenamefont{{Abazajian},
  {Addison}, {Adshead}, {Ahmed}, {Allen}, {Alonso}, {Alvarez}, {Anderson},
  {Arnold}, {Baccigalupi} et~al.}}]{CMBS4DSR}
\bibinfo{author}{\bibfnamefont{K.}~\bibnamefont{{Abazajian}}},
  \bibinfo{author}{\bibfnamefont{G.}~\bibnamefont{{Addison}}},
  \bibinfo{author}{\bibfnamefont{P.}~\bibnamefont{{Adshead}}},
  \bibinfo{author}{\bibfnamefont{Z.}~\bibnamefont{{Ahmed}}},
  \bibinfo{author}{\bibfnamefont{S.~W.} \bibnamefont{{Allen}}},
  \bibinfo{author}{\bibfnamefont{D.}~\bibnamefont{{Alonso}}},
  \bibinfo{author}{\bibfnamefont{M.}~\bibnamefont{{Alvarez}}},
  \bibinfo{author}{\bibfnamefont{A.}~\bibnamefont{{Anderson}}},
  \bibinfo{author}{\bibfnamefont{K.~S.} \bibnamefont{{Arnold}}},
  \bibinfo{author}{\bibfnamefont{C.}~\bibnamefont{{Baccigalupi}}},
  \bibnamefont{et~al.}, \bibinfo{journal}{arXiv e-prints}
  \bibinfo{eid}{arXiv:1907.04473} (\bibinfo{year}{2019}), \eprint{1907.04473}.

\bibitem[{\citenamefont{{Maniyar}
  et~al.}(2021{\natexlab{b}})\citenamefont{{Maniyar}, {Ali-Ha{\"\i}moud},
  {Carron}, {Lewis}, and {Madhavacheril}}}]{Maniyar_2021}
\bibinfo{author}{\bibfnamefont{A.~S.} \bibnamefont{{Maniyar}}},
  \bibinfo{author}{\bibfnamefont{Y.}~\bibnamefont{{Ali-Ha{\"\i}moud}}},
  \bibinfo{author}{\bibfnamefont{J.}~\bibnamefont{{Carron}}},
  \bibinfo{author}{\bibfnamefont{A.}~\bibnamefont{{Lewis}}}, \bibnamefont{and}
  \bibinfo{author}{\bibfnamefont{M.~S.} \bibnamefont{{Madhavacheril}}},
  \bibinfo{journal}{\prd} \textbf{\bibinfo{volume}{103}}, \bibinfo{eid}{083524}
  (\bibinfo{year}{2021}{\natexlab{b}}), \eprint{2101.12193}.

\bibitem[{\citenamefont{{Carron}
  et~al.}(2022{\natexlab{b}})\citenamefont{{Carron}, {Mirmelstein}, and
  {Lewis}}}]{2022JCAP...09..039C}
\bibinfo{author}{\bibfnamefont{J.}~\bibnamefont{{Carron}}},
  \bibinfo{author}{\bibfnamefont{M.}~\bibnamefont{{Mirmelstein}}},
  \bibnamefont{and} \bibinfo{author}{\bibfnamefont{A.}~\bibnamefont{{Lewis}}},
  \bibinfo{journal}{\jcap} \textbf{\bibinfo{volume}{2022}}, \bibinfo{eid}{039}
  (\bibinfo{year}{2022}{\natexlab{b}}), \eprint{2206.07773}.

\bibitem[{\citenamefont{{Lenz} et~al.}(2019{\natexlab{b}})\citenamefont{{Lenz},
  {Dor{\'e}}, and {Lagache}}}]{Lenz:2019ugy}
\bibinfo{author}{\bibfnamefont{D.}~\bibnamefont{{Lenz}}},
  \bibinfo{author}{\bibfnamefont{O.}~\bibnamefont{{Dor{\'e}}}},
  \bibnamefont{and}
  \bibinfo{author}{\bibfnamefont{G.}~\bibnamefont{{Lagache}}},
  \bibinfo{journal}{\apj} \textbf{\bibinfo{volume}{883}}, \bibinfo{eid}{75}
  (\bibinfo{year}{2019}{\natexlab{b}}), \eprint{1905.00426}.

\bibitem[{\citenamefont{{Mueller} et~al.}(2021)\citenamefont{{Mueller},
  {Rezaie}, {Percival}, {Ross}, {Ruggeri}, {Seo}, {Gil-Mar{\i}n}, {Bautista},
  {Brownstein}, {Dawson} et~al.}}]{2021arXiv210613725M}
\bibinfo{author}{\bibfnamefont{E.-M.} \bibnamefont{{Mueller}}},
  \bibinfo{author}{\bibfnamefont{M.}~\bibnamefont{{Rezaie}}},
  \bibinfo{author}{\bibfnamefont{W.~J.} \bibnamefont{{Percival}}},
  \bibinfo{author}{\bibfnamefont{A.~J.} \bibnamefont{{Ross}}},
  \bibinfo{author}{\bibfnamefont{R.}~\bibnamefont{{Ruggeri}}},
  \bibinfo{author}{\bibfnamefont{H.-J.} \bibnamefont{{Seo}}},
  \bibinfo{author}{\bibfnamefont{H.}~\bibnamefont{{Gil-Mar{\i}n}}},
  \bibinfo{author}{\bibfnamefont{J.}~\bibnamefont{{Bautista}}},
  \bibinfo{author}{\bibfnamefont{J.~R.} \bibnamefont{{Brownstein}}},
  \bibinfo{author}{\bibfnamefont{K.}~\bibnamefont{{Dawson}}},
  \bibnamefont{et~al.}, \bibinfo{journal}{arXiv e-prints}
  \bibinfo{eid}{arXiv:2106.13725} (\bibinfo{year}{2021}), \eprint{2106.13725}.

\bibitem[{\citenamefont{{Barreira} et~al.}(2020)\citenamefont{{Barreira},
  {Cabass}, {Schmidt}, {Pillepich}, and {Nelson}}}]{2006.09368}
\bibinfo{author}{\bibfnamefont{A.}~\bibnamefont{{Barreira}}},
  \bibinfo{author}{\bibfnamefont{G.}~\bibnamefont{{Cabass}}},
  \bibinfo{author}{\bibfnamefont{F.}~\bibnamefont{{Schmidt}}},
  \bibinfo{author}{\bibfnamefont{A.}~\bibnamefont{{Pillepich}}},
  \bibnamefont{and} \bibinfo{author}{\bibfnamefont{D.}~\bibnamefont{{Nelson}}},
  \bibinfo{journal}{\jcap} \textbf{\bibinfo{volume}{2020}}, \bibinfo{eid}{013}
  (\bibinfo{year}{2020}), \eprint{2006.09368}.

\bibitem[{\citenamefont{{Barreira}}(2020)}]{2009.06622}
\bibinfo{author}{\bibfnamefont{A.}~\bibnamefont{{Barreira}}},
  \bibinfo{journal}{\jcap} \textbf{\bibinfo{volume}{2020}}, \bibinfo{eid}{031}
  (\bibinfo{year}{2020}), \eprint{2009.06622}.

\bibitem[{\citenamefont{{Barreira}}(2022{\natexlab{a}})}]{2107.06887}
\bibinfo{author}{\bibfnamefont{A.}~\bibnamefont{{Barreira}}},
  \bibinfo{journal}{\jcap} \textbf{\bibinfo{volume}{2022}}, \bibinfo{eid}{033}
  (\bibinfo{year}{2022}{\natexlab{a}}), \eprint{2107.06887}.

\bibitem[{\citenamefont{{Barreira}}(2022{\natexlab{b}})}]{2022arXiv220505673B}
\bibinfo{author}{\bibfnamefont{A.}~\bibnamefont{{Barreira}}},
  \bibinfo{journal}{arXiv e-prints} \bibinfo{eid}{arXiv:2205.05673}
  (\bibinfo{year}{2022}{\natexlab{b}}), \eprint{2205.05673}.

\bibitem[{\citenamefont{{Lazeyras} et~al.}(2022)\citenamefont{{Lazeyras},
  {Barreira}, {Schmidt}, and {Desjacques}}}]{2209.07251}
\bibinfo{author}{\bibfnamefont{T.}~\bibnamefont{{Lazeyras}}},
  \bibinfo{author}{\bibfnamefont{A.}~\bibnamefont{{Barreira}}},
  \bibinfo{author}{\bibfnamefont{F.}~\bibnamefont{{Schmidt}}},
  \bibnamefont{and}
  \bibinfo{author}{\bibfnamefont{V.}~\bibnamefont{{Desjacques}}},
  \bibinfo{journal}{arXiv e-prints} \bibinfo{eid}{arXiv:2209.07251}
  (\bibinfo{year}{2022}), \eprint{2209.07251}.

\bibitem[{\citenamefont{{Reid} et~al.}(2010)\citenamefont{{Reid}, {Verde},
  {Dolag}, {Matarrese}, and {Moscardini}}}]{1004.1637}
\bibinfo{author}{\bibfnamefont{B.~A.} \bibnamefont{{Reid}}},
  \bibinfo{author}{\bibfnamefont{L.}~\bibnamefont{{Verde}}},
  \bibinfo{author}{\bibfnamefont{K.}~\bibnamefont{{Dolag}}},
  \bibinfo{author}{\bibfnamefont{S.}~\bibnamefont{{Matarrese}}},
  \bibnamefont{and}
  \bibinfo{author}{\bibfnamefont{L.}~\bibnamefont{{Moscardini}}},
  \bibinfo{journal}{\jcap} \textbf{\bibinfo{volume}{2010}}, \bibinfo{eid}{013}
  (\bibinfo{year}{2010}), \eprint{1004.1637}.

\end{thebibliography}

\end{document}